\pdfoutput=1
\documentclass{article}

\usepackage{arxiv}

\usepackage[utf8]{inputenc} 
\usepackage[T1]{fontenc}    
\usepackage{url}            
\usepackage{booktabs}       
\usepackage{amsfonts,amsmath,amssymb,amsbsy,latexsym,amsthm}       
\usepackage{nicefrac}       
\usepackage{microtype}      
\usepackage{lipsum}
\usepackage{graphicx}
 \usepackage[table]{xcolor}
\usepackage[flushleft]{threeparttable}
\usepackage{xr-hyper}
\usepackage[bookmarks=false]{hyperref}
\usepackage{import}
\usepackage[squaren,Gray]{SIunits}

\usepackage{graphicx}
\graphicspath{{Figures/}}
\graphicspath{{SuppFigures/}}
\usepackage{caption}
\usepackage{tikz,pgfplots}
\pgfplotsset{
        colormap={parula}{%
        rgb=(0.2081,0.1663,0.5292)rgb=(0.2116,0.1898,0.5777)rgb=(0.2123,0.2138,0.627)
        rgb=(0.2081,0.2386,0.6771)rgb=(0.1959,0.2645,0.7279)rgb=(0.1707,0.2919,0.7792)
        rgb=(0.1253,0.3242,0.8303)rgb=(0.0591,0.3598,0.8683)rgb=(0.0117,0.3875,0.882)
        rgb=(0.006,0.4086,0.8828) rgb=(0.0165,0.4266,0.8786)rgb=(0.0329,0.443,0.872)
        rgb=(0.0498,0.4586,0.8641)rgb=(0.0629,0.4737,0.8554)rgb=(0.0723,0.4887,0.8467)
        rgb=(0.0779,0.504,0.8384) rgb=(0.0793,0.52,0.8312)  rgb=(0.0749,0.5375,0.8263)
        rgb=(0.0641,0.557,0.824)  rgb=(0.0488,0.5772,0.8228)rgb=(0.0343,0.5966,0.8199)
        rgb=(0.0265,0.6137,0.8135)rgb=(0.0239,0.6287,0.8038)rgb=(0.0231,0.6418,0.7913)
        rgb=(0.0228,0.6535,0.7768)rgb=(0.0267,0.6642,0.7607)rgb=(0.0384,0.6743,0.7436)
        rgb=(0.059,0.6838,0.7254) rgb=(0.0843,0.6928,0.7062)rgb=(0.1133,0.7015,0.6859)
        rgb=(0.1453,0.7098,0.6646)rgb=(0.1801,0.7177,0.6424)rgb=(0.2178,0.725,0.6193)
        rgb=(0.2586,0.7317,0.5954)rgb=(0.3022,0.7376,0.5712)rgb=(0.3482,0.7424,0.5473)
        rgb=(0.3953,0.7459,0.5244)rgb=(0.442,0.7481,0.5033) rgb=(0.4871,0.7491,0.484)
        rgb=(0.53,0.7491,0.4661)  rgb=(0.5709,0.7485,0.4494)rgb=(0.6099,0.7473,0.4337)
        rgb=(0.6473,0.7456,0.4188)rgb=(0.6834,0.7435,0.4044)rgb=(0.7184,0.7411,0.3905)
        rgb=(0.7525,0.7384,0.3768)rgb=(0.7858,0.7356,0.3633)rgb=(0.8185,0.7327,0.3498)
        rgb=(0.8507,0.7299,0.336) rgb=(0.8824,0.7274,0.3217)rgb=(0.9139,0.7258,0.3063)
        rgb=(0.945,0.7261,0.2886) rgb=(0.9739,0.7314,0.2666)rgb=(0.9938,0.7455,0.2403)
        rgb=(0.999,0.7653,0.2164) rgb=(0.9955,0.7861,0.1967)rgb=(0.988,0.8066,0.1794)
        rgb=(0.9789,0.8271,0.1633)rgb=(0.9697,0.8481,0.1475)rgb=(0.9626,0.8705,0.1309)
        rgb=(0.9589,0.8949,0.1132)rgb=(0.9598,0.9218,0.0948)rgb=(0.9661,0.9514,0.0755)
        rgb=(0.9763,0.9831,0.0538)
        }
,
        colormap/parula/.style={
        colormap name=parula}
    }

\usepgfplotslibrary{groupplots}
\usetikzlibrary{intersections,calc,arrows,matrix,spy,math,calc,fit}

\usetikzlibrary{arrows.meta}
\usepackage{color}
\definecolor{darkred}{rgb}{0.6,0,0}
\definecolor{darkgreen}{rgb}{0,0.5,0}
\definecolor{darkblue}{rgb}{0,0,0.5}
\pgfplotsset{compat=newest}
\usepackage{bm}

\usepackage[skins]{tcolorbox}

\newcommand{\R}{\mathbb{R}}
\newcommand{\Z}{\mathbb{Z}}
\newcommand{\btheta}{\boldsymbol{\theta}}

\makeatletter
\newcommand*{\addFileDependency}[1]{
  \typeout{(#1)}
  \@addtofilelist{#1}
  \IfFileExists{#1}{}{\typeout{No file #1.}}
}

\definecolor{urlblue}{RGB}{46,46,177}

\newcommand\addplotgraphicsnatural[2][]{%
    \begingroup
    \pgfqkeys{/pgfplots/plot graphics}{#1}%
    \setbox0=\hbox{\includegraphics{#2}}%
    \pgfmathparse{\wd0/(\pgfkeysvalueof{/pgfplots/plot graphics/xmax} - \pgfkeysvalueof{/pgfplots/plot graphics/xmin})}%
    \let\xunit=\pgfmathresult
    \pgfmathparse{\ht0/(\pgfkeysvalueof{/pgfplots/plot graphics/ymax} - \pgfkeysvalueof{/pgfplots/plot graphics/ymin})}%
    \let\yunit=\pgfmathresult
    \xdef\marshal{%
        \noexpand\pgfplotsset{unit vector ratio={\xunit\space \yunit}}%
    }%
    \endgroup
    \marshal
    \addplot graphics[#1] {#2};
}   

\makeatother

\newcommand*{\myexternaldocument}[1]{%
    \externaldocument{#1}%
    \addFileDependency{#1.tex}%
    \addFileDependency{#1.aux}%
}
\myexternaldocument{PUDIP-suppl}

\title{Robust Phase Unwrapping via Deep Image Prior for Quantitative Phase Imaging}

\author{
 Fangshu Yang \\
  School of Mathematics\\
  Harbin Institute of Technology\\
  Biomedical Imaging Group, EPFL\\
  \texttt{yfs2016@hit.edu.cn} \\
   \And
 Thanh-an Pham \\
  Biomedical Imaging Group, EPFL\\
  \texttt{thanh-an.pham@epfl.ch} \\
  \And
 Nathalie Brandenberg \\
  Laboratory of Stem Cell Bioengineering, EPFL\\
  \texttt{nathalie.brandenberg@epfl.ch} \\
  \And
  Matthias P. Lutolf \\
  Laboratory of Stem Cell Bioengineering, EPFL\\
  \texttt{matthias.lutolf@epfl.ch} \\
  \And
  Jianwei Ma \\
  School of Mathematics\\
  Harbin Institute of Technology\\
  \texttt{jma@hit.edu.cn} \\
  \And
  Michael Unser \\
  Biomedical Imaging Group, EPFL\\
  \texttt{michael.unser@epfl.ch} \\%
}

\begin{document}
\maketitle
\begin{abstract}
Quantitative phase imaging (QPI) is an emerging label-free technique that produces images containing morphological and dynamical information without contrast agents.
Unfortunately, the phase is wrapped in most imaging system.
Phase unwrapping is the computational process that recovers a more informative image.
It is particularly challenging with thick and complex samples such as organoids.
Recent works that rely on supervised training show that deep learning is a powerful method to unwrap the phase; however, supervised approaches 
require large and representative datasets which are difficult to obtain for complex biological samples.
Inspired by the concept of deep image priors, we propose a deep-learning-based method that does not need any training set. Our framework relies on an untrained convolutional neural network to accurately unwrap the phase while ensuring the consistency of the measurements.
We experimentally demonstrate that the proposed method faithfully recovers the phase of complex samples on both real and simulated data.
Our work paves the way to reliable phase imaging of thick and complex samples with QPI.
\end{abstract}


\section{Introduction}
In recent years, three-dimensional stem-cell cultures, called organoids, have emerged as an ideal \textit{ex vivo} model in regenerative medicine, disease modeling, and studies of biological tissues~\cite{rios2018imaging, rossi2018progress}.
For such samples, the privileged imaging modalities are fluorescence-based techniques~\cite{rios2018imaging}.
Recent works have shown that quantitative phase imaging~(QPI)~\cite{mir2012quantitative} can be used to complement fluorescence-based techniques~\cite{park2006diffraction,pavillon2010cell,quan2015phase,chowdhury2015spatial,park2018quantitative} or to monitor the rates of growth and morphological changes over an extended period of time~\cite{park2018quantitative,hu2019quantitative}.
The abundant literature as well as the existence of commercial modules 
suggest that QPI and multimodal imaging are mature and relevant approaches to study biological samples.
In practice, the measured phase suffers from wrapping~($i.e.,$ modulo $2\pi$ of the original phase),
which introduces non-representative discontinuities in its distribution.
Once recovered from the measurements,
the unwrapped version provides quantitative information on the sample~\cite{ghiglia1998two}.
This process, known as phase unwrapping, is an important step for phase imaging.
However, its application to biological specimens such as organoids is challenging; in particular, 
the advent of thick and complex samples calls for advanced methods. Classical methods, largely optimized for the analysis of two-dimensional~(2D) samples, exhibit important unwrapping artifacts and thus remain challenging to use reliably for these complex samples (see Fig.~\ref{fig:illus}).
In this work, we propose a method with untrained convolutional neural networks to solve this challenging task.

\newcommand{\szx}{350}
\newcommand{\szy}{200}
\newcommand{\szsub}{0.18}
\newcommand{\szax}{5*\szsub}
\newcommand{\foldoi}{RealRK_C05_12_12_SingleComparison_200_350}
\newcommand{\vmin}{-12.12}
\newcommand{\vmax}{17.37}
\pgfmathsetmacro{\step}{\vmin + (\vmax - \vmin)/4}
\pgfmathsetmacro{\piv}{3.14159265359}
\newcommand{\zoomx}{245}
\newcommand{\zoomy}{\szy - 25}
\newcommand{\insep}{5} 
\newcommand{\insepspy}{0.1}

\newcommand{\szxp}{\szx}

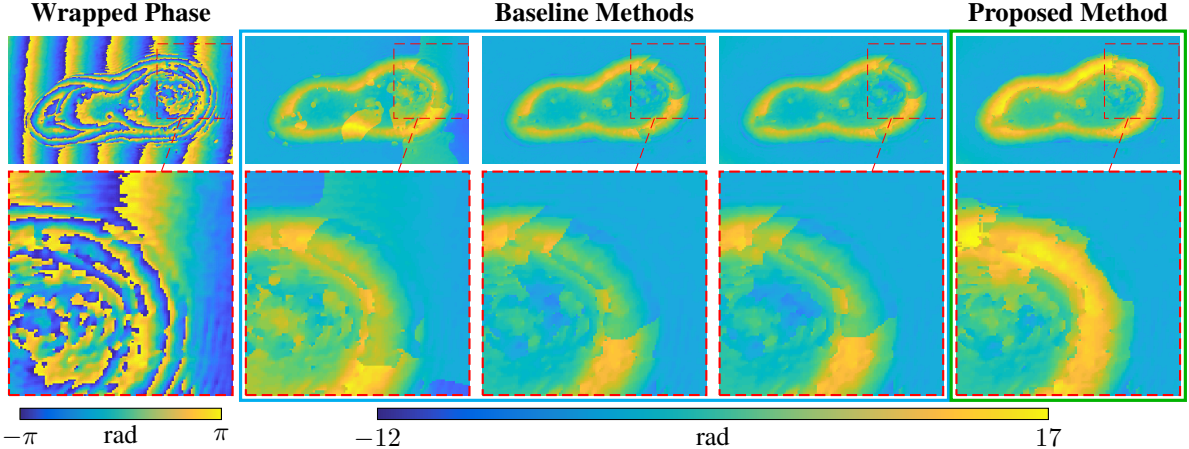
\begin{figure*}[!tbp]
\centering
\begin{tikzpicture}
\begin{scope}[spy using outlines={rectangle,magnification=3,width={\szsub\textwidth},height={\szsub\textwidth},
connect spies,line width=4pt,dotted,inner sep = 2*\insep},
]
\begin{groupplot}[
     group style = {group size = 5 by 1,
     vertical sep=\insep, horizontal sep=\insep},
     xmin = 0,xmax = \szx, ymin = 0, ymax = \szy,
    enlargelimits=false,
    axis equal image,
    scale only axis,
    width = {\szsub*\textwidth},
    hide axis,
    title style = {anchor=base,yshift=0em}  
    ]
    \nextgroupplot[title = \textbf{Wrapped Phase}]
    \addplot graphics[xmin = 0,xmax = \szx, ymin = 0, ymax = \szy] {{Figures/\foldoi/wrapPhase}};
    \coordinate (WP) at (\szx/2,0);
    \coordinate (WPz) at (\zoomx + 45,\zoomy - 45);
    
    \nextgroupplot 
    \addplot graphics[xmin = 0,xmax = \szx, ymin = 0, ymax = \szy] {{Figures/\foldoi/LS}};
    
    \coordinate (LS) at (\szx/2,0);
    \coordinate (LSz) at (\zoomx + 45,\zoomy-45);
 
 \nextgroupplot[title={\textbf{Baseline Methods}}]
 \addplot graphics[xmin = 0,xmax = \szx, ymin = 0, ymax = \szy]{{Figures/\foldoi/IRTV}};
    \coordinate (IRTV) at (\szx/2,0);    
    \coordinate (IRTVz) at (\zoomx + 45,\zoomy-45);
    
    \nextgroupplot 
    \addplot graphics[xmin = 0,xmax = \szx, ymin = 0, ymax = \szy]{Figures/\foldoi/PUMA};
    \coordinate (PUMA) at (\szx/2,0);
    \coordinate (PUMAz) at (\zoomx + 45,\zoomy-45);
    \coordinate (PUMAtr) at (\szx,\szy);
    
    \nextgroupplot[title={\textbf{Proposed Method}}]
    \addplot graphics[xmin = 0,xmax = \szx, ymin = 0, ymax = \szy]{Figures/\foldoi/DIP};
    \coordinate (DIP) at (\szx/2,0);
    \coordinate (DIPz) at (\zoomx + 45,\zoomy-45);
    \coordinate (DIPtr) at (\szx,\szy);
    \end{groupplot}
    
    \spy[red] on (WPz) in node (a) [below] at ($(WP.south) - (0,\insepspy)$);
    \spy[red] on (LSz) in node (LSi) [below] at ($(LS.south) - (0,\insepspy)$);
    \spy[red] on (IRTVz) in node (b) [below] at ($(IRTV.south) - (0,\insepspy)$);
    \spy[red] on (PUMAz) in node (c) [below] at ($(PUMA.south) - (0,\insepspy)$);
    \spy[red] on (DIPz) in node (DIPi) [below] at ($(DIP.south) - (0,\insepspy)$);
    \end{scope}
   
   \draw[cyan, very thick] ($(LSi.south west) - (0.2em,0.2em)$) rectangle ($(PUMAtr.north east) + (0.2em,0.2em)$);
   
   \draw[black!30!green, very thick] ($(DIPi.south west) - (0.2em,0.2em)$) rectangle ($(DIPtr.north east) + (0.2em,0.2em)$);
    
    \begin{axis}[at = {($(a.south) - (0,0.5em)$)}, 
    anchor = north,
    xmin = {-\piv}, xmax = \piv, ymin = 0, ymax = 0,
    width = 0pt, 
    height = 0pt,
    scale only axis,
    enlargelimits = false,
    hide axis,
    axis equal image,
    colorbar horizontal,
    colormap/parula,
    colorbar style={
        at = {(0,1)},
        anchor = {north},
        xticklabel pos = lower,
        point meta min = {-\piv},
        point meta max = {\piv},
        scale only axis,
        enlargelimits = false,
        scaled x ticks = true,
        width = {0.9*\szsub\textwidth},
        samples = 200,
        height = 0.15cm,
        xticklabel style={/pgf/number format/fixed, /pgf/number format/precision=0},
        xtick = {-\piv,\piv},
        xticklabels={$-\pi$,$\pi$},
        xlabel={rad},
        xlabel style={xshift=0cm,yshift=1.5em}
            }]
    \end{axis}
    
    \begin{axis}[at = {($0.5*(b.south) + 0.5*(c.south) - (0,0.5em)$)}, 
    anchor = north,
    xmin = \vmin, xmax = \vmax, ymin = 0, ymax = 0,
    width=0pt, 
    height=0pt,
    scale only axis,
    enlargelimits=false,
    hide axis,
    axis equal image,
    colorbar horizontal,
    colormap/parula,
    colorbar style={
        at = {(0,1)},
        anchor = {north},
        xticklabel pos = lower,
        xmin = \vmin, xmax = \vmax,
        point meta min = \vmin,
        point meta max = \vmax,
        scale only axis,
        enlargelimits = false,
        scaled x ticks = true,
        width = {3*\szsub\textwidth},
        samples = 200,
        height = 0.15cm,
        xticklabel style={/pgf/number format/fixed, /pgf/number format/precision=0},
        xtick = {\vmin,\vmax},
        xlabel={rad},
        xlabel style={xshift=0cm,yshift=1.5em},
        }]
    \end{axis}
\end{tikzpicture}
\caption{Example of phase image of organoids. First column: measured (wrapped) phase image. Second to fifth columns: baseline methods (LS, IRTV, and PUMA) and the proposed method (PUDIP). First row: reconstructed phase. Second row: zoomed inset.
The size of the unwrapped phase image is ($200 \times 350$).}
\label{fig:illus}
\end{figure*}

\subsection{Classical Methods}

In the past decades, numerous 2D phase-unwrapping algorithms have been proposed.
These approaches generally fall into four categories: path following~\cite{goldstein1988satellite,su2004reliability}, minimum $L_{p}$-norm~\cite{ghiglia1996minimum,ghiglia1994robust,he2014sparse}, Bayesian/regularization~\cite{nico2000bayesian,ying2005unwrapping}, and parametric modeling~\cite{liang1996model}.

Most of the path-following algorithms perform a line integration along some path established by techniques such as
the branch-cut algorithm~\cite{goldstein1988satellite}. 
Generally, the path-following methods encounter issues of consistency as the resulting unwrapped phase depends on the path.

By contrast, the minimum-norm methods are global. They estimate the unwrapped phase by minimizing an $L_{p}$-norm.
When $p=2$ (least-squares methods)~\cite{takajo1988least}, there exist approximate solutions which can be obtained by fast Fourier transforms or discrete cosine transforms~\cite{ghiglia1994robust}. However, the $L_{2}$-norm tends to smooth image edges, especially at the discontinuities~\cite{ghiglia1996minimum}. 
The drawback associated to $p=2$ can be overcome by setting $0 \leq p \leq 1$, 
which usually increases the computational cost.
Bioucas-Dias and Valadao~\cite{bioucas2007phase} introduced a specific energy-minimization framework for phase unwrapping that is solved via graph-cut optimization~(PUMA).
Recent works have extended this method for other imaging modalities~\cite{zhang2018precise,zhou2019extended}.
In~\cite{kamilov2015isotropic}, the authors describe a weighted energy function combined with a Hessian-Schatten-norm regularization~\cite{lefkimmiatis2013hessian}. They optimize the minimization problem with an iterative algorithm~(IRTV) based on the alternating direction method of multipliers~(ADMM)~\cite{boyd2011distributed}.

Bayesian approaches take into account a data-acquisition model and prior knowledge on the phase.

The parametric-modeling algorithms constrain the unwrapped phase to a parametric surface, usually a low-order polynomial. These approaches yield excellent performance only if the parametric model accurately represents the true phase.

Importantly, an assumption considered by most phase-unwrapping approaches is that the absolute value of the unwrapped phase difference between neighboring pixels is less than $\pi$, the so-called Itoh condition~\cite{itoh1982analysis}.

It is worthy to note that there exist alternative methods for quantitative phase-imaging methods that rely on multiple wavelengths or broadband sources~\cite{bhaduri2012diffraction,wang2011spatial,mann2008quantitative,li2014multiple}.
An imaging system with multiple wavelength sources typically acquires several images so that the wrapping events occur at different locations, thus facilitating the unwrapping task.
While our work mainly focuses on a single-wavelength source,
our proposed framework can be adapted to the multi-wavelength setting.
We refer to the recent reviews on QPI and their detailed description~\cite{park2018quantitative,hu2019neuro}.

\subsection{Deep-Learning-Based Approaches}

Recently, deep-learning methods, in particular, convolutional neural networks~(CNN), have achieved unprecedented performance in a variety of applications.
They surpass conventional methods in diverse fields such as image reconstruction~\cite{zhu2018image,mccann2017convolutional}, superresolution~\cite{wang2019deep}, x-ray computed tomography~\cite{jin2017deep}, and others~\cite{yan2020deep,yan20203d}.
Overall, deep learning in computational imaging is an emerging and promising field of research~\cite{jo2018quantitative,barbastathis2019use}.

To address the 2D phase-unwrapping problem, several works based on deep learning have been proposed. In~\cite{schwartzkopf2000two}, the authors used a supervised feedforward multilayer perceptron to detect the phase discontinuities in optical Doppler tomography images.
More recently, a residual neural network using supervised learning~\cite{he2016deep} was adopted in~\cite{dardikman2018phase} to approximate the mapping between the wrapped and the unwrapped phase in the presence of steep gradients. In~\cite{spoorthi2018phasenet}, a CNN-based framework, termed PhaseNet, has been designed. It predicts the wrap-count (integer multiple of $2\pi$) at each pixel, similar to the task of semantic segmentation. Furthermore, a clustering-based postprocessing enforces smoothness by incorporating complementary information.
Similar works were proposed in~\cite{zhang2019rapid,wang2019one}.  In~\cite{zhang2019phase}, the authors improved upon~\cite{spoorthi2018phasenet} by integrating a network to denoise the noisy wrapped phase. In~\cite{li2019method}, a generative adversarial network was introduced to effectively suppress the influence of noise. In addition, a framework composed of a residual neural network and the objective function in~\cite{kamilov2015isotropic} was proposed in~\cite{dardikman2020phun} to unwrap quantitative phase images of biological cells.

The aforementioned works rely on supervised learning to learn the mapping between the input-output data pairs.
This paradigm needs a large representative training dataset composed of the measured phase and the corresponding ground-truth, which may not be available in many practical applications.
In addition, the solutions obtained by direct feedforward networks might be inconsistent with the measurements due to the lack of a feedback mechanism~\cite{Chang_2017_ICCV, gupta2018cnn, yang2020deep}. Nevertheless, these works still suggest that CNN is an appealing solution to the specific challenges of phase unwrapping.

\begin{figure*}[t]
\centering
\includegraphics[width=1.0\textwidth]{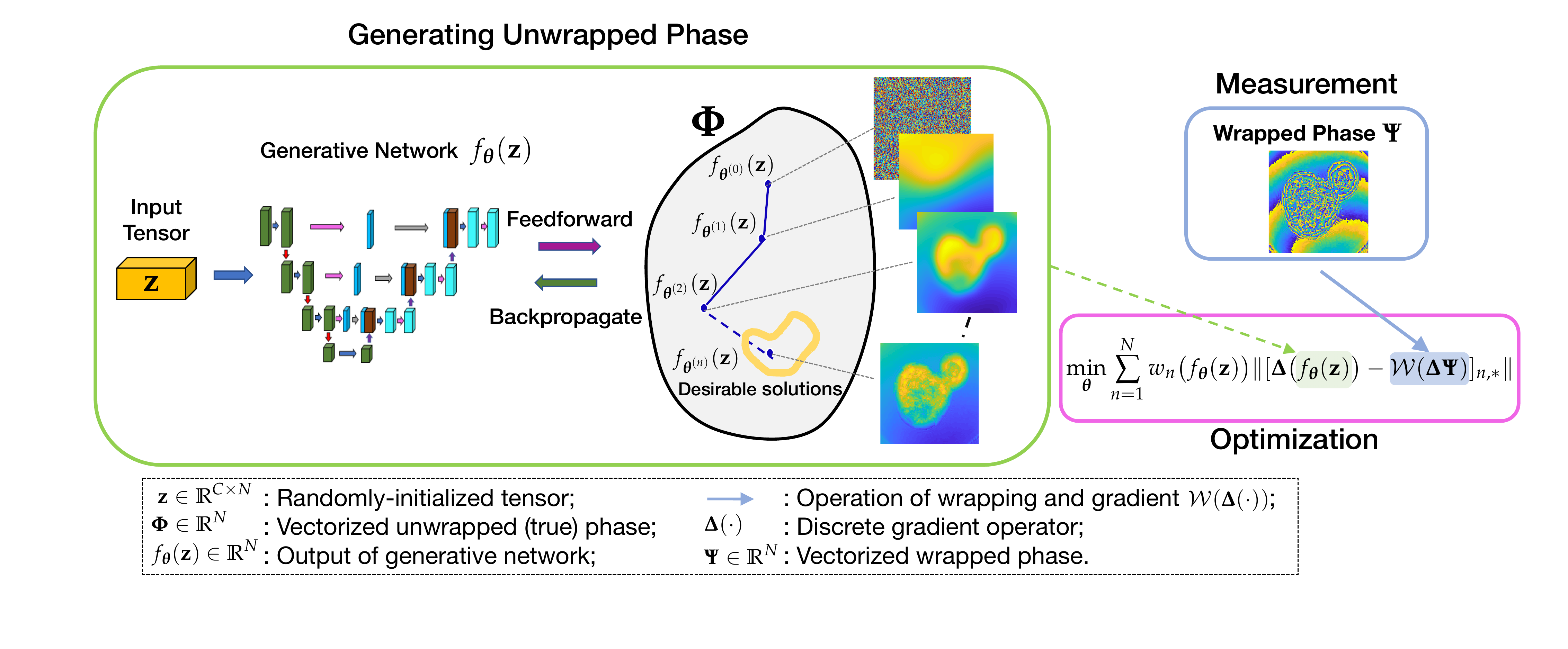}
\caption{Schematic diagram of the proposed PUDIP, for two-dimensional phase unwrapping.
The architecture of the generative network is fully described in the Supplementary Note S$1$ of \textcolor{urlblue}{Supplement~$1$}.} 
\label{fig:flowchart}
\end{figure*}

\subsection{Contributions}
In this paper, we introduce a framework with an untrained CNN for 2D phase unwrapping.
Our approach uses the concept of deep image prior recently introduced by Ulyanov~\textit{et al.}~\cite{ulyanov2018deep}.
We incorporate an explicit feedback mechanism and do not require prior training of the neural network.
Taking advantage of these features, we propose a robust and versatile method for phase unwrapping with deep image prior (PUDIP).

In Section~\ref{sec:physical}, we introduce the physical model and formulate the computational problem in a variational framework.
In Section~\ref{sec:method}, we describe the proposed scheme with an untrained deep neural network.
In Section~\ref{sec:results}, we compare the proposed method against other state-of-the-art ($e.g.,$ IRTV, PUMA) approaches on experimental data of organoids.
In Section~\ref{sec:simulate},
we quantitatively assess PUDIP on several simulated data with diverse configurations.
We extensively compare our framework with other methods such as the recent deep-learning-based PhaseNet method.
The results show that PUDIP improves upon other approaches by taking advantage of model-based and deep-learning worlds.
Our work shows that QPI can be applied to large and complex three-dimensional samples with higher reliability.

\section{Problem Formulation}
\label{sec:physical}
Let the region of interest $\Omega \subset \R^2$ be discretized into $N$ pixels.
To represent the phase of our specimen, we consider the observation model
\begin{equation}\label{eq:model}
    \mathbf{\Phi} = \boldsymbol{\Psi} + 2\pi \mathbf{k},
\end{equation}
where $\mathbf{\Phi} = (\phi_n) \in \R^N$ and $\mathbf{\Psi} = (\psi_n) \in [-\pi,\pi)^N$ denote the vectorized unwrapped and wrapped phase images, respectively; $\mathbf{k} \in \Z^N$ represents the integer multiple of $2\pi$ referred to as "wrap-count" to be added to the wrapped phase to recover the unwrapped phase. The wrapping process is represented by a function $\mathcal{W}$ applied on the $n$th component of \eqref{eq:model} as 
\begin{equation}\label{eq:wrapmodel}
    \psi_n = \mathcal{W}(\phi_n) = \left((\phi_n + \pi) \, \mathrm{mod} (2\pi) \right) - \pi \in [-\pi,\pi).
\end{equation}
The discrete gradient operator $\boldsymbol{\Delta}: \R^N \mapsto \R^{N\times2}$ is given by
\begin{equation}\label{eq:gradient}
    \boldsymbol{\Delta}\mathbf{\Phi} = \begin{bmatrix}
                             \boldsymbol{\Delta}_{\mathrm{x}}\mathbf{\Phi} \quad
                             \boldsymbol{\Delta}_{\mathrm{y}}\mathbf{\Phi} 
                             \end{bmatrix},
\end{equation}
where $ \boldsymbol{\Delta}_{\mathrm{x}}:\R^N\mapsto\R^N$ and $ \boldsymbol{\Delta}_{\mathrm{y}}:\R^N\mapsto\R^N$ denote the horizontal and vertical finite-difference operations, respectively.
The phases~$\mathbf{\Phi}$ and $\mathbf{\Psi}$ are related by the equality
\begin{equation}\label{eq:equalitynoItoh}
    \mathcal{W}([\boldsymbol{\Delta}\mathbf{\Phi}]_{n,*}) = \mathcal{W}([\boldsymbol{\Delta}\mathbf{\Psi}]_{n,*}), \quad n \in [1 \ldots N].
\end{equation}
For 2D phase-unwrapping problems, the phase $\mathbf{\Phi}$ satisfies the Itoh continuity condition~\cite{itoh1982analysis} if
\begin{equation}\label{eq:Itoh}
    \|[\boldsymbol{\Delta}\mathbf{\Phi}]_{n,*}\|^2   \leq \pi^2, \quad n \in [1 \ldots N],
\end{equation}
where $[\boldsymbol{\Delta}\mathbf{\Phi}]_{n,*} \triangleq ([\boldsymbol{\Delta}_{\mathrm{x}}\mathbf{\Phi}]_n,[\boldsymbol{\Delta}_{\mathrm{y}}\mathbf{\Phi}]_n)$ represents the $n$th component 2D vector of the discrete gradient~($i.e.,$ the $n$th row of $\boldsymbol{\Delta}\mathbf{\Phi}$).
If~\eqref{eq:Itoh} is satisfied, then \eqref{eq:equalitynoItoh} simplifies as
\begin{equation}\label{eq:equality}
    [\boldsymbol{\Delta}\mathbf{\Phi}]_{n,*} = \mathcal{W}([\boldsymbol{\Delta}\mathbf{\Psi}]_{n,*}), \quad n \in [1 \ldots N].
\end{equation}

Under the hypothesis that a great majority of pixels in $\mathbf{\Phi}$ satisfy the constraint condition in~\eqref{eq:Itoh}, we can reconstruct the unwrapped phase by minimizing the weighted energy function~\cite{kamilov2015isotropic}
\begin{equation}\label{eq:loss}
    \begin{aligned}
       \hat{\mathbf{\Phi}} = \arg\min\limits_{\mathbf{\Phi}\in \R^N} \sum _{n=1}^{N}w_n(\mathbf{\Phi})\|[\boldsymbol{\Delta}\mathbf{\Phi}-\mathcal{W}(\boldsymbol{\Delta}\mathbf{\Psi})]_{n,*}\|, 
   \end{aligned}
\end{equation}
where $w_n(\mathbf{\Phi}) \in \R_{\geq 0}$ is the adaptive nonnegative weight for the $n$th component of the cost to relax the restriction. It is defined as

\begin{equation}\label{eq:weights}
    w_n(\mathbf{\Phi}) = \left\{
    \begin{array}{cl}
         \frac{1}{\|[\boldsymbol{\epsilon}]_{n,*}\|}, &  \epsilon_{\mathrm{min}} \leq \|[\boldsymbol{\epsilon}]_{n,*}\| \leq \epsilon_{\mathrm{max}}\\
         \frac{1}{\epsilon_{\mathrm{max}}}, & \|[\boldsymbol{\epsilon}]_{n,*}\| \geq \epsilon_{\mathrm{max}}\\
         \frac{1}{\epsilon_{\mathrm{min}}}, & \|[\boldsymbol{\epsilon}]_{n,*}\| \leq \epsilon_{\mathrm{min}},
    \end{array} \right.
\end{equation}
where $\boldsymbol{\epsilon} = (\boldsymbol{\Delta}\mathbf{\Phi}-\mathcal{W}(\boldsymbol{\Delta}\mathbf{\Psi}))$, and where $\epsilon_{\mathrm{min}}$ and $\epsilon_{\mathrm{max}}$ are the user-defined minimum and maximum boundary weights, respectively. In addition, the solutions can be improved by imposing prior knowledge ($i.e.,$ a regularization term) such as a total-variation (TV)~\cite{rudin1992nonlinear} or a Hessian-Schatten norm (HS)~\cite{lefkimmiatis2013hessian} in an attempt to compensate for the ill-posed nature of the problem.

It is worthy to note that the solution obtained by iteratively minimizing the objective function~\eqref{eq:loss} offers no guarantee regarding the consistency between the rewrapped phase $\mathcal{W}(\hat{\mathbf{\Phi}})$ and the wrapped phase~$\mathbf{\Psi}$~\cite{kamilov2015isotropic}. This is because~\eqref{eq:loss} relies on continuous optimization to solve the discrete-optimization problem~\eqref{eq:model}.
Therefore, we adopt the single postprocessing step~\cite{pritt1997congruence}
\begin{equation}\label{eq:postprocess}
    \tilde{\mathbf{\Phi}} = \hat{\mathbf{\Phi}} + \mathcal{W}(\mathbf{\Psi}-\hat{\mathbf{\Phi}}),
\end{equation}
where $\tilde{\mathbf{\Phi}}$ is the final solution, congruent with the measurement~$\mathbf{\Psi}$.

\section{Phase unwrapping with deep image prior}
\label{sec:method}

Deep image prior (DIP) is a scheme recently introduced in~\cite{ulyanov2018deep}.
Rather than learning the mapping between input and output with a large training dataset, DIP handles the inverse problem by assuming that the unknown image can be represented well by the output of an untrained generative network.
Recent works have shown the effectiveness of DIP for computational imaging~\cite{gong2018pet,zhou2020diffraction, bostan2020deep,wang2020phase}.
In the spirit of this approach, we propose a framework where we restore the unwrapped phase based on this implicit prior.

The unwrapped phase is generated by the CNN given by
\begin{equation}\label{eq:dip}
    \mathbf{\Phi} = f_{\btheta}(\mathbf{z}),
\end{equation}
where $f$ denotes the neural network and $\btheta$ stands for the network parameters to be learned.
The fixed randomly-initialized vector $\mathbf{z} \in \R^{C\times N}$ acts as input to the generative network, while $C$ is the number of input channels. 

Plugging~\eqref{eq:dip} in~\eqref{eq:loss} leads to the optimization problem
\begin{equation}\label{eq:NNloss}
    \begin{aligned}
        \hat{\btheta} =\arg\min\limits_{\btheta} \sum_{n=1}^{N}w_n\big(f_{\btheta}(\mathbf{z})\big)\|[\boldsymbol{\Delta}\big(f_{\btheta}(\mathbf{z})\big)-\mathcal{W}(\boldsymbol{\Delta}\mathbf{\Psi})]_{n,*}\|.
    \end{aligned}    
\end{equation}
In our optimization approach, we aim at minimizing this loss function by taking advantage of the family of stochastic gradient-descent methods. The schematic diagram of PUDIP is shown in Fig.~\ref{fig:flowchart}.

Finally, we achieve congruence with the single step
\begin{equation}\label{eq:finalsolution}
    \tilde{\mathbf{\Phi}} = f_{\hat{\btheta}}(\mathbf{z}) + \mathcal{W}\big(\mathbf{\Psi}- f_{\hat{\btheta}}(\mathbf{z})\big).
\end{equation}

\subsection{Architecture}

We design a CNN based on the U-Net-like encoder-decoder architecture~\cite{ronneberger2015u, ulyanov2018deep}.
The setup includes skip connections with convolution and concatenation.
This enables the network to reconstruct the feature maps with both local details and global texture.
We set a constant number of channels~($i.e., 128$) in all the convolutional layers, except for those included in the skip connection whose channel number is $4$.
We chose the parametric rectified linear unit~\cite{he2015delving} as the nonlinear activation function.
Furthermore, the downsampling operation is implemented by convolutional modules with strides of 2, so that the size of the feature map is halved in the contracting path.
The upsampling operation doubles the size through bilinear interpolation.
The scaling-expanding structure makes the effective receptive field increase at deeper layers~\cite{ronneberger2015u}. We added one last layer to implement the offset subtraction in such a way that the bias of background is removed~(see Supplementary Note S$1$ of~\textcolor{urlblue}{Supplement~$1$}).
\subsection{Optimization Strategy}

In our experiments, we adopt the following strategy: The input variable $\mathbf{z}$ is a random vector filled with the uniform noise ~$\mathcal{U}(0,0.1)$.
To avoid undetermined gradients with respect to $\mathbf{\btheta}$ in~\eqref{eq:NNloss}, we offset the norm there by the small constant~\mbox{$\delta=10^{-18}$}.
In practice, the adaptive weights~$w_n$ are updated every $N_{w}$ iterations to enforce sparsity in the loss function~\eqref{eq:NNloss}~\cite{kamilov2015isotropic}. We optimize~\eqref{eq:NNloss} by using the adaptive moment-estimation algorithm (Adam,~$\beta_1=0.9$ and $\beta_2=0.999$)~\cite{KingmaB14}.
The optimization is performed on a desktop workstation (Nvidia Titan X GPU, Ubuntu operating system) and implemented on PyTorch~\cite{ketkar2017introduction}.
All the parameters used in the experiments are detailed in the Supplementary Notes S$2$ and~S$6$ of \textcolor{urlblue}{Supplement~$1$}.
In our experiments, the random initialization of the input variable did significantly impact neither the performance, nor the time of computation.

\begin{table}[t]
\setlength{\tabcolsep}{3.5pt}
\centering
\caption{Baseline methods. CNN$^{1}$ denotes the supervised-learning method, while CNN$^{2}$ denotes our method with untrained network.}
\begin{tabular} {lllll}
\arrayrulecolor{black}
\cmidrule[\heavyrulewidth]{1-4}
\cmidrule[\heavyrulewidth]{1-4}
Method & Reference & Regularization  & Optimization \\  
\cmidrule{1-4} 
\rowcolor[gray]{.9}
GA &\cite{goldstein1988satellite} &$-$  & branch-cut &\cellcolor{white} \\ 
\rowcolor[gray]{.9}
LS &\cite{ghiglia1994robust} & $-$  & least-squares  &\cellcolor{white}\rotatebox[origin=c]{90}{-based}\\ 
\rowcolor[gray]{.9}
IRTV &\cite{kamilov2015isotropic} &HS~\cite{lefkimmiatis2013hessian}  & ADMM~\cite{boyd2011distributed} &\cellcolor{white}\rotatebox[origin=c]{90}{model}\\ 
\rowcolor[gray]{.9}
PUMA  &\cite{bioucas2007phase}  &$-$  & graph cut &\cellcolor{white} \\
\\
\rowcolor[rgb]{0.88,1,1}
PhaseNet &\cite{spoorthi2018phasenet} &$-$  &CNN$^{1}$ &\cellcolor{white}\rotatebox[origin=c]{90}{based}\\
\rowcolor[rgb]{0.88,1,1}
PUDIP & &$-$   &CNN$^{2}$ &\cellcolor{white}\rotatebox[origin=c]{90}{CNN-} \\
\cline{1-4}
\cmidrule[\heavyrulewidth]{1-4}
\cmidrule[\heavyrulewidth]{1-4}
\end{tabular}
\label{tb:methods}
\end{table}

\section{Experiments}
\label{sec:results}

Thick and complex samples present complicated wrapping events and
potentially contain a few sharp edges at which the Itoh condition may not hold in the true phase.
These combined factors increase the difficulty to unwrap their phase.
To illustrate these challenges, we acquired images of organoids with digital holography microscopy and unwrapped their phase using the proposed method as well as other baseline methods.
The quality of unwrapped images will impact the subsequent steps of image analysis.
Hence, we additionally illustrate how segmentation---a typical image processing for QPI~\cite{vicar2019cell}---can be altered by the outcome of phase unwrapping.


\renewcommand{\szx}{450}
\renewcommand{\szy}{350}
\renewcommand{\szsub}{0.18}
\renewcommand{\szax}{5*\szsub}
\renewcommand{\foldoi}{RealRK_C05_01_08_SingleComparison_350_450}
\renewcommand{\vmin}{-20}
\renewcommand{\vmax}{20}
\pgfmathsetmacro{\step}{\vmin + (\vmax - \vmin)/4}
\pgfmathsetmacro{\piv}{3.14159265359}
\renewcommand{\zoomx}{280}
\renewcommand{\zoomy}{\szy - 135}
\renewcommand{\insep}{5} 
\renewcommand{\insepspy}{0.1}

\renewcommand{\szxp}{\szx}

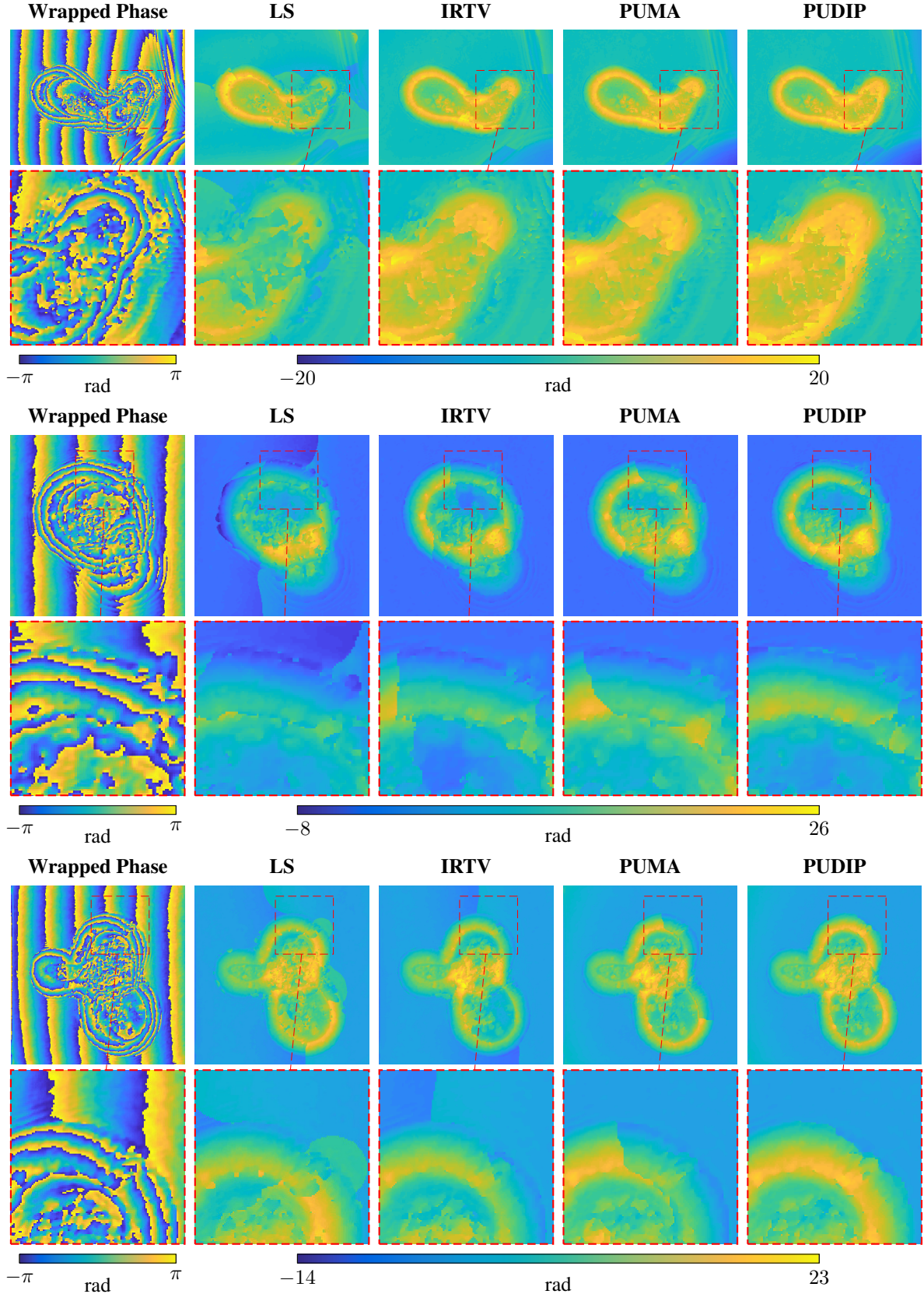
\begin{figure*}[!tbp]
\centering
\begin{tikzpicture}
\begin{scope}[spy using outlines={rectangle,magnification=3,width={\szsub\textwidth},height={\szsub\textwidth},
connect spies,line width=4pt,dotted,inner sep = 2*\insep},
]
\begin{groupplot}[
     group style = {group size = 5 by 1,
     vertical sep=\insep, horizontal sep=\insep},
     every plot/.style={width = {\szsub\textwidth},height = {\szy/\szx*\szsub\textwidth},axis equal image},
     xmin = 0,xmax = \szx, ymin = 0, ymax = \szy,
    enlargelimits=false,
    axis equal image,
    scale only axis,
    width = {\szsub\textwidth},
    height = {\szy/\szx*\szsub\textwidth},
    hide axis,
    title style = {anchor=base,yshift=0}   
    ]
    
    \nextgroupplot[title = \textbf{Wrapped Phase}]
    \addplot graphics[xmin = 0,xmax = \szx, ymin = 0, ymax = \szy] {{Figures/\foldoi/wrapPhase}};
    \coordinate (WP) at (\szx/2,0);
    \coordinate (WPz) at (\zoomx + 45,\zoomy - 45);
    
    \nextgroupplot[title={\textbf{LS}}]
    \addplot graphics[xmin = 0,xmax = \szx, ymin = 0, ymax = \szy] {{Figures/\foldoi/LS}};
    \coordinate (LS) at (\szx/2,0);
    \coordinate (LSz) at (\zoomx + 45,\zoomy-45);
 
 \nextgroupplot[title={\textbf{IRTV}}]
 \addplot graphics[xmin = 0,xmax = \szx, ymin = 0, ymax = \szy]{{Figures/\foldoi/IRTV}};
    \coordinate (IRTV) at (\szx/2,0);    
    \coordinate (IRTVz) at (\zoomx + 45,\zoomy-45);
    
    \nextgroupplot[title={\textbf{PUMA}}]
    \addplot graphics[xmin = 0,xmax = \szx, ymin = 0, ymax = \szy]{Figures/\foldoi/PUMA};
    \coordinate (PUMA) at (\szx/2,0);
    \coordinate (PUMAz) at (\zoomx + 45,\zoomy-45);
    \coordinate (PUMAi) at (\szx,\szy);
    
    \nextgroupplot[title={\textbf{PUDIP}}]
    \addplot graphics[xmin = 0,xmax = \szx, ymin = 0, ymax = \szy]{Figures/\foldoi/DIP};
    \coordinate (DIP) at (\szx/2,0);
    \coordinate (DIPz) at (\zoomx + 45,\zoomy-45);
    \end{groupplot}
    
    \spy[red] on (WPz) in node (a) [below] at ($(WP.south) - (0,\insepspy)$);
    \spy[red] on (LSz) in node (LSi) [below] at ($(LS.south) - (0,\insepspy)$);
    \spy[red] on (IRTVz) in node (b) [below] at ($(IRTV.south) - (0,\insepspy)$);
    \spy[red] on (PUMAz) in node (c) [below] at ($(PUMA.south) - (0,\insepspy)$);
    \spy[red] on (DIPz) in node (DIPi) [below] at ($(DIP.south) - (0,\insepspy)$);
    \end{scope}

    \begin{axis}[at = {($(a.south) - (0,0.5em)$)}, 
    anchor = north,
    xmin = {-\piv}, xmax = \piv, ymin = 0, ymax = 0,
    width = 0pt, 
    height = 0pt,
    scale only axis,
    enlargelimits = false,
    hide axis,
    axis equal image,
    colorbar horizontal,
    colormap/parula,
    colorbar style={
        at = {(0,1)},
        anchor = {north},
        xticklabel pos = lower,
        point meta min = {-\piv},
        point meta max = {\piv},
        scale only axis,
        enlargelimits = false,
        scaled x ticks = true,
        width = {0.9*\szsub\textwidth},
        samples = 200,
        height = 0.15cm,
        xticklabel style={/pgf/number format/fixed, /pgf/number format/precision=0},
        xtick = {-\piv,\piv},
        xticklabels={$-\pi$,$\pi$},
        xlabel={rad},
        xlabel style={xshift=0cm,yshift=1em}
            }]
    \end{axis}
    
    \begin{axis}[at = {($0.5*(b.south) + 0.5*(c.south) - (0,0.5em)$)}, 
    anchor = north,
    xmin = \vmin, xmax = \vmax, ymin = 0, ymax = 0,
    width=0pt, 
    height=0pt,
    scale only axis,
    enlargelimits=false,
    hide axis,
    axis equal image,
    colorbar horizontal,
    colormap/parula,
    colorbar style={
        at = {(0,1)},
        anchor = {north},
        xticklabel pos = lower,
        xmin = \vmin, xmax = \vmax,
        point meta min = \vmin,
        point meta max = \vmax,
        scale only axis,
        enlargelimits = false,
        scaled x ticks = true,
        width = {3*\szsub\textwidth},
        samples = 200,
        height = 0.15cm,
        xticklabel style={/pgf/number format/fixed, /pgf/number format/precision=0},
        xtick = {\vmin,\vmax},
        xlabel={rad},
        xlabel style={xshift=0cm,yshift=1.1em},
        }]
    \end{axis}
\end{tikzpicture}

\renewcommand{\szx}{250}
\renewcommand{\szy}{260}
\renewcommand{\foldoi}{RealRK_C05_15_07_SingleComparison_260_250}
\renewcommand{\vmin}{-7.65}
\renewcommand{\vmax}{26.34}
\pgfmathsetmacro{\step}{\vmin + (\vmax - \vmin)/4}
\renewcommand{\zoomx}{90}
\renewcommand{\zoomy}{\szy - 20}
\renewcommand{\szxp}{\szx}

\begin{tikzpicture}[at = {(current bounding box.south west)}]
\begin{scope}[spy using outlines={rectangle,magnification=3,width={\szsub\textwidth},height={\szsub\textwidth},
connect spies,line width=4pt,dotted,inner sep = 2*\insep},
]
\begin{groupplot}[
     group style = {group size = 5 by 1,
     vertical sep=\insep, horizontal sep=\insep},
     every plot/.style={width = {\szsub\textwidth},height = {\szy/\szx*\szsub\textwidth},axis equal image},
     xmin = 0,xmax = \szx, ymin = 0, ymax = \szy,
    enlargelimits=false,
    axis equal image,
    scale only axis,
    width = {\szsub\textwidth},
    height = {\szy/\szx*\szsub\textwidth},
    hide axis,
    title style = {anchor=base,yshift=0}  
    ]
    
    \nextgroupplot[title = \textbf{Wrapped Phase}]
    \addplot graphics[xmin = 0, xmax = \szx, ymin = 0, ymax = \szy] {{Figures/\foldoi/wrapPhase}};
    \coordinate (WP) at (\szx/2,0);
    \coordinate (WPz) at (\zoomx + 45,\zoomy - 45);
    
    \nextgroupplot[title={\textbf{LS}}, width = \szsub\textwidth]
    \addplot graphics[xmin = 0,xmax = \szx, ymin = 0, ymax = \szy] {{Figures/\foldoi/LS}};
    \coordinate (LS) at (\szx/2,0);
    \coordinate (LSz) at (\zoomx + 45,\zoomy-45);
 
 \nextgroupplot[title={\textbf{IRTV}}, width = \szsub\textwidth]
 \addplot graphics[xmin = 0,xmax = \szx, ymin = 0, ymax = \szy]{{Figures/\foldoi/IRTV}};
    \coordinate (IRTV) at (\szx/2,0);    
    \coordinate (IRTVz) at (\zoomx + 45,\zoomy-45);
    
    \nextgroupplot[title={\textbf{PUMA}}, width = \szsub\textwidth]
    \addplot graphics[xmin = 0,xmax = \szx, ymin = 0, ymax = \szy]{Figures/\foldoi/PUMA};
    \coordinate (PUMA) at (\szx/2,0);
    \coordinate (PUMAz) at (\zoomx + 45,\zoomy-45);
    \coordinate (PUMAi) at (\szx,\szy);
    
    \nextgroupplot[title={\textbf{PUDIP}}, width = \szsub\textwidth]
    \addplot graphics[xmin = 0,xmax = \szx, ymin = 0, ymax = \szy]{Figures/\foldoi/DIP};
    \coordinate (DIP) at (\szx/2,0);
    \coordinate (DIPz) at (\zoomx + 45,\zoomy-45);
    \end{groupplot}
    
    \spy[red] on (WPz) in node (a) [below] at ($(WP.south) - (0,\insepspy)$);
    \spy[red] on (LSz) in node (LSi) [below] at ($(LS.south) - (0,\insepspy)$);
    \spy[red] on (IRTVz) in node (b) [below] at ($(IRTV.south) - (0,\insepspy)$);
    \spy[red] on (PUMAz) in node (c) [below] at ($(PUMA.south) - (0,\insepspy)$);
    \spy[red] on (DIPz) in node (DIPi) [below] at ($(DIP.south) - (0,\insepspy)$);
    \end{scope}

    \begin{axis}[at = {($(a.south) - (0,0.5em)$)}, 
    anchor = north,
    xmin = {-\piv}, xmax = \piv, ymin = 0, ymax = 0,
    width = 0pt, 
    height = 0pt,
    scale only axis,
    enlargelimits = false,
    hide axis,
    axis equal image,
    colorbar horizontal,
    colormap/parula,
    colorbar style={
        at = {(0,1)},
        anchor = {north},
        xticklabel pos = lower,
        point meta min = {-\piv},
        point meta max = {\piv},
        scale only axis,
        enlargelimits = false,
        scaled x ticks = true,
        width = {0.9*\szsub\textwidth},
        samples = 200,
        height = 0.15cm,
        xticklabel style={/pgf/number format/fixed, /pgf/number format/precision=0},
        xtick = {-\piv,\piv},
        xticklabels={$-\pi$,$\pi$},
        xlabel={rad},
        xlabel style={xshift=0cm,yshift=1em}
            }]
    \end{axis}
    
    \begin{axis}[at = {($0.5*(b.south) + 0.5*(c.south) - (0,0.5em)$)}, 
    anchor = north,
    xmin = \vmin, xmax = \vmax, ymin = 0, ymax = 0,
    width=0pt, 
    height=0pt,
    scale only axis,
    enlargelimits=false,
    hide axis,
    axis equal image,
    colorbar horizontal,
    colormap/parula,
    colorbar style={
        at = {(0,1)},
        anchor = {north},
        xticklabel pos = lower,
        xmin = \vmin, xmax = \vmax,
        point meta min = \vmin,
        point meta max = \vmax,
        scale only axis,
        enlargelimits = false,
        scaled x ticks = true,
        width = {3*\szsub\textwidth},
        samples = 200,
        height = 0.15cm,
        xticklabel style={/pgf/number format/fixed, /pgf/number format/precision=0},
        xtick = {\vmin,\vmax},
        xlabel={rad},
        xlabel style={xshift=0cm,yshift=1.1em},
        }]
    \end{axis}
\end{tikzpicture}

\renewcommand{\szx}{350}
\renewcommand{\szy}{360}
\renewcommand{\foldoi}{RealRK_C05_17_07_SingleComparison_360_350}
\renewcommand{\vmin}{-13.76}
\renewcommand{\vmax}{23.29}
\pgfmathsetmacro{\step}{\vmin + (\vmax - \vmin)/4}
\renewcommand{\zoomx}{175}
\renewcommand{\zoomy}{\szy - 35}
\renewcommand{\szxp}{\szx}

\begin{tikzpicture}[at = {(current bounding box.south west)}]
\begin{scope}[spy using outlines={rectangle,magnification=3,width={\szsub\textwidth},height={\szsub\textwidth},
connect spies,line width=4pt,dotted,inner sep = 2*\insep},
]
\begin{groupplot}[
     group style = {group size = 5 by 1,
     vertical sep=\insep, horizontal sep=\insep},
     every plot/.style={width = {\szsub\textwidth},height = {\szy/\szx*\szsub\textwidth},axis equal image},
     xmin = 0,xmax = \szx, ymin = 0, ymax = \szy,
    enlargelimits=false,
    axis equal image,
    scale only axis,
    width = {\szsub\textwidth},
    height = {\szy/\szx*\szsub\textwidth},
    hide axis,
    title style = {anchor=base,yshift=0}    
    ]
    
    \nextgroupplot[title = \textbf{Wrapped Phase}]
    \addplot graphics[xmin = 0,xmax = \szx, ymin = 0, ymax = \szy] {{Figures/\foldoi/wrapPhase}};
    \coordinate (WP) at (\szx/2,0);
    \coordinate (WPz) at (\zoomx + 45,\zoomy - 45);
    
    \nextgroupplot[title={\textbf{LS}}]
    \addplot graphics[xmin = 0,xmax = \szx, ymin = 0, ymax = \szy] {{Figures/\foldoi/LS}};
    \coordinate (LS) at (\szx/2,0);
    \coordinate (LSz) at (\zoomx + 45,\zoomy-45);
 
 \nextgroupplot[title={\textbf{IRTV}}]
 \addplot graphics[xmin = 0,xmax = \szx, ymin = 0, ymax = \szy]{{Figures/\foldoi/IRTV}};
    \coordinate (IRTV) at (\szx/2,0);    
    \coordinate (IRTVz) at (\zoomx + 45,\zoomy-45);
    
    \nextgroupplot[title={\textbf{PUMA}}]
    \addplot graphics[xmin = 0,xmax = \szx, ymin = 0, ymax = \szy]{Figures/\foldoi/PUMA};
    \coordinate (PUMA) at (\szx/2,0);
    \coordinate (PUMAz) at (\zoomx + 45,\zoomy-45);
    \coordinate (PUMAi) at (\szx,\szy);
    
    \nextgroupplot[title={\textbf{PUDIP}}]
    \addplot graphics[xmin = 0,xmax = \szx, ymin = 0, ymax = \szy]{Figures/\foldoi/DIP};
    \coordinate (DIP) at (\szx/2,0);
    \coordinate (DIPz) at (\zoomx + 45,\zoomy-45);
    \end{groupplot}
    
    \spy[red] on (WPz) in node (a) [below] at ($(WP.south) - (0,\insepspy)$);
    \spy[red] on (LSz) in node (LSi) [below] at ($(LS.south) - (0,\insepspy)$);
    \spy[red] on (IRTVz) in node (b) [below] at ($(IRTV.south) - (0,\insepspy)$);
    \spy[red] on (PUMAz) in node (c) [below] at ($(PUMA.south) - (0,\insepspy)$);
    \spy[red] on (DIPz) in node (DIPi) [below] at ($(DIP.south) - (0,\insepspy)$);
    \end{scope}

    \begin{axis}[at = {($(a.south) - (0,0.5em)$)}, 
    anchor = north,
    xmin = {-\piv}, xmax = \piv, ymin = 0, ymax = 0,
    width = 0pt, 
    height = 0pt,
    scale only axis,
    enlargelimits = false,
    hide axis,
    axis equal image,
    colorbar horizontal,
    colormap/parula,
    colorbar style={
        at = {(0,1)},
        anchor = {north},
        xticklabel pos = lower,
        point meta min = {-\piv},
        point meta max = {\piv},
        scale only axis,
        enlargelimits = false,
        scaled x ticks = true,
        width = {0.9*\szsub\textwidth},
        samples = 200,
        height = 0.15cm,
        xticklabel style={/pgf/number format/fixed, /pgf/number format/precision=0},
        xtick = {-\piv,\piv},
        xticklabels={$-\pi$,$\pi$},
        xlabel={rad},
        xlabel style={xshift=0cm,yshift=1em}
            }]
    \end{axis}
    
    \begin{axis}[at = {($0.5*(b.south) + 0.5*(c.south) - (0,0.5em)$)}, 
    anchor = north,
    xmin = \vmin, xmax = \vmax, ymin = 0, ymax = 0,
    width=0pt, 
    height=0pt,
    scale only axis,
    enlargelimits=false,
    hide axis,
    axis equal image,
    colorbar horizontal,
    colormap/parula,
    colorbar style={
        at = {(0,1)},
        anchor = {north},
        xticklabel pos = lower,
        xmin = \vmin, xmax = \vmax,
        point meta min = \vmin,
        point meta max = \vmax,
        scale only axis,
        enlargelimits = false,
        scaled x ticks = true,
        width = {3*\szsub\textwidth},
        samples = 200,
        height = 0.15cm,
        xticklabel style={/pgf/number format/fixed, /pgf/number format/precision=0},
        xtick = {\vmin,\vmax},
        xlabel={rad},
        xlabel style={xshift=0cm,yshift=1.1em},
        }]
    \end{axis}
\end{tikzpicture}

\caption{Reconstructed phase images of organoids. First column:  measured (wrapped) phase image. Second to fifth columns: algorithms using LS, IRTV, PUMA, and the proposed method (PUDIP). First row: reconstructed phase. Second row:  zoomed inset.
The size of the unwrapped phase image is ($350 \times 450$),  ($260 \times 250$),
and ($360 \times 350$), respectively.}
\label{fig:real}
\end{figure*}

\subsection{Experimental Setup}
Mouse organoids of the small intestine were released from Matrigel\textregistered{} (Corning) and dissociated into single cells. After centrifugation, the cells were re-suspended at the appropriate density in ENR-CV medium supplemented with Thiazovivin (ReproCell) and seeded to deposit about 100 cells per microwell onto imaging bottom Gri3D hydrogel microwell array plates (SUN bioscience) of 300 micrometer in diameter. The cells were then let to sediment for 30 minutes as such and 150$\mu$L of self-renewal medium supplemented with 2$\%$ Matrigel. The stem cells were expanded in self-renewal for 3~days, and the organoids were differentiated for another 3~days in differentiation medium (ENR)
~\cite{brandenberg2020high}. Once the stem cells underwent morphogenesis and formed fully matured organoids, the organoids were imaged using a digital holographic microscope (T1000-Fluo, LynceeTec). The holograms, phases, and amplitudes were acquired for downstream reconstruction with a pixel of physical length of~{6.45}{\micro\metre}~($\mathrm{NA}=0.3$, magnification~$10\times$, and wavelength $684.6$nm).
The time interval between each frame was $1$~minute for the time-lapse measurements.

\renewcommand{\szx}{390}
\renewcommand{\szy}{280}
\renewcommand{\szsub}{0.115}
\renewcommand{\szax}{4*\szsub}
\renewcommand{\foldoi}{RealRK_B05_03_StackComparison_280_390}

\renewcommand{\vmin}{-8}
\renewcommand{\vmax}{25}
\pgfmathsetmacro{\step}{\vmin + (\vmax - \vmin)/4}
\renewcommand{\insep}{1} 

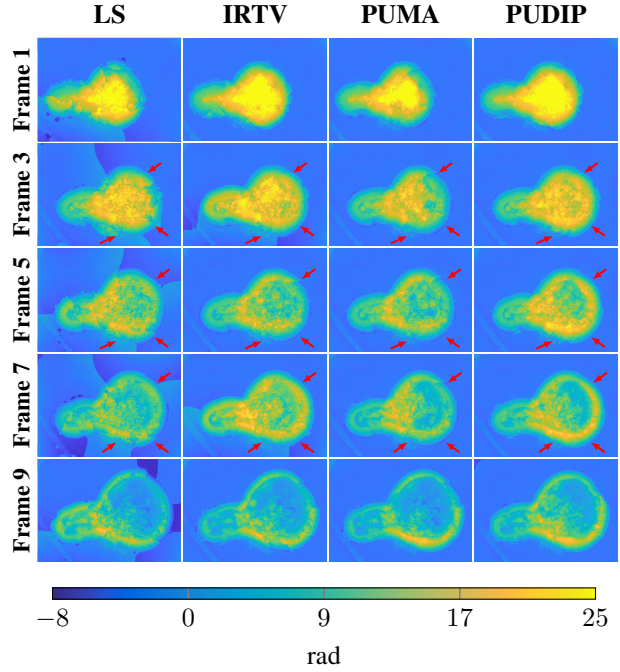
\begin{figure}[t]
    \centering
    \begin{tikzpicture}
    \begin{groupplot}[name = mygroup,
     group style = {group size = 4 by 5,
     vertical sep=\insep, horizontal sep=\insep},
     every plot/.style={width = {\szsub\textwidth},height = {\szy/\szx*\szsub\textwidth},axis equal image, scale only axis,
     }
    xmin = 0,xmax = \szx, ymin = 0, ymax = \szy,
    enlargelimits=false,
    axis equal image,
    scale only axis,
    width = {\szsub\textwidth},
    height = {\szy/\szx*\szsub\textwidth},
    hide axis,
    title style = {anchor=base}
    ]
    
    \nextgroupplot[ylabel = {\footnotesize \textbf{Frame 1}},
    y label style={at={(axis description cs:0.00001*\szx,.5)},anchor=south},
    yticklabels = {,,},
    xticklabels = {,,},
    hide axis = false,
    title = {\textbf{LS}},
        ]
    \addplot graphics[xmin = 0, xmax = \szx, ymin = 0, ymax = \szy] {Figures/\foldoi/LS1_1};
    
   \nextgroupplot[title = {\textbf{IRTV}},
        ]
    \addplot graphics[xmin = 0, xmax = \szx, ymin = 0, ymax = \szy] {Figures/\foldoi/IRTV1_1};
    \nextgroupplot[title = {\textbf{PUMA}},
        ]
    \addplot graphics[xmin = 0, xmax = \szx, ymin = 0, ymax = \szy] {Figures/\foldoi/PUMA1_1};
    \nextgroupplot[title = {\textbf{PUDIP}},
        ]
    \addplot graphics[xmin = 0, xmax = \szx, ymin = 0, ymax = \szy] {Figures/\foldoi/DIP1_1};
    \nextgroupplot[
    ylabel = {\footnotesize \textbf{Frame 3}},
    y label style={at={(axis description cs:0.00001*\szx,.5)},anchor=south},
    yticklabels = {,,},
    xticklabels = {,,},
    hide axis = false,
        ]
    \addplot graphics[xmin = 0, xmax = \szx, ymin = 0, ymax = \szy] {Figures/\foldoi/LS1_3};
    \draw[-{Latex[length=1.25mm]}, thick,red] ( 0.43*\szx, 0.025*\szy)-- (0.55*\szx, 0.1*\szy);
    \draw[-{Latex[length=1.25mm]}, thick,red] ( 0.91*\szx, 0.1*\szy)-- (0.81*\szx, 0.2*\szy);
    \draw[-{Latex[length=1.25mm]}, thick,red] ( 0.86*\szx, 0.81*\szy)-- (0.76*\szx, 0.71*\szy);
    \nextgroupplot
    \addplot graphics[xmin = 0, xmax = \szx, ymin = 0, ymax = \szy] {Figures/\foldoi/IRTV1_3};
    \draw[-{Latex[length=1.25mm]}, thick,red] ( 0.43*\szx, 0.025*\szy)-- (0.55*\szx, 0.1*\szy);
    \draw[-{Latex[length=1.25mm]}, thick,red] ( 0.91*\szx, 0.1*\szy)-- (0.81*\szx, 0.2*\szy);
    \draw[-{Latex[length=1.25mm]}, thick,red] ( 0.86*\szx, 0.81*\szy)-- (0.76*\szx, 0.71*\szy);
    \nextgroupplot
    \addplot graphics[xmin = 0, xmax = \szx, ymin = 0, ymax = \szy] {Figures/\foldoi/PUMA1_3};
    \draw[-{Latex[length=1.25mm]}, thick,red] ( 0.43*\szx, 0.025*\szy)-- (0.55*\szx, 0.1*\szy);
    \draw[-{Latex[length=1.25mm]}, thick,red] ( 0.91*\szx, 0.1*\szy)-- (0.81*\szx, 0.2*\szy);
    \draw[-{Latex[length=1.25mm]}, thick,red] ( 0.86*\szx, 0.81*\szy)-- (0.76*\szx, 0.71*\szy);
    
    \nextgroupplot
    \addplot graphics[xmin = 0, xmax = \szx, ymin = 0, ymax = \szy] {Figures/\foldoi/DIP1_3};
    \draw[-{Latex[length=1.25mm]}, thick,red] ( 0.43*\szx, 0.025*\szy)-- (0.55*\szx, 0.1*\szy);
    \draw[-{Latex[length=1.25mm]}, thick,red] ( 0.91*\szx, 0.1*\szy)-- (0.81*\szx, 0.2*\szy);
    \draw[-{Latex[length=1.25mm]}, thick,red] ( 0.86*\szx, 0.81*\szy)-- (0.76*\szx, 0.71*\szy);
    
    \nextgroupplot[
    ylabel = {\footnotesize \textbf{Frame 5}},
    y label style={at={(axis description cs:0.00001*\szx,.5)},anchor=south},
    yticklabels = {,,},
    xticklabels = {,,},
    hide axis = false,
        ]
    \addplot graphics[xmin = 0, xmax = \szx, ymin = 0, ymax = \szy] {Figures/\foldoi/LS1_5};
    \draw[-{Latex[length=1.25mm]}, thick,red] ( 0.43*\szx, 0.025*\szy)-- (0.55*\szx, 0.1*\szy);
    \draw[-{Latex[length=1.25mm]}, thick,red] ( 0.91*\szx, 0.04*\szy)-- (0.81*\szx, 0.14*\szy);
    \draw[-{Latex[length=1.25mm]}, thick,red] ( 0.92*\szx, 0.8*\szy)-- (0.82*\szx, 0.7*\szy);
    \nextgroupplot
    \addplot graphics[xmin = 0, xmax = \szx, ymin = 0, ymax = \szy] {Figures/\foldoi/IRTV1_5};
    \draw[-{Latex[length=1.25mm]}, thick,red] ( 0.43*\szx, 0.025*\szy)-- (0.55*\szx, 0.1*\szy);
    \draw[-{Latex[length=1.25mm]}, thick,red] ( 0.91*\szx, 0.04*\szy)-- (0.81*\szx, 0.14*\szy);
    \draw[-{Latex[length=1.25mm]}, thick,red] ( 0.92*\szx, 0.8*\szy)-- (0.82*\szx, 0.7*\szy);
    \nextgroupplot
    \addplot graphics[xmin = 0, xmax = \szx, ymin = 0, ymax = \szy] {Figures/\foldoi/PUMA1_5};
    \draw[-{Latex[length=1.25mm]}, thick,red] ( 0.43*\szx, 0.025*\szy)-- (0.55*\szx, 0.1*\szy);
    \draw[-{Latex[length=1.25mm]}, thick,red] ( 0.91*\szx, 0.04*\szy)-- (0.81*\szx, 0.14*\szy);
    \draw[-{Latex[length=1.25mm]}, thick,red] ( 0.92*\szx, 0.8*\szy)-- (0.82*\szx, 0.7*\szy);
    \nextgroupplot
    \addplot graphics[xmin = 0, xmax = \szx, ymin = 0, ymax = \szy] {Figures/\foldoi/DIP1_5};
    \draw[-{Latex[length=1.25mm]}, thick,red] ( 0.43*\szx, 0.025*\szy)-- (0.55*\szx, 0.1*\szy);
    \draw[-{Latex[length=1.25mm]}, thick,red] ( 0.91*\szx, 0.04*\szy)-- (0.81*\szx, 0.14*\szy);
    \draw[-{Latex[length=1.25mm]}, thick,red] ( 0.92*\szx, 0.8*\szy)-- (0.82*\szx, 0.7*\szy);

    \nextgroupplot[
    ylabel = {\footnotesize \textbf{Frame 7}},
    y label style={at={(axis description cs:0.00001*\szx,.5)},anchor=south},
    yticklabels = {,,},
    xticklabels = {,,},
    hide axis = false,
        ]
    \addplot graphics[xmin = 0, xmax = \szx, ymin = 0, ymax = \szy] {Figures/\foldoi/LS1_7};
    \draw[-{Latex[length=1.25mm]}, thick,red] ( 0.45*\szx, 0.04*\szy)-- (0.55*\szx, 0.14*\szy);
    \draw[-{Latex[length=1.25mm]}, thick,red] ( 0.91*\szx, 0.04*\szy)-- (0.81*\szx, 0.14*\szy);
    \draw[-{Latex[length=1.25mm]}, thick,red] ( 0.925*\szx, 0.815*\szy)-- (0.825*\szx, 0.715*\szy);
    \nextgroupplot
    \addplot graphics[xmin = 0, xmax = \szx, ymin = 0, ymax = \szy] {Figures/\foldoi/IRTV1_7};
    \draw[-{Latex[length=1.25mm]}, thick,red] ( 0.45*\szx, 0.04*\szy)-- (0.55*\szx, 0.14*\szy);
    \draw[-{Latex[length=1.25mm]}, thick,red] ( 0.91*\szx, 0.04*\szy)-- (0.81*\szx, 0.14*\szy);
    \draw[-{Latex[length=1.25mm]}, thick,red] ( 0.925*\szx, 0.815*\szy)-- (0.825*\szx, 0.715*\szy);
    \nextgroupplot\addplot graphics[xmin = 0, xmax = \szx, ymin = 0, ymax = \szy] {Figures/\foldoi/PUMA1_7};
    \draw[-{Latex[length=1.25mm]}, thick,red] ( 0.45*\szx, 0.04*\szy)-- (0.55*\szx, 0.14*\szy);
    \draw[-{Latex[length=1.25mm]}, thick,red] ( 0.91*\szx, 0.04*\szy)-- (0.81*\szx, 0.14*\szy);
    \draw[-{Latex[length=1.25mm]}, thick,red] ( 0.925*\szx, 0.815*\szy)-- (0.825*\szx, 0.715*\szy);
    \nextgroupplot\addplot graphics[xmin = 0, xmax = \szx, ymin = 0, ymax = \szy] {Figures/\foldoi/DIP1_7};
    \draw[-{Latex[length=1.25mm]}, thick,red] ( 0.45*\szx, 0.04*\szy)-- (0.55*\szx, 0.14*\szy);
    \draw[-{Latex[length=1.25mm]}, thick,red] ( 0.91*\szx, 0.04*\szy)-- (0.81*\szx, 0.14*\szy);
    \draw[-{Latex[length=1.25mm]}, thick,red] ( 0.925*\szx, 0.815*\szy)-- (0.825*\szx, 0.715*\szy);
    \nextgroupplot[
    ylabel = {\footnotesize \textbf{Frame 9}},
    y label style={at={(axis description cs:0.00001*\szx,.5)},anchor=south},
    yticklabels = {,,},
    xticklabels = {,,},
    hide axis = false,
        ]
    \addplot graphics[xmin = 0, xmax = \szx, ymin = 0, ymax = \szy] {Figures/\foldoi/LS1_9};
    \nextgroupplot
    \addplot graphics[xmin = 0, xmax = \szx, ymin = 0, ymax = \szy] {Figures/\foldoi/IRTV1_9};
    \nextgroupplot
    \addplot graphics[xmin = 0, xmax = \szx, ymin = 0, ymax = \szy] {Figures/\foldoi/PUMA1_9};
    \nextgroupplot
    \addplot graphics[xmin = 0, xmax = \szx, ymin = 0, ymax = \szy] {Figures/\foldoi/DIP1_9};
\end{groupplot}

\begin{axis}[at={($(group c1r5.south west) - (0,1em)$)}, anchor = north west,
    xmin = \vmin, xmax = \vmax, ymin = 0, ymax = 0.5,
    width=0,
    height = 0,
    scale only axis,
    enlargelimits=false,
    hide axis,
    axis equal image,
    colorbar horizontal,
    colormap/parula,
    colorbar style={
    at = {(0.025*\szax*\textwidth,0)},
        anchor = north west,
        xticklabel pos = lower,
        xmin = \vmin, xmax = \vmax,
        point meta min = \vmin,
        point meta max = \vmax,
        scale only axis,
        enlargelimits = false,
        scaled x ticks = true,
        width = {0.95*\szax*\textwidth},
    samples = 200,
    height = 0.15cm,
    xticklabel style={/pgf/number format/fixed, /pgf/number format/precision=0},
    xtick = {\vmin,\step,...,\vmax},
    xlabel={rad},
    xlabel style={yshift=0cm,xshift=0cm},
        }]
        \end{axis}
    \end{tikzpicture}
    
    \caption{Time-lapse reconstructions. The images were saturated for visualization purpose. The size of the unwrapped phase image is ($280 \times 390$).
    }
    \label{fig:timelapse}
\end{figure}

\renewcommand{\szx}{390}
\renewcommand{\szy}{280}
\renewcommand{\szsub}{0.115}
\renewcommand{\szax}{4*\szsub}
\renewcommand{\foldoi}{RealRK_B05_03_StackComparison_280_390_seg_0.2}

\renewcommand{\vmin}{-8}
\renewcommand{\vmax}{25}
\pgfmathsetmacro{\step}{\vmin + (\vmax - \vmin)/4}

\begin{figure}[t]
    \centering
    \begin{tikzpicture}
    \begin{groupplot}[name = mygroup,
     group style = {group size = 4 by 5,
     vertical sep=\insep, horizontal sep=\insep},
     every plot/.style={width = {\szsub\textwidth},height = {\szy/\szx*\szsub\textwidth},axis equal image, scale only axis,
     }
    xmin = 0,xmax = \szx, ymin = 0, ymax = \szy,
    enlargelimits=false,
    axis equal image,
    scale only axis,
    width = {\szsub\textwidth},
    height = {\szy/\szx*\szsub\textwidth},
    hide axis,
    title style = {anchor=base}
    ]
    
    \nextgroupplot[ylabel = {\footnotesize \textbf{Frame 1}},
    y label style={at={(axis description cs:0.00001*\szx,.5)},anchor=south},
    yticklabels = {,,},
    xticklabels = {,,},
    hide axis = false,
    title = {\textbf{LS}},
        ]
    \addplot graphics[xmin = 0, xmax = \szx, ymin = 0, ymax = \szy] {Figures/\foldoi/LS1_1};
   \nextgroupplot[title = {\textbf{IRTV}},
        ]
    \addplot graphics[xmin = 0, xmax = \szx, ymin = 0, ymax = \szy] {Figures/\foldoi/IRTV1_1};
    \nextgroupplot[title = {\textbf{PUMA}},
        ]
    \addplot graphics[xmin = 0, xmax = \szx, ymin = 0, ymax = \szy] {Figures/\foldoi/PUMA1_1};
    \nextgroupplot[title = {\textbf{PUDIP}},
        ]
    \addplot graphics[xmin = 0, xmax = \szx, ymin = 0, ymax = \szy] {Figures/\foldoi/DIP1_1};
    \nextgroupplot[
    ylabel = {\footnotesize \textbf{Frame 3}},
    y label style={at={(axis description cs:0.00001*\szx,.5)},anchor=south},
    yticklabels = {,,},
    xticklabels = {,,},
    hide axis = false,
        ]
    \addplot graphics[xmin = 0, xmax = \szx, ymin = 0, ymax = \szy] {Figures/\foldoi/LS1_3};
    \draw[-{Latex[length=1.25mm]}, thick,red] ( 0.43*\szx, 0.025*\szy)-- (0.55*\szx, 0.1*\szy);
    \draw[-{Latex[length=1.25mm]}, thick,red] ( 0.91*\szx, 0.1*\szy)-- (0.81*\szx, 0.2*\szy);
    \draw[-{Latex[length=1.25mm]}, thick,red] ( 0.86*\szx, 0.81*\szy)-- (0.76*\szx, 0.71*\szy);
    \nextgroupplot
    \addplot graphics[xmin = 0, xmax = \szx, ymin = 0, ymax = \szy] {Figures/\foldoi/IRTV1_3};
    \draw[-{Latex[length=1.25mm]}, thick,red] ( 0.43*\szx, 0.025*\szy)-- (0.55*\szx, 0.1*\szy);
    \draw[-{Latex[length=1.25mm]}, thick,red] ( 0.91*\szx, 0.1*\szy)-- (0.81*\szx, 0.2*\szy);
    \draw[-{Latex[length=1.25mm]}, thick,red] ( 0.86*\szx, 0.81*\szy)-- (0.76*\szx, 0.71*\szy);
    \nextgroupplot
    \addplot graphics[xmin = 0, xmax = \szx, ymin = 0, ymax = \szy] {Figures/\foldoi/PUMA1_3};
    \draw[-{Latex[length=1.25mm]}, thick,red] ( 0.43*\szx, 0.025*\szy)-- (0.55*\szx, 0.1*\szy);
    \draw[-{Latex[length=1.25mm]}, thick,red] ( 0.91*\szx, 0.1*\szy)-- (0.81*\szx, 0.2*\szy);
    \draw[-{Latex[length=1.25mm]}, thick,red] ( 0.86*\szx, 0.81*\szy)-- (0.76*\szx, 0.71*\szy);
    \nextgroupplot
    \addplot graphics[xmin = 0, xmax = \szx, ymin = 0, ymax = \szy] {Figures/\foldoi/DIP1_3};
    \draw[-{Latex[length=1.25mm]}, thick,red] ( 0.43*\szx, 0.025*\szy)-- (0.55*\szx, 0.1*\szy);
    \draw[-{Latex[length=1.25mm]}, thick,red] ( 0.91*\szx, 0.1*\szy)-- (0.81*\szx, 0.2*\szy);
    \draw[-{Latex[length=1.25mm]}, thick,red] ( 0.86*\szx, 0.81*\szy)-- (0.76*\szx, 0.71*\szy);
    \nextgroupplot[
    ylabel = {\footnotesize \textbf{Frame 5}},
    y label style={at={(axis description cs:0.00001*\szx,.5)},anchor=south},
    yticklabels = {,,},
    xticklabels = {,,},
    hide axis = false,
        ]
    \addplot graphics[xmin = 0, xmax = \szx, ymin = 0, ymax = \szy] {Figures/\foldoi/LS1_5};
    \draw[-{Latex[length=1.25mm]}, thick,red] ( 0.43*\szx, 0.025*\szy)-- (0.55*\szx, 0.1*\szy);
    \draw[-{Latex[length=1.25mm]}, thick,red] ( 0.91*\szx, 0.04*\szy)-- (0.81*\szx, 0.14*\szy);
    \draw[-{Latex[length=1.25mm]}, thick,red] ( 0.92*\szx, 0.8*\szy)-- (0.82*\szx, 0.7*\szy);
    \nextgroupplot
    \addplot graphics[xmin = 0, xmax = \szx, ymin = 0, ymax = \szy] {Figures/\foldoi/IRTV1_5};
    \draw[-{Latex[length=1.25mm]}, thick,red] ( 0.43*\szx, 0.025*\szy)-- (0.55*\szx, 0.1*\szy);
    \draw[-{Latex[length=1.25mm]}, thick,red] ( 0.91*\szx, 0.04*\szy)-- (0.81*\szx, 0.14*\szy);
    \draw[-{Latex[length=1.25mm]}, thick,red] ( 0.92*\szx, 0.8*\szy)-- (0.82*\szx, 0.7*\szy);
    \nextgroupplot
    \addplot graphics[xmin = 0, xmax = \szx, ymin = 0, ymax = \szy] {Figures/\foldoi/PUMA1_5};
     \draw[-{Latex[length=1.25mm]}, thick,red] ( 0.43*\szx, 0.025*\szy)-- (0.55*\szx, 0.1*\szy);
    \draw[-{Latex[length=1.25mm]}, thick,red] ( 0.91*\szx, 0.04*\szy)-- (0.81*\szx, 0.14*\szy);
    \draw[-{Latex[length=1.25mm]}, thick,red] ( 0.92*\szx, 0.8*\szy)-- (0.82*\szx, 0.7*\szy);
    \nextgroupplot
    \addplot graphics[xmin = 0, xmax = \szx, ymin = 0, ymax = \szy] {Figures/\foldoi/DIP1_5};
    \draw[-{Latex[length=1.25mm]}, thick,red] ( 0.43*\szx, 0.025*\szy)-- (0.55*\szx, 0.1*\szy);
    \draw[-{Latex[length=1.25mm]}, thick,red] ( 0.91*\szx, 0.04*\szy)-- (0.81*\szx, 0.14*\szy);
    \draw[-{Latex[length=1.25mm]}, thick,red] ( 0.92*\szx, 0.8*\szy)-- (0.82*\szx, 0.7*\szy);
    \nextgroupplot[
    ylabel = {\footnotesize \textbf{Frame 7}},
    y label style={at={(axis description cs:0.00001*\szx,.5)},anchor=south},
    yticklabels = {,,},
    xticklabels = {,,},
    hide axis = false,
        ]
    \addplot graphics[xmin = 0, xmax = \szx, ymin = 0, ymax = \szy] {Figures/\foldoi/LS1_7};
     \draw[-{Latex[length=1.25mm]}, thick,red] ( 0.45*\szx, 0.04*\szy)-- (0.55*\szx, 0.14*\szy);
    \draw[-{Latex[length=1.25mm]}, thick,red] ( 0.91*\szx, 0.04*\szy)-- (0.81*\szx, 0.14*\szy);
    \draw[-{Latex[length=1.25mm]}, thick,red] ( 0.925*\szx, 0.815*\szy)-- (0.825*\szx, 0.715*\szy);
    \nextgroupplot
    \addplot graphics[xmin = 0, xmax = \szx, ymin = 0, ymax = \szy] {Figures/\foldoi/IRTV1_7};
    \draw[-{Latex[length=1.25mm]}, thick,red] ( 0.45*\szx, 0.04*\szy)-- (0.55*\szx, 0.14*\szy);
    \draw[-{Latex[length=1.25mm]}, thick,red] ( 0.91*\szx, 0.04*\szy)-- (0.81*\szx, 0.14*\szy);
    \draw[-{Latex[length=1.25mm]}, thick,red] ( 0.925*\szx, 0.815*\szy)-- (0.825*\szx, 0.715*\szy);
    \nextgroupplot\addplot graphics[xmin = 0, xmax = \szx, ymin = 0, ymax = \szy] {Figures/\foldoi/PUMA1_7};
    \draw[-{Latex[length=1.25mm]}, thick,red] ( 0.45*\szx, 0.04*\szy)-- (0.55*\szx, 0.14*\szy);
    \draw[-{Latex[length=1.25mm]}, thick,red] ( 0.91*\szx, 0.04*\szy)-- (0.81*\szx, 0.14*\szy);
    \draw[-{Latex[length=1.25mm]}, thick,red] ( 0.925*\szx, 0.815*\szy)-- (0.825*\szx, 0.715*\szy);
    \nextgroupplot\addplot graphics[xmin = 0, xmax = \szx, ymin = 0, ymax = \szy] {Figures/\foldoi/DIP1_7};
    \draw[-{Latex[length=1.25mm]}, thick,red] ( 0.45*\szx, 0.04*\szy)-- (0.55*\szx, 0.14*\szy);
    \draw[-{Latex[length=1.25mm]}, thick,red] ( 0.91*\szx, 0.04*\szy)-- (0.81*\szx, 0.14*\szy);
    \draw[-{Latex[length=1.25mm]}, thick,red] ( 0.925*\szx, 0.815*\szy)-- (0.825*\szx, 0.715*\szy);
    \nextgroupplot[
    ylabel = {\footnotesize \textbf{Frame 9}},
    y label style={at={(axis description cs:0.00001*\szx,.5)},anchor=south},
    yticklabels = {,,},
    xticklabels = {,,},
    hide axis = false,
        ]
    \addplot graphics[xmin = 0, xmax = \szx, ymin = 0, ymax = \szy] {Figures/\foldoi/LS1_9};
    \nextgroupplot
    \addplot graphics[xmin = 0, xmax = \szx, ymin = 0, ymax = \szy] {Figures/\foldoi/IRTV1_9};
    \nextgroupplot
    \addplot graphics[xmin = 0, xmax = \szx, ymin = 0, ymax = \szy] {Figures/\foldoi/PUMA1_9};
    \nextgroupplot
    \addplot graphics[xmin = 0, xmax = \szx, ymin = 0, ymax = \szy] {Figures/\foldoi/DIP1_9};
\end{groupplot}
\end{tikzpicture}

\caption{Segmentation of time-lapse reconstructions.
We thresholded at $20\%$ of the maximum value of the image.
}
\label{fig:seg}
\end{figure}
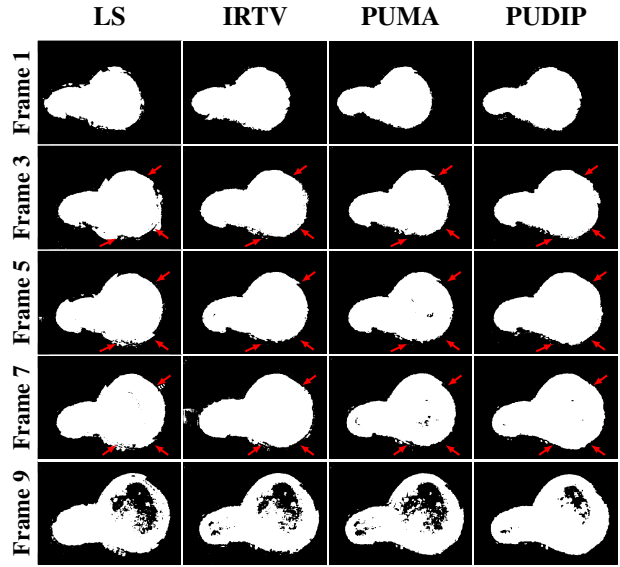

\subsection{Baseline Methods}
We compare the proposed method with other state-of-the-art conventional or CNN-based methods such as Goldstein’s algorithm (GA)~\cite{goldstein1988satellite}, unweighted least-squares algorithm (LS)~\cite{ghiglia1994robust}, IRTV\footnote{The source code for IRTV is available from \url{https://cigroup.wustl.edu/publications/open-source/}}~\cite{kamilov2015isotropic}, PUMA\footnote{The source code for PUMA is available from \url{http://www.lx.it.pt/~bioucas/code.htm}}~\cite{bioucas2007phase}, and PhaseNet~\cite{spoorthi2018phasenet} (see Table~\ref{tb:methods}).
Goldstein’s algorithm is a path-following method that adopts the branch-cut strategy based on the phase residues and needs the knowledge of a phase-reference point.
By contrast, the LS, IRTV, and PUMA approaches aim at minimizing an objective function and belong to the minimum-norm category.
Note that the original LS method, which relies on a continuous optimization, may result in an inconsistent solution, while GA, IRTV, and PUMA always return consistent solutions.
To enforce measurement consistency for LS, we adopted the strategy defined by~\eqref{eq:postprocess}.
We also compare PUDIP to the recently proposed PhaseNet~\cite{spoorthi2018phasenet}.
We adopted the strategy of~\cite{wang2019one} to generate a training dataset.
In addition, we only kept the central disk of the generated phase images and filled the background with~$0$. 
The training dataset is composed of 9,600 samples; the size of each image is $(256\times 256)$. The wrap-count in the training data varies between $0$ and $20$, which makes it a $21$-class problem~(see Supplementary Note S$3$ of \textcolor{urlblue}{Supplement~$1$}).
We set the other hyperparameters as in~\cite{spoorthi2018phasenet} and trained PhaseNet with this generated dataset for all the experiments.

All model-based methods were run on a desktop computer (Intel XeonE5-1650 CPU, 3.5 GHz, 32 GB of RAM) and implemented in MATLAB R2019a.
For their implementation, we initialized the unwrapped phase with $\mathbf{0} \in \R^N$. All parameters were set and optimized according to the guidelines provided by the authors. Specifically, the regularization parameter for the Hessian-Schatten-norm regularization in IRTV was set between $10^{-3}$ and $10^{-1}$. 
In PUMA, we set the non-convex quantized potential of exponent $p=0.5$, the quadratic region threshold as~$0.5$, and the high-order cliques $[1, 0], [0, 1], [1, 1]$, and~$[-1, 1]$.

PUDIP takes about 100 seconds on GPU to unwrap a \mbox{($256 \times  256$)} image with $1000$ iterations. In comparison, PUMA and IRTV take about 2 and 380 seconds on CPU, respectively.


\renewcommand{\szx}{159}
\renewcommand{\szy}{159}
\renewcommand{\szsub}{0.15}
\renewcommand{\szax}{4*\szsub}
\renewcommand{\foldoi}{FakeOrganoid1_original}

\newcommand{\nvmin}{1.333}
\newcommand{\nvmax}{1.393}
\pgfmathsetmacro{\nstep}{\nvmin + (\nvmax - \nvmin)/1}
\renewcommand{\vmin}{0}
\renewcommand{\vmax}{18.84}
\pgfmathsetmacro{\step}{\vmin + (\vmax - \vmin)/3}
\renewcommand{\insep}{1} 

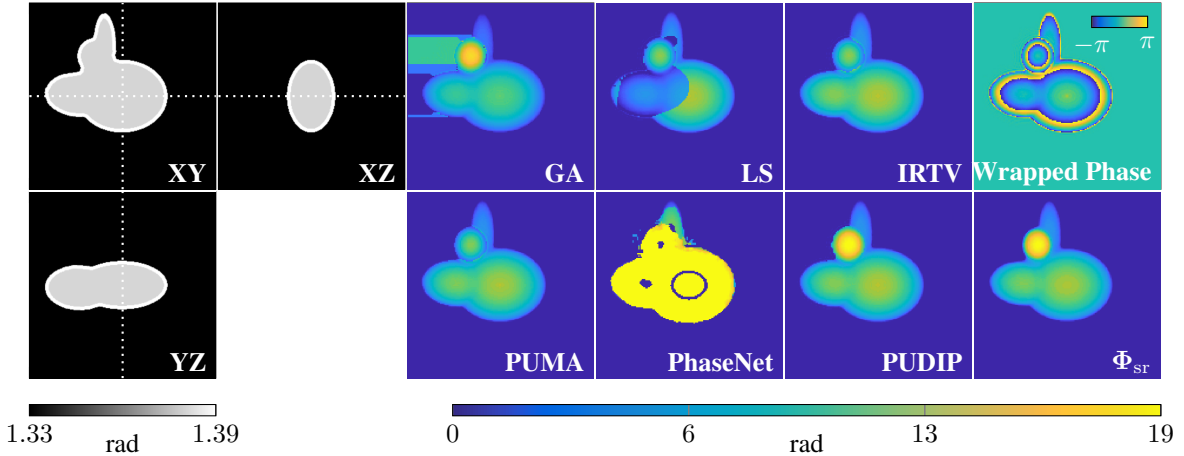
\begin{figure*}[t]
    \centering
    \begin{tikzpicture}
    \begin{groupplot}[
     group style = {group size = 6 by 2,
     vertical sep=\insep, horizontal sep=\insep},
     every plot/.style={width = {\szsub\textwidth},height = {\szy/\szx*\szsub\textwidth},axis equal image, scale only axis,
     }
    xmin = 0,xmax = \szx, ymin = 0, ymax = \szy,
    enlargelimits=false,
    axis equal image,
    scale only axis,
    width = {\szsub\textwidth},
    height = {\szy/\szx*\szsub\textwidth},
    hide axis,
    title style = {anchor=east,
    at={(axis cs:\szx,0)},
    color=white}
    ]
    \nextgroupplot[
    y label style={at={(axis description cs:0.00001*\szx,.5)},anchor=south},
    yticklabels = {,,},
    xticklabels = {,,},
    hide axis = false,
    title = {\textbf{XY}},
        ]
    \addplot graphics[xmin = 0, xmax = \szx, ymin = 0, ymax = \szy,] {Figures/\foldoi/nx};
    \draw[white,dotted,thick](0.5*\szx,0)--(0.5*\szx,\szy);
    \draw[white,dotted,thick](0,0.5*\szy)--(\szx,0.5*\szy);
    
    \nextgroupplot[
    y label style={at={(axis description cs:0.00001*\szx,.5)},anchor=south},
    yticklabels = {,,},
    xticklabels = {,,},
    hide axis = false,
    title = {\textbf{XZ}},
        ]
    \addplot graphics[xmin = 0, xmax = \szx, ymin = 0, ymax = \szy] {Figures/\foldoi/ny};
    \draw[white,dotted,thick](0,0.5*\szy)--(\szx,0.5*\szy);

    \nextgroupplot[
    y label style={at={(axis description cs:0.00001*\szx,.5)},anchor=south},
    yticklabels = {,,},
    xticklabels = {,,},
    hide axis = false,
    title = {\textbf{GA}},
        ]
    \addplot graphics[xmin = 0, xmax = \szx, ymin = 0, ymax = \szy] {Figures/\foldoi/GA};
    
    
    \nextgroupplot[title = {\textbf{LS}},
        ]
    \addplot graphics[xmin = 0, xmax = \szx, ymin = 0, ymax = \szy] {Figures/\foldoi/LS};
   \nextgroupplot[title = {\textbf{IRTV}},
        ]
    \addplot graphics[xmin = 0, xmax = \szx, ymin = 0, ymax = \szy] {Figures/\foldoi/IRTV};
    
    \nextgroupplot[
    y label style={at={(axis description cs:0.00001*\szx,.5)},anchor=south},
    yticklabels = {,,},
    xticklabels = {,,},
    hide axis = false,
    title = {\textbf{Wrapped Phase}},
        ]
    \addplot graphics[xmin = 0, xmax = \szx, ymin = 0, ymax = \szy] {Figures/\foldoi/wrapPhase};
    
    
    \nextgroupplot[
    y label style={at={(axis description cs:0.00001*\szx,.5)},anchor=south},
    yticklabels = {,,},
    xticklabels = {,,},
    hide axis = false,
    title = {\textbf{YZ}},
        ]
    \addplot graphics[xmin = 0, xmax = \szx, ymin = 0, ymax = \szy] {Figures/\foldoi/nz};
    \draw[white,dotted,thick](0.5*\szx,0)--(0.5*\szx,\szy);
    
    \nextgroupplot;
    
    \nextgroupplot[title = {\textbf{PUMA}},
        ]
    \addplot graphics[xmin = 0, xmax = \szx, ymin = 0, ymax = \szy] {Figures/\foldoi/PUMA};
    
    \nextgroupplot[title = {\textbf{PhaseNet}},
        ]
    \addplot graphics[xmin = 0, xmax = \szx, ymin = 0, ymax = \szy] {Figures/\foldoi/PhaseNet};

    \nextgroupplot[title = {\textbf{PUDIP}},
        ]
    \addplot graphics[xmin = 0, xmax = \szx, ymin = 0, ymax = \szy] {Figures/\foldoi/PUDIP};
    
    \nextgroupplot[title = {$\mathbf{\Phi}_{\mathrm{sr}}$},
        ]
    \addplot graphics[xmin = 0, xmax = \szx, ymin = 0, ymax = \szy] {Figures/\foldoi/linearap};
\end{groupplot}

\begin{axis}[at={($(group c6r1.north east) - (0,0.5em)$)}, anchor = north west,
    ymin = 0, ymax = 0.5, xmin = -3.14, xmax = 3.14,
    width = 0,
    height = 0,
    scale only axis,
    enlargelimits=false,
    hide axis,
    axis equal image,
    colorbar horizontal,
    colormap/parula,
    colorbar style={
    at = {(-0.5em,0.0)},
        anchor = north east,
        xticklabel pos = lower,
        xmin = -3.14, xmax = 3.14,
        point meta min = -3.14,
        point meta max = 3.14,
        scale only axis,
        enlargelimits = false,
        scaled x ticks = true,
        width = {0.075*\szax*\textwidth},
    samples = 200,
    height = 0.15cm,
    xticklabel style={/pgf/number format/fixed, /pgf/number format/precision=0},
    xtick = {-3.14,3.14},
    xticklabels = {{\color{white}$-\pi$}, {\color{white}$\pi$}},
    xlabel style={yshift=1em,xshift=0em}
        }]
\end{axis}

\begin{axis}[at={($(group c1r2.south east) - (0,1em)$)}, anchor = north west,
    xmin = \nvmin, xmax = \nvmax, ymin = 0, ymax = 0.5,
    width = 0,
    height = 0,
    scale only axis,
    enlargelimits=false,
    hide axis,
    axis equal image,
    colorbar horizontal,
    colormap/blackwhite,
    colorbar style={
    at = {(0.0*\szax*\textwidth,0.5)},
        anchor = north east,
        xticklabel pos = lower,
        xmin = \nvmin, xmax = \nvmax,
        point meta min = \nvmin,
        point meta max = \nvmax,
        scale only axis,
        enlargelimits = false,
        scaled x ticks = true,
        width = {0.25*\szax*\textwidth},
    samples = 200,
    height = 0.15cm,
    xticklabel style={/pgf/number format/fixed, /pgf/number format/precision=2},
    xtick = {\nvmin,\nstep,...,\nvmax},
    xlabel={rad},
    xlabel style={yshift=1em,xshift=0cm},
        }]
        \end{axis}

\begin{axis}[at={($(group c6r2.south east) - (0,1em)$)}, anchor = north west,
    xmin = \vmin, xmax = \vmax, ymin = 0, ymax = 0.5,
    width=0,
    height = 0,
    scale only axis,
    enlargelimits=false,
    hide axis,
    axis equal image,
    colorbar horizontal,
    colormap/parula,
    colorbar style={
    at = {(0.0*\szax*\textwidth,0.5)},
        anchor = north east,
        xticklabel pos = lower,
        xmin = \vmin, xmax = \vmax,
        point meta min = \vmin,
        point meta max = \vmax,
        scale only axis,
        enlargelimits = false,
        scaled x ticks = true,
        width = {0.95*\szax*\textwidth},
    samples = 200,
    height = 0.15cm,
    xticklabel style={/pgf/number format/fixed, /pgf/number format/precision=0},
    xtick = {\vmin,\step,...,\vmax},
    xlabel={rad},
    xlabel style={yshift=1em,xshift=0cm},
        }]
        \end{axis}
    \end{tikzpicture}
    
    \caption{Organoid-like reconstructions. The images were saturated for visualization purpose. The size of the unwrapped phase image is ($159 \times 159$).
    The first two columns are orthographic slices of the three-dimensional~(3D) distribution of refractive indices. All slices include the center of the volume. From the third to fifth column, the text gives the method used to unwrap. The wrapped phase resulting from 3D simulation and the ground-truth~$\mathbf{\Phi}_{\mathrm{sr}}$ are displayed in the last column (from top to bottom).
    }
    \label{fig:fakeorganoid}
\end{figure*}

\subsection{Phase Unwrapping of Organoids}

The results of various methods are shown in Fig.~\ref{fig:real}.
The LS method yields inaccurate results over large areas, such as non-flat background or disrupted structures.
In comparison, the three other approaches perform better.
However, some areas pointed out by the rectangle exhibit sudden breaks in the phase unwrapped by IRTV and PUMA.
The phase is expected to be relatively smooth since the epithelium of the organoids consists in a continuous layer of cells, forming then the border of the sample~\cite{yin2016engineering}.
By contrast, PUDIP better recovers it for all samples.

PhaseNet failed to reconstruct the unwrapped phase in all cases~(see Supplementary Note S$2$ of \textcolor{urlblue}{Supplement~$1$}), most probably because the training set is not adequate for our experimental data.
Likewise, GA was unable to recover the samples.
The solutions found by PhaseNet and GA exhibit several areas with values higher than their surrounding, which does not accurately represent the characteristic features found in intestinal organoids, such as the epithelium and the lumen.

In the first row of Fig.~3, the unwrapped phase might deviate from the phase image
predicted by the straight-ray approximation~\cite{kak2002principles} in the center part where it is non-smooth.
The approximation is accurate if the wavelength is much smaller than the features of the sample~($e.g.,$ local inhomogeneity of the refractive index).
The mismatches are then likely to occur in the areas where the features are, which suggests that local inhomogeneities are present in the inner part.

It is noteworthy that all displayed methods are consistent with the measurements as their relative errors are similar~(see Supplementary Note S$2$ of \textcolor{urlblue}{Supplement~$1$}).

\subsection{Phase Unwrapping of Time-Lapse Measurements}

Further, we acquired time-lapse measurements of organoids to validate the benefits of our approach in sequential imaging.
In the last frames, the size of the organoids increases and the intra-organoid composition becomes visibly more heterogeneous.
It is noteworthy that the intestinal organoids are absorbing water as they grow over time~\cite{fair2018intestinal}, which explains that the phase value gets closer to the background value.
As a consequence, the unwrapping task becomes even more challenging.
By using PUDIP, we show here that the borders as well as the flatness of the background are well preserved (see Fig.~\ref{fig:timelapse} and Supplementary Note S$2$ of \textcolor{urlblue}{Supplement~$1$}). 
On the contrary, the unwrapped phase of the other methods either result in a background with unlikely $2\pi$ jumps or borders with sudden breaks.

\subsection{Segmentation of Time-Lapse Measurements}
Image segmentation is a step that one would usually perform on the unwrapped phase~\cite{vicar2019cell}.
Our aim now is to illustrate how unwrapping can affect the segmentation results.
To that end, we simply thresholded the images obtained from the different methods with a threshold set at $20\%$ of the maximal value.

In Fig.~\ref{fig:seg}, we observe that the segmentation is especially impacted at the borders where sudden breaks occur in the unwrapped phase.
In all frames, the segmentation of PUDIP solutions preserves the integrity of the boundaries better than the other methods~(see Supplementary Note S$2$ of \textcolor{urlblue}{Supplement~$1$}).

\section{Simulated data}
\label{sec:simulate}

While the results on experimental data are encouraging, we want to quantitatively assess the quality of our proposed method.
To that end, we simulated the acquisition of phase images of organoid-like samples~(see Supplementary Note S$5$ of \textcolor{urlblue}{Supplement~$1$}).
In addition, we generated diverse data which are similar to those found in~\cite{bioucas2007phase} and ~\cite{wang2019one}~(see Supplementary Note S$5$ of \textcolor{urlblue}{Supplement~$1$}).
The parameter setting of PUDIP is detailed in Supplementary Note S$6$ of \textcolor{urlblue}{Supplement~$1$}.

\subsection{Quantitative Evaluation}

We quantitatively evaluate the quality of the reconstructed phase $\tilde{\mathbf{\Phi}}$ with respect to the ground-truth ${\mathbf{\Phi}}$. Our first metric is the regressed signal-to-noise ratio (RSNR) defined as
\begin{equation}
\mathrm{RSNR}(\tilde{\mathbf{\Phi}},\mathbf{\Phi})= \max\limits_{b\in \R^+} \Bigg(20 \mathrm{log}_{10}\Bigg(\frac{\|\mathbf{\Phi}\|_{2}}{\|(\tilde{\mathbf{\Phi}}+b)-\mathbf{\Phi}\|_{2}}\Bigg)\Bigg), 
\label{eq:RSNR}
\end{equation}
where $\|\cdot \|_{2}$ denotes the $L_{2}$ norm and where~$b$ adjusts for a potential global offset.
This adjustment is used in the interest of fairness, because phase unwrapping can only recover the phase up to a constant.
When the RSNR is more than 100 dB, the recovered phase image differs from the ground truth because of numerical imprecision and not because of wrong unwrapping.
We therefore set the corresponding value to infinity.
In addition, we compute the structural similarity~(SSIM) as an additional metric~(see Supplementary Note S$4$ of \textcolor{urlblue}{Supplement~$1$}).

\subsection{Simulated Phase Images of Organoid-Like Sample}

\begin{figure*}[t]
    \centering
    \includegraphics[width=\textwidth]{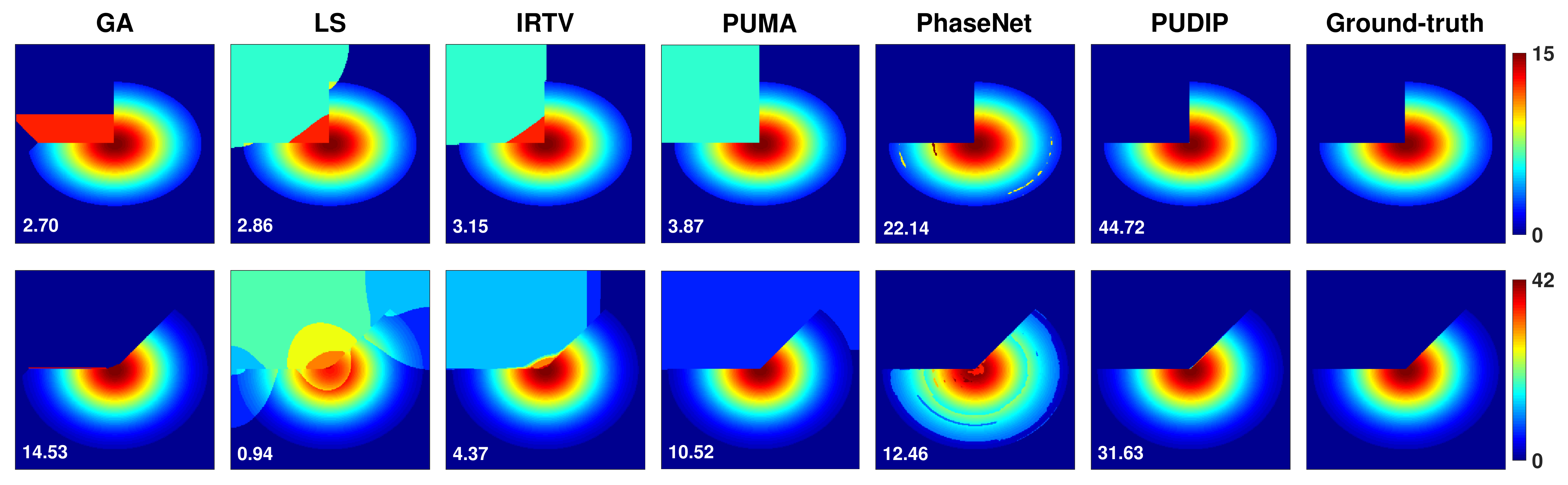}
    \caption{Unwrapped phases of two simulated samples. From left to right, the results are obtained by GA, LS, IRTV, PUMA, PhaseNet, and our approach (PUDIP).
    The ground-truth images are presented in the last column. The corresponding RSNR [dB] is showed at the left bottom of each subfigure.}
    \label{fig:mainsim}
\end{figure*}

In order to obtain a physically-realistic ground-truth, we simulated the wave propagation through the sample with the beam-propagation method~\cite{feit1988beam}.
From the three-dimensional simulation, we directly obtain the wrapped phase.
Under the straight-ray approximation~\cite{kak2002principles}, we expect that the unwrapped phase is proportional to the integral of the refractive index differences.
We therefore refer to the straight-ray approximation~~$\mathbf{\Phi}_\mathrm{sr}$ as the ground-truth.
As shown in Fig.~\ref{fig:fakeorganoid}, the phase unwrapped by PUDIP is consistent with~$\mathbf{\Phi}_\mathrm{sr}$.
The solutions of the other methods have wrongly unwrapped areas.
The entanglement of several elements complicates the wrapping events in those areas~(Fig.~\ref{fig:fakeorganoid} top right panel).
The fact that some parts are defocused adds to the challenge since ripples are present around the border~(see Supplementary Note S$7$ of \textcolor{urlblue}{Supplement~$1$}). It is worthy to note that real data also have ripples around the border, which might partially explain the difficulty to unwrap phase images of organoids.

\subsection{Reconstruction of Artificial Images}

When the unwrapping task is relatively simple, all the baseline methods, as well as our method, perform well (see Supplementary Notes S$8$-S$9$ of \textcolor{urlblue}{Supplement~$1$}).
When the phase images are more complex~($e.g.,$ when a few pixels violate the Itoh condition),
all the conventional methods lead to blocky errors~(see Supplementary Notes S$8$-S$10$ of \textcolor{urlblue}{Supplement~$1$}).
By contrast, PhaseNet correctly recovers the unwrapped phase when the training and testing sets match~(see Supplementary Note S$10$ of \textcolor{urlblue}{Supplement~$1$}).
As expected, PhaseNet wrongly estimates the unwrapped phases when they differ from the training set.
On the contrary, our framework with untrained CNN
faithfully unwraps the phase for nearly all configurations.
In Fig.~\ref{fig:mainsim}, one can observe some typical unwrapping behavior of the different methods, as well as the obtained RNSR.
The RSNR and SSIM (see Supplementary Notes S$8$-S$10$ of \textcolor{urlblue}{Supplement~$1$}) corroborate the observations we made on both real and simulated data.
In addition, PUDIP remains stable when structured noise is added to the phase image~(see Supplementary Notes S$5$ and S$11$ of \textcolor{urlblue}{Supplement~$1$}).
The comparisons suggest that our method is robust and versatile and is able to
cope with phase unwrapping of diverse complexity without prior knowledge.

Let us observe that the results of PUDIP are still imperfect, in the sense that a few pixels of the output deviate from the ground-truth. However, these are inconspicuous.
Based on our experiments, it appears that the results of PUDIP are generally superior to those of the other methods when the conditions are difficult, and otherwise equivalent, which should make PUDIP of interest for practitioners.

\section{Conclusion}
\label{sec:conclusion}
We proposed a general iterative framework PUDIP that takes advantage of model-based approaches and deep priors for two-dimensional phase unwrapping.
The iterative inversion algorithm is based on a forward model that ensures consistency with the measurements and a generative network that learns the implicit knowledge of the image automatically. Further, the prior generated by the convolutional neural network without ground-truth overcomes the limitation of conventional supervised-learning strategies which need large-scale or tailored training datasets.
We have validated our approach on simulated data with diverse challenging settings in which the unwrapped phase has many discontinuities. Numerical experiments have shown that the proposed method outperforms state-of-the-art conventional or network-based methods in many configurations.
In addition, we have also applied our framework to single and time-lapse measurements of organoids, which are particularly large and complex samples.
PUDIP can help in all instances of optical imaging that acquire wrapped phase data, quantitative phase imaging as well as more sophisticated tomographic schemes~\cite{jin2017tomographic}.
We believe that PUDIP should be of interest to practitioners.
The substantial improvement of our method
and the quality of reconstruction effectively allow the application of quantitative phase imaging to thick and complex three-dimensional samples, from which subsequent image processing can be carried on with higher reliability.

\section*{Funding}
National Key Research and Development Program of China (2017YFB0202902);
Natural Science Foundation of China (41625017);
European Research Council (ERC 692726).

\section*{Acknowledgments}
The authors would like to thank the China Scholarship Council for supporting the visit of the first co-author. This research was supported by  the European Research Council (ERC) under the European Union's Horizon 2020 research and innovation program, Grant Agreement no 692726 ``GlobalBioIm: Global integrative framework for computational bio-imaging.''


\section*{Disclosures}
The authors declare no conflict of interest.

\bigskip \noindent See \textcolor{urlblue}{Supplement~$1$} for supporting content.

\bibliographystyle{unsrt}  
\bibliography{pudip}
\end{document}


\maketitle
\begin{abstract}
This document contains supplementary information to ''Robust Phase Unwrapping via Deep Image Prior for Quantitative Phase Imaging,”  We detail the architecture of the generative network, experimental data, the reconstructions of additional organoids,
the training dataset of PhaseNet, the additional metric for quantitative evaluation,  the simulated acquisition of organoid-like samples by digital holography microscopy, the simulation setup of synthetic data, the parameter setting of simulated experiments, and the reconstructions of diverse simulated data.

\end{abstract}
\section{Architecture of the Generative Network}
\label{sec:Architecture}
\begin{figure*}[!tbp]
\centering
\includegraphics[width=0.9\textwidth]{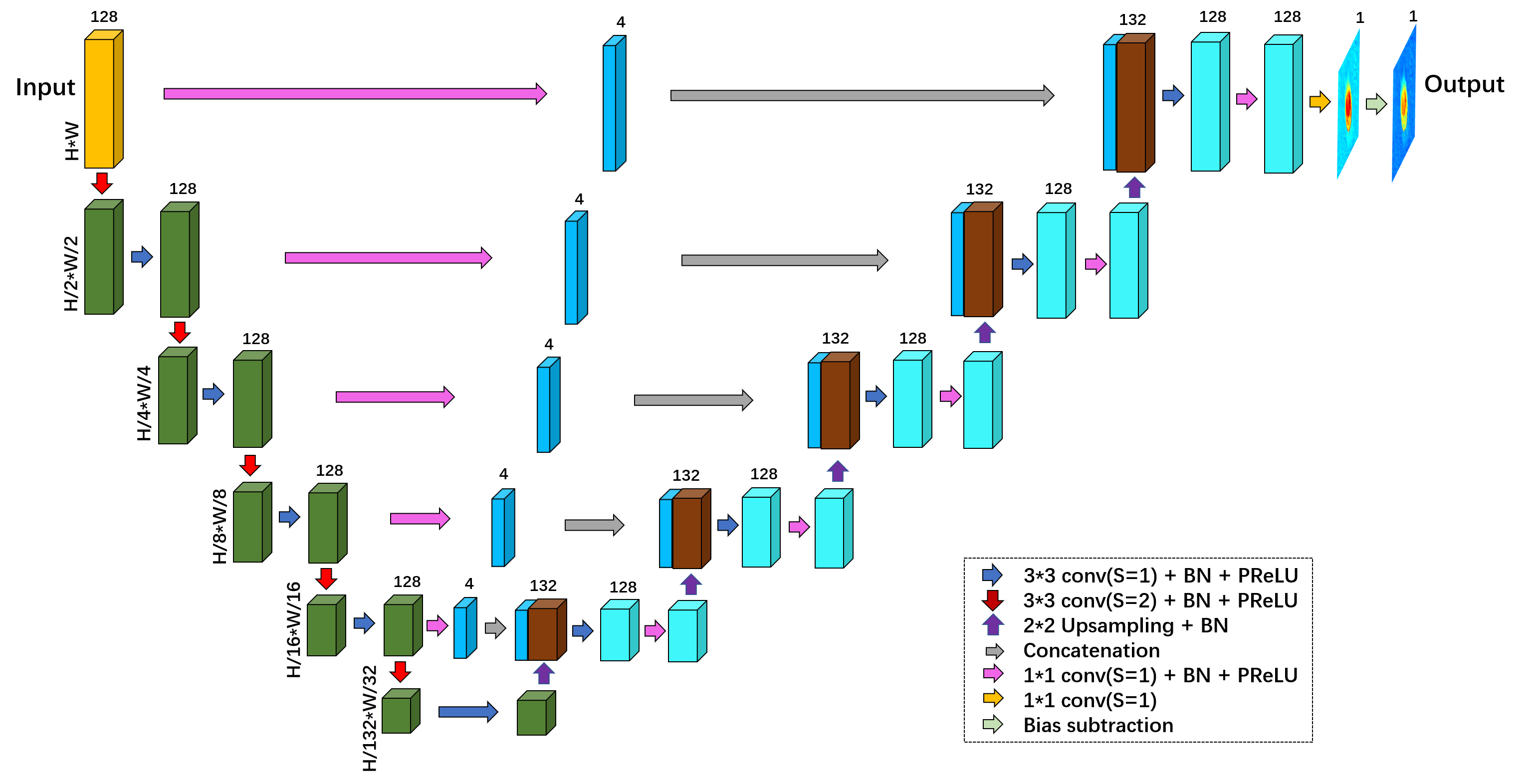}
\caption{Architecture of the network. Each box corresponds to a multichannel feature map. The number of channels is shown at the top of the cube. The height of the output is as same as that of the input. The  size of the single-channel feature map is halved after the downsampling and doubled after the upsampling. The skip connections combine convolution and concatenation, which differs from a traditional U-Net~\cite{ronneberger2015u}.
}
\label{fig:network}
\end{figure*}

We choose a U-Net-like architecture based on the work of deep image prior~\cite{ulyanov2018deep} (Fig.~\ref{fig:network}). It consists of repeated applications of four blocks of operations.
\begin{enumerate}
    \item A ($3 \times 3$)  2D convolutional layer with stride ($2 \times 2$) for downsampling followed by a batch normalization (BN)~\cite{ioffe2015batch} layer and a parametric rectified linear unit (PReLU)~\cite{he2015delving} layer.
    \item A ($3 \times 3$) 2D convolutional layer with stride ($1 \times 1$) followed by a BN and a PReLU layer.
    \item A ($2 \times 2$) bilinear interpolation layer for upsampling followed by a BN layer.
    \item A skip connection which contains a ($1 \times 1$) 2D convolutional layer that concatenates the left-side encoder path to the right-side decoder path.
\end{enumerate}

As last stage, we have set one layer that subtracts a scalar value from the image. This scalar takes care of the bias intrinsic to phase unwrapping, which can recover phase only up to a constant.
For simulated data, we subtracted the minimum value of the entire image to enforce nonnegativity.
For real data, we subtracted the mean value of a top-left area whose dimension is ($30 \times 30$) and corresponds to a background region.

\section{Experimental data}
\label{sec:expermental}
\subsection{Post-Processing of the Unwrapped Phase}

The microwells in which the organoids are loaded induce a non-flat (smooth) background.
For the sake of clarity, we removed the background of each unwrapped phase.
We estimated the background by fitting a polynomial of degree $3$ in background areas.
To detect the background, we applied a ($3 \times 3$) standard-deviation filter on the unwrapped image.
We defined the background as any pixel below a certain threshold~$T_\sigma \in [0.5,1]$.

\subsection{Parameter Setting}

For the optimization of the four real samples shown in Fig.~\ref{fig:realGAPhaseNet}, we set the number of iterations as 300, 1000, 300, and 800. 
The hyperparameters of the network were initialized to default values by PyTorch.
We used a learning rate of $0.01$. The weights $w_n$ were updated every $N_{w}=100$ iterations with $[\epsilon_{\mathrm{min}},\epsilon_{\mathrm{max}}]=[0.1, 8]$.
During a typical optimization, the weights~$w_n$ will be large in the area around sharp edges~\cite{kamilov2015isotropic}.
The parameter $\epsilon_\mathrm{min}$ prevents that the weights from becoming too large in the early iterations of the global optimization, which would force the corresponding pixels to be irreversibly set to zero.
Similarly, $\epsilon_\mathrm{max}$ ensures that the weights do not become too small.

\subsection{Reconstructions by GA and PhaseNet}

\begin{figure*}[!tbp]
\centering
\includegraphics[width=0.9\textwidth]{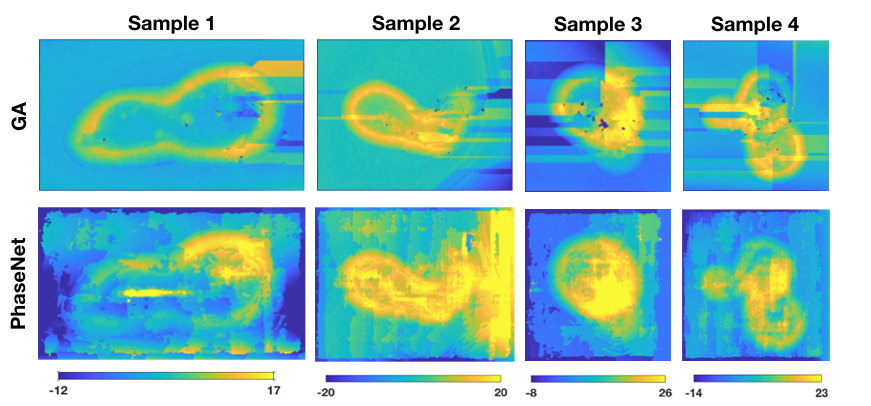}
\caption{Reconstructions of experimental data obtained by GA and PhaseNet.}
\label{fig:realGAPhaseNet}
\end{figure*}

In Fig.~\ref{fig:realGAPhaseNet}, GA~\cite{goldstein1988satellite} and PhaseNet failed to reconstruct the unwrapped phase for all real data.
GA solutions exhibit several rectangular areas that cover both the background and the organoids.
Their phase differs from their surrounding, which is inconsistent with the expected features of the sample.
PhaseNet solutions similarly show jumps along vertical stripes and are likely to be artifacts of unwrapping.
For PhaseNet, this behavior is expected since the network was trained on (mismatched) simulated data.

\subsection{Surface Plot of Reconstructions of Organoids}

\begin{figure*}[!tbp]
\centering
\includegraphics[width=0.95\textwidth]{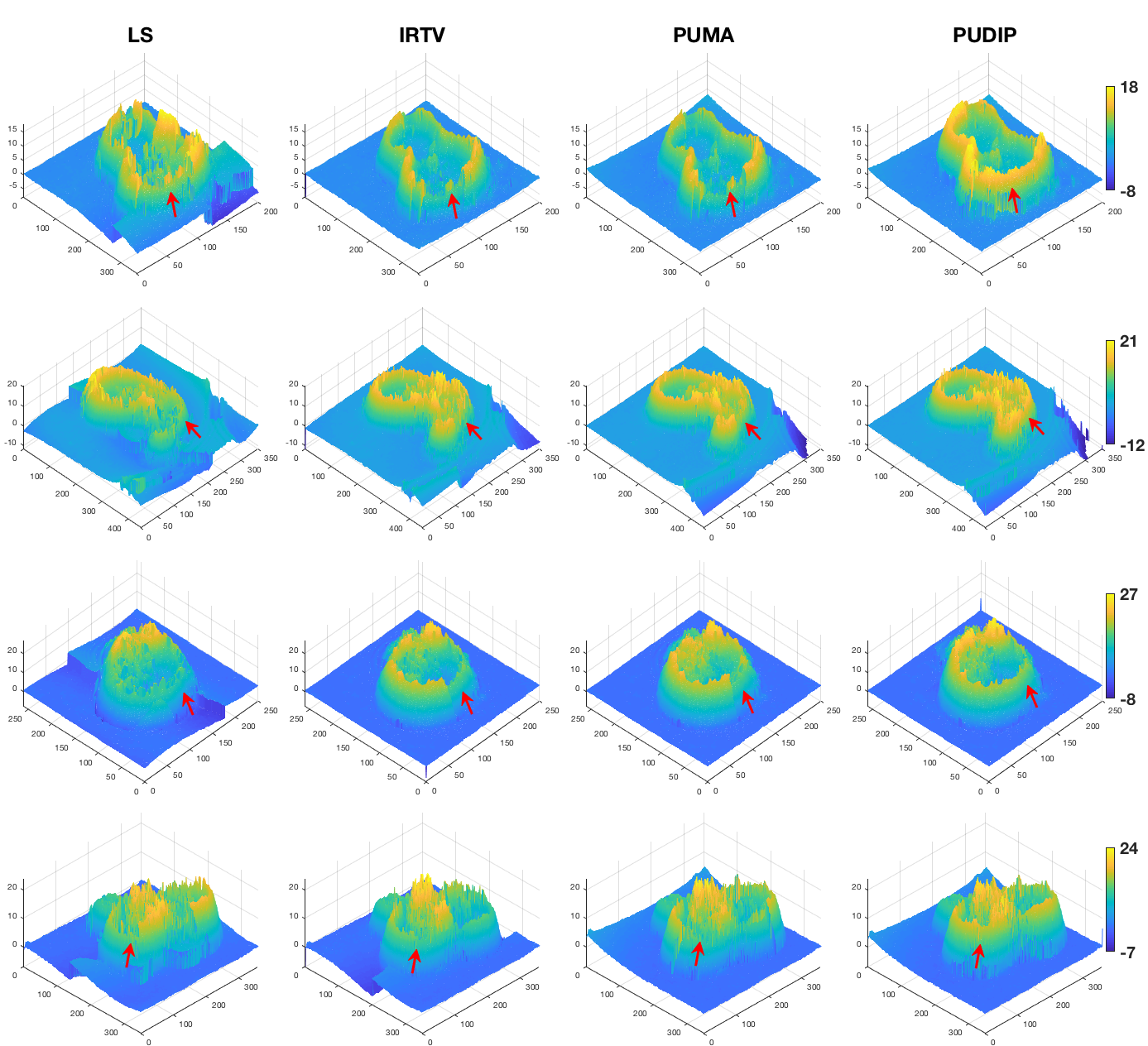}
\caption{Surface plots of the reconstructions of experimental data obtained by LS, IRTV, PUMA, and our method (from left to right).
Our framework leads to better visual performance at the area highlighted by the arrow.}
\label{fig:realmesh}
\end{figure*}

We emphasize in Fig.~\ref{fig:realmesh} the structural difference between the reconstructions by displaying a surface plot, especially the region highlighted by an arrow.

\subsection{Supplementary Time-Lapse Measurements and Segmentations and Reconstructions by PhaseNet}
%

\newcommand{\szx}{270}
\newcommand{\szy}{380}
\newcommand{\szsub}{0.115}
\newcommand{\szax}{4*\szsub}
\newcommand{\foldoi}{RealRK_B04_05_StackComparison_380_270}
\newcommand{\vmin}{-6.97}
\newcommand{\vmax}{18.32}
\pgfmathsetmacro{\step}{\vmin + (\vmax - \vmin)/4}
\newcommand{\insep}{1} 
%
\begin{figure*}[t]
\centering
    \begin{tikzpicture}[baseline=0pt]
    \begin{groupplot}[name = mygroupunw,
     group style = {group size = 4 by 5,
     vertical sep=\insep, horizontal sep=\insep},
     every plot/.style={width = {\szsub\textwidth},height = {\szy/\szx*\szsub\textwidth},axis equal image, scale only axis,
     }
    xmin = 0,xmax = \szx, ymin = 0, ymax = \szy,
    enlargelimits=false,
    axis equal image,
    scale only axis,
    width = {\szsub\textwidth},
    height = {\szy/\szx*\szsub\textwidth},
    hide axis,
    title style = {anchor=base}
    ]
    
    \nextgroupplot[ylabel = {\footnotesize \textbf{Frame 1}},
    y label style={at={(axis description cs:0.00001*\szx,.5)},anchor=south},
    yticklabels = {,,},
    xticklabels = {,,},
    hide axis = false,
    title = {\textbf{LS}},
        ]
    \addplot graphics[xmin = 0, xmax = \szx, ymin = 0, ymax = \szy] {Figures/\foldoi/LS1_1};

    \draw[-{Latex[length=1.25mm]}, thick,red] (0.05*\szx, 0.02*\szy)-- (0.15*\szx, 0.12*\szy);
    \draw[-{Latex[length=1.25mm]}, thick,red] (0.9*\szx, 0.05*\szy)-- (0.8*\szx, 0.15*\szy);
   \nextgroupplot[title = {\textbf{IRTV}},
        ]
    \addplot graphics[xmin = 0, xmax = \szx, ymin = 0, ymax = \szy] {Figures/\foldoi/IRTV1_1};
    \draw[-{Latex[length=1.25mm]}, thick,red] (0.05*\szx, 0.02*\szy)-- (0.15*\szx, 0.12*\szy);
    \draw[-{Latex[length=1.25mm]}, thick,red] (0.9*\szx, 0.05*\szy)-- (0.8*\szx, 0.15*\szy);
    \nextgroupplot[title = {\textbf{PUMA}},
        ]
    \addplot graphics[xmin = 0, xmax = \szx, ymin = 0, ymax = \szy] {Figures/\foldoi/PUMA1_1};
    \draw[-{Latex[length=1.25mm]}, thick,red] (0.05*\szx, 0.02*\szy)-- (0.15*\szx, 0.12*\szy);
    \draw[-{Latex[length=1.25mm]}, thick,red] (0.9*\szx, 0.05*\szy)-- (0.8*\szx, 0.15*\szy);
    \nextgroupplot[title = {\textbf{PUDIP}},
        ]
    \addplot graphics[xmin = 0, xmax = \szx, ymin = 0, ymax = \szy] {Figures/\foldoi/DIP1_1};
    \draw[-{Latex[length=1.25mm]}, thick,red] (0.05*\szx, 0.02*\szy)-- (0.15*\szx, 0.12*\szy);
    \draw[-{Latex[length=1.25mm]}, thick,red] (0.9*\szx, 0.05*\szy)-- (0.8*\szx, 0.15*\szy);
    \nextgroupplot[
    ylabel = {\footnotesize \textbf{Frame 3}},
    y label style={at={(axis description cs:0.00001*\szx,.5)},anchor=south},
    yticklabels = {,,},
    xticklabels = {,,},
    hide axis = false,
        ]
    \addplot graphics[xmin = 0, xmax = \szx, ymin = 0, ymax = \szy] {Figures/\foldoi/LS1_3};
    \draw[-{Latex[length=1.25mm]}, thick,red] (0.05*\szx, 0.02*\szy)-- (0.15*\szx, 0.12*\szy);
    \draw[-{Latex[length=1.25mm]}, thick,red] (0.9*\szx, 0.05*\szy)-- (0.8*\szx, 0.15*\szy);
    \nextgroupplot
    \addplot graphics[xmin = 0, xmax = \szx, ymin = 0, ymax = \szy] {Figures/\foldoi/IRTV1_3};
    \draw[-{Latex[length=1.25mm]}, thick,red] (0.05*\szx, 0.02*\szy)-- (0.15*\szx, 0.12*\szy);
    \draw[-{Latex[length=1.25mm]}, thick,red] (0.9*\szx, 0.05*\szy)-- (0.8*\szx, 0.15*\szy);
    \nextgroupplot
    \addplot graphics[xmin = 0, xmax = \szx, ymin = 0, ymax = \szy] {Figures/\foldoi/PUMA1_3};
    \draw[-{Latex[length=1.25mm]}, thick,red] (0.05*\szx, 0.02*\szy)-- (0.15*\szx, 0.12*\szy);
    \draw[-{Latex[length=1.25mm]}, thick,red] (0.9*\szx, 0.05*\szy)-- (0.8*\szx, 0.15*\szy);
    
    \nextgroupplot
    \addplot graphics[xmin = 0, xmax = \szx, ymin = 0, ymax = \szy] {Figures/\foldoi/DIP1_3};
    \draw[-{Latex[length=1.25mm]}, thick,red] (0.05*\szx, 0.02*\szy)-- (0.15*\szx, 0.12*\szy);
    \draw[-{Latex[length=1.25mm]}, thick,red] (0.9*\szx, 0.05*\szy)-- (0.8*\szx, 0.15*\szy);
    \nextgroupplot[
    ylabel = {\footnotesize \textbf{Frame 5}},
    y label style={at={(axis description cs:0.00001*\szx,.5)},anchor=south},
    yticklabels = {,,},
    xticklabels = {,,},
    hide axis = false,
        ]
    \addplot graphics[xmin = 0, xmax = \szx, ymin = 0, ymax = \szy] {Figures/\foldoi/LS1_5};
    \draw[-{Latex[length=1.25mm]}, thick,red] (0.05*\szx, 0.02*\szy)-- (0.15*\szx, 0.12*\szy);
    \draw[-{Latex[length=1.25mm]}, thick,red] (0.9*\szx, 0.05*\szy)-- (0.8*\szx, 0.15*\szy);
    \nextgroupplot
    \addplot graphics[xmin = 0, xmax = \szx, ymin = 0, ymax = \szy] {Figures/\foldoi/IRTV1_5};
    \draw[-{Latex[length=1.25mm]}, thick,red] (0.05*\szx, 0.02*\szy)-- (0.15*\szx, 0.12*\szy);
    \draw[-{Latex[length=1.25mm]}, thick,red] (0.9*\szx, 0.05*\szy)-- (0.8*\szx, 0.15*\szy);
    \nextgroupplot
    \addplot graphics[xmin = 0, xmax = \szx, ymin = 0, ymax = \szy] {Figures/\foldoi/PUMA1_5};
    \draw[-{Latex[length=1.25mm]}, thick,red] (0.05*\szx, 0.02*\szy)-- (0.15*\szx, 0.12*\szy);
    \draw[-{Latex[length=1.25mm]}, thick,red] (0.9*\szx, 0.05*\szy)-- (0.8*\szx, 0.15*\szy);
    \nextgroupplot
    \addplot graphics[xmin = 0, xmax = \szx, ymin = 0, ymax = \szy] {Figures/\foldoi/DIP1_5};
    \draw[-{Latex[length=1.25mm]}, thick,red] (0.05*\szx, 0.02*\szy)-- (0.15*\szx, 0.12*\szy);
    \draw[-{Latex[length=1.25mm]}, thick,red] (0.9*\szx, 0.05*\szy)-- (0.8*\szx, 0.15*\szy);

    \nextgroupplot[
    ylabel = {\footnotesize \textbf{Frame 7}},
    y label style={at={(axis description cs:0.00001*\szx,.5)},anchor=south},
    yticklabels = {,,},
    xticklabels = {,,},
    hide axis = false,
        ]
    \addplot graphics[xmin = 0, xmax = \szx, ymin = 0, ymax = \szy] {Figures/\foldoi/LS1_7};
    \draw[-{Latex[length=1.25mm]}, thick,red] (0.05*\szx, 0.02*\szy)-- (0.15*\szx, 0.12*\szy);
    \draw[-{Latex[length=1.25mm]}, thick,red] (0.9*\szx, 0.05*\szy)-- (0.8*\szx, 0.15*\szy);
    \nextgroupplot
    \addplot graphics[xmin = 0, xmax = \szx, ymin = 0, ymax = \szy] {Figures/\foldoi/IRTV1_7};
    \draw[-{Latex[length=1.25mm]}, thick,red] (0.05*\szx, 0.02*\szy)-- (0.15*\szx, 0.12*\szy);
    \draw[-{Latex[length=1.25mm]}, thick,red] (0.9*\szx, 0.05*\szy)-- (0.8*\szx, 0.15*\szy);
    \nextgroupplot\addplot graphics[xmin = 0, xmax = \szx, ymin = 0, ymax = \szy] {Figures/\foldoi/PUMA1_7};
    \draw[-{Latex[length=1.25mm]}, thick,red] (0.05*\szx, 0.02*\szy)-- (0.15*\szx, 0.12*\szy);
    \draw[-{Latex[length=1.25mm]}, thick,red] (0.9*\szx, 0.05*\szy)-- (0.8*\szx, 0.15*\szy);
    \nextgroupplot\addplot graphics[xmin = 0, xmax = \szx, ymin = 0, ymax = \szy] {Figures/\foldoi/DIP1_7};
    \draw[-{Latex[length=1.25mm]}, thick,red] (0.05*\szx, 0.02*\szy)-- (0.15*\szx, 0.12*\szy);
    \draw[-{Latex[length=1.25mm]}, thick,red] (0.9*\szx, 0.05*\szy)-- (0.8*\szx, 0.15*\szy);
    \nextgroupplot[
    ylabel = {\footnotesize \textbf{Frame 9}},
    y label style={at={(axis description cs:0.00001*\szx,.5)},anchor=south},
    yticklabels = {,,},
    xticklabels = {,,},
    hide axis = false,
        ]
    \addplot graphics[xmin = 0, xmax = \szx, ymin = 0, ymax = \szy] {Figures/\foldoi/LS1_9};
    \draw[-{Latex[length=1.25mm]}, thick,red] (0.05*\szx, 0.02*\szy)-- (0.15*\szx, 0.12*\szy);
    \draw[-{Latex[length=1.25mm]}, thick,red] (0.9*\szx, 0.05*\szy)-- (0.8*\szx, 0.15*\szy);
    \nextgroupplot
    \addplot graphics[xmin = 0, xmax = \szx, ymin = 0, ymax = \szy] {Figures/\foldoi/IRTV1_9};
    \draw[-{Latex[length=1.25mm]}, thick,red] (0.05*\szx, 0.02*\szy)-- (0.15*\szx, 0.12*\szy);
    \draw[-{Latex[length=1.25mm]}, thick,red] (0.9*\szx, 0.05*\szy)-- (0.8*\szx, 0.15*\szy);
    \nextgroupplot
    \addplot graphics[xmin = 0, xmax = \szx, ymin = 0, ymax = \szy] {Figures/\foldoi/PUMA1_9};
    \draw[-{Latex[length=1.25mm]}, thick,red] (0.05*\szx, 0.02*\szy)-- (0.15*\szx, 0.12*\szy);
    \draw[-{Latex[length=1.25mm]}, thick,red] (0.9*\szx, 0.05*\szy)-- (0.8*\szx, 0.15*\szy);
    \nextgroupplot
    \addplot graphics[xmin = 0, xmax = \szx, ymin = 0, ymax = \szy] {Figures/\foldoi/DIP1_9};
    \draw[-{Latex[length=1.25mm]}, thick,red] (0.05*\szx, 0.02*\szy)-- (0.15*\szx, 0.12*\szy);
    \draw[-{Latex[length=1.25mm]}, thick,red] (0.9*\szx, 0.05*\szy)-- (0.8*\szx, 0.15*\szy);
\end{groupplot}

\begin{axis}[at={($(group c1r5.south west) - (0,1em)$)}, anchor = north west,
    xmin = \vmin, xmax = \vmax, ymin = 0, ymax = 0.5,
    width=0,
    height = 0,
    scale only axis,
    enlargelimits=false,
    hide axis,
    axis equal image,
    colorbar horizontal,
    colormap/parula,
    colorbar style={
    at = {(0.025*\szax*\textwidth,0)},
        anchor = north west,
        xticklabel pos = lower,
        xmin = \vmin, xmax = \vmax,
        point meta min = \vmin,
        point meta max = \vmax,
        scale only axis,
        enlargelimits = false,
        scaled x ticks = true,
        width = {0.95*\szax*\textwidth},
    samples = 200,
    height = 0.15cm,
    xticklabel style={/pgf/number format/fixed, /pgf/number format/precision=0},
    xtick = {\vmin,\step,...,\vmax},
    xlabel={rad},
    xlabel style={yshift=0cm,xshift=0cm},
        }]
        \end{axis}
    \end{tikzpicture}
    \begin{tikzpicture}[baseline=0pt]
     \renewcommand{\foldoi}{RealRK_B04_05_StackComparison_380_270_seg_0.2}
    \begin{groupplot}[name = mygroupseg,
     group style = {group size = 4 by 5,
     vertical sep=\insep, horizontal sep=\insep},
     every plot/.style={width = {\szsub\textwidth},height = {\szy/\szx*\szsub\textwidth},axis equal image, scale only axis,
     }
    xmin = 0,xmax = \szx, ymin = 0, ymax = \szy,
    enlargelimits=false,
    axis equal image,
    scale only axis,
    width = {\szsub\textwidth},
    height = {\szy/\szx*\szsub\textwidth},
    hide axis,
    title style = {anchor=base}
    ]
    
    \nextgroupplot[
    y label style={at={(axis description cs:0.00001*\szx,.5)},anchor=south},
    yticklabels = {,,},
    xticklabels = {,,},
    hide axis = false,
    title = {\textbf{LS}},
        ]
    \addplot graphics[xmin = 0, xmax = \szx, ymin = 0, ymax = \szy] {Figures/\foldoi/LS1_1};
    \draw[-{Latex[length=1.25mm]}, thick,red] (0.05*\szx, 0.02*\szy)-- (0.15*\szx, 0.12*\szy);
    \draw[-{Latex[length=1.25mm]}, thick,red] (0.9*\szx, 0.05*\szy)-- (0.8*\szx, 0.15*\szy);
   \nextgroupplot[title = {\textbf{IRTV}},
        ]
    \addplot graphics[xmin = 0, xmax = \szx, ymin = 0, ymax = \szy] {Figures/\foldoi/IRTV1_1};
    \draw[-{Latex[length=1.25mm]}, thick,red] (0.05*\szx, 0.02*\szy)-- (0.15*\szx, 0.12*\szy);
    \draw[-{Latex[length=1.25mm]}, thick,red] (0.9*\szx, 0.05*\szy)-- (0.8*\szx, 0.15*\szy);
    \nextgroupplot[title = {\textbf{PUMA}},
        ]
    \addplot graphics[xmin = 0, xmax = \szx, ymin = 0, ymax = \szy] {Figures/\foldoi/PUMA1_1};
    \draw[-{Latex[length=1.25mm]}, thick,red] (0.05*\szx, 0.02*\szy)-- (0.15*\szx, 0.12*\szy);
    \draw[-{Latex[length=1.25mm]}, thick,red] (0.9*\szx, 0.05*\szy)-- (0.8*\szx, 0.15*\szy);
    \nextgroupplot[title = {\textbf{PUDIP}},
        ]
    \addplot graphics[xmin = 0, xmax = \szx, ymin = 0, ymax = \szy] {Figures/\foldoi/DIP1_1};
    \draw[-{Latex[length=1.25mm]}, thick,red] (0.05*\szx, 0.02*\szy)-- (0.15*\szx, 0.12*\szy);
    \draw[-{Latex[length=1.25mm]}, thick,red] (0.9*\szx, 0.05*\szy)-- (0.8*\szx, 0.15*\szy);
    \nextgroupplot[
    y label style={at={(axis description cs:0.00001*\szx,.5)},anchor=south},
    yticklabels = {,,},
    xticklabels = {,,},
    hide axis = false,
        ]
    \addplot graphics[xmin = 0, xmax = \szx, ymin = 0, ymax = \szy] {Figures/\foldoi/LS1_3};
    \draw[-{Latex[length=1.25mm]}, thick,red] (0.05*\szx, 0.02*\szy)-- (0.15*\szx, 0.12*\szy);
    \draw[-{Latex[length=1.25mm]}, thick,red] (0.9*\szx, 0.05*\szy)-- (0.8*\szx, 0.15*\szy);
    \nextgroupplot
    \addplot graphics[xmin = 0, xmax = \szx, ymin = 0, ymax = \szy] {Figures/\foldoi/IRTV1_3};
    \draw[-{Latex[length=1.25mm]}, thick,red] (0.05*\szx, 0.02*\szy)-- (0.15*\szx, 0.12*\szy);
    \draw[-{Latex[length=1.25mm]}, thick,red] (0.9*\szx, 0.05*\szy)-- (0.8*\szx, 0.15*\szy);
    \nextgroupplot
    \addplot graphics[xmin = 0, xmax = \szx, ymin = 0, ymax = \szy] {Figures/\foldoi/PUMA1_3};
    \draw[-{Latex[length=1.25mm]}, thick,red] (0.05*\szx, 0.02*\szy)-- (0.15*\szx, 0.12*\szy);
    \draw[-{Latex[length=1.25mm]}, thick,red] (0.9*\szx, 0.05*\szy)-- (0.8*\szx, 0.15*\szy);
    \nextgroupplot
    \addplot graphics[xmin = 0, xmax = \szx, ymin = 0, ymax = \szy] {Figures/\foldoi/DIP1_3};
    \draw[-{Latex[length=1.25mm]}, thick,red] (0.05*\szx, 0.02*\szy)-- (0.15*\szx, 0.12*\szy);
    \draw[-{Latex[length=1.25mm]}, thick,red] (0.9*\szx, 0.05*\szy)-- (0.8*\szx, 0.15*\szy);
    \nextgroupplot[
    y label style={at={(axis description cs:0.00001*\szx,.5)},anchor=south},
    yticklabels = {,,},
    xticklabels = {,,},
    hide axis = false,
        ]
    \addplot graphics[xmin = 0, xmax = \szx, ymin = 0, ymax = \szy] {Figures/\foldoi/LS1_5};
    \draw[-{Latex[length=1.25mm]}, thick,red] (0.05*\szx, 0.02*\szy)-- (0.15*\szx, 0.12*\szy);
    \draw[-{Latex[length=1.25mm]}, thick,red] (0.9*\szx, 0.05*\szy)-- (0.8*\szx, 0.15*\szy);
    \nextgroupplot
    \addplot graphics[xmin = 0, xmax = \szx, ymin = 0, ymax = \szy] {Figures/\foldoi/IRTV1_5};
    \draw[-{Latex[length=1.25mm]}, thick,red] (0.05*\szx, 0.02*\szy)-- (0.15*\szx, 0.12*\szy);
    \draw[-{Latex[length=1.25mm]}, thick,red] (0.9*\szx, 0.05*\szy)-- (0.8*\szx, 0.15*\szy);
    \nextgroupplot
    \addplot graphics[xmin = 0, xmax = \szx, ymin = 0, ymax = \szy] {Figures/\foldoi/PUMA1_5};
    \draw[-{Latex[length=1.25mm]}, thick,red] (0.05*\szx, 0.02*\szy)-- (0.15*\szx, 0.12*\szy);
    \draw[-{Latex[length=1.25mm]}, thick,red] (0.9*\szx, 0.05*\szy)-- (0.8*\szx, 0.15*\szy);
    \nextgroupplot
    \addplot graphics[xmin = 0, xmax = \szx, ymin = 0, ymax = \szy] {Figures/\foldoi/DIP1_5};
    \draw[-{Latex[length=1.25mm]}, thick,red] (0.05*\szx, 0.02*\szy)-- (0.15*\szx, 0.12*\szy);
    \draw[-{Latex[length=1.25mm]}, thick,red] (0.9*\szx, 0.05*\szy)-- (0.8*\szx, 0.15*\szy);
    \nextgroupplot[
    y label style={at={(axis description cs:0.00001*\szx,.5)},anchor=south},
    yticklabels = {,,},
    xticklabels = {,,},
    hide axis = false,
        ]
    \addplot graphics[xmin = 0, xmax = \szx, ymin = 0, ymax = \szy] {Figures/\foldoi/LS1_7};
    \draw[-{Latex[length=1.25mm]}, thick,red] (0.05*\szx, 0.02*\szy)-- (0.15*\szx, 0.12*\szy);
    \draw[-{Latex[length=1.25mm]}, thick,red] (0.9*\szx, 0.05*\szy)-- (0.8*\szx, 0.15*\szy);
    \nextgroupplot
    \addplot graphics[xmin = 0, xmax = \szx, ymin = 0, ymax = \szy] {Figures/\foldoi/IRTV1_7};
    \draw[-{Latex[length=1.25mm]}, thick,red] (0.05*\szx, 0.02*\szy)-- (0.15*\szx, 0.12*\szy);
    \draw[-{Latex[length=1.25mm]}, thick,red] (0.9*\szx, 0.05*\szy)-- (0.8*\szx, 0.15*\szy);
    \nextgroupplot\addplot graphics[xmin = 0, xmax = \szx, ymin = 0, ymax = \szy] {Figures/\foldoi/PUMA1_7};
    \draw[-{Latex[length=1.25mm]}, thick,red] (0.05*\szx, 0.02*\szy)-- (0.15*\szx, 0.12*\szy);
    \draw[-{Latex[length=1.25mm]}, thick,red] (0.9*\szx, 0.05*\szy)-- (0.8*\szx, 0.15*\szy);
    \nextgroupplot\addplot graphics[xmin = 0, xmax = \szx, ymin = 0, ymax = \szy] {Figures/\foldoi/DIP1_7};
    \draw[-{Latex[length=1.25mm]}, thick,red] (0.05*\szx, 0.02*\szy)-- (0.15*\szx, 0.12*\szy);
    \draw[-{Latex[length=1.25mm]}, thick,red] (0.9*\szx, 0.05*\szy)-- (0.8*\szx, 0.15*\szy);
    \nextgroupplot[
    y label style={at={(axis description cs:0.00001*\szx,.5)},anchor=south},
    yticklabels = {,,},
    xticklabels = {,,},
    hide axis = false,
        ]
    \addplot graphics[xmin = 0, xmax = \szx, ymin = 0, ymax = \szy] {Figures/\foldoi/LS1_9};
    \draw[-{Latex[length=1.25mm]}, thick,red] (0.05*\szx, 0.02*\szy)-- (0.15*\szx, 0.12*\szy);
    \draw[-{Latex[length=1.25mm]}, thick,red] (0.9*\szx, 0.05*\szy)-- (0.8*\szx, 0.15*\szy);
    \nextgroupplot
    \addplot graphics[xmin = 0, xmax = \szx, ymin = 0, ymax = \szy] {Figures/\foldoi/IRTV1_9};
    \draw[-{Latex[length=1.25mm]}, thick,red] (0.05*\szx, 0.02*\szy)-- (0.15*\szx, 0.12*\szy);
    \draw[-{Latex[length=1.25mm]}, thick,red] (0.9*\szx, 0.05*\szy)-- (0.8*\szx, 0.15*\szy);
    \nextgroupplot
    \addplot graphics[xmin = 0, xmax = \szx, ymin = 0, ymax = \szy] {Figures/\foldoi/PUMA1_9};
    \draw[-{Latex[length=1.25mm]}, thick,red] (0.05*\szx, 0.02*\szy)-- (0.15*\szx, 0.12*\szy);
    \draw[-{Latex[length=1.25mm]}, thick,red] (0.9*\szx, 0.05*\szy)-- (0.8*\szx, 0.15*\szy);
    \nextgroupplot
    \addplot graphics[xmin = 0, xmax = \szx, ymin = 0, ymax = \szy] {Figures/\foldoi/DIP1_9};
    \draw[-{Latex[length=1.25mm]}, thick,red] (0.05*\szx, 0.02*\szy)-- (0.15*\szx, 0.12*\szy);
    \draw[-{Latex[length=1.25mm]}, thick,red] (0.9*\szx, 0.05*\szy)-- (0.8*\szx, 0.15*\szy);
\end{groupplot}

\end{tikzpicture}
\caption{Time-lapse reconstructions for supplementary real data of size ($380 \times 270$) and their corresponding segmentation. Left: the images were saturated for visualization purpose. 
Right: segmentation of time-lapse reconstructions.
We thresholded at $20\%$ of the maximum value of the image.}\label{fig:timelapsesupp1}
\end{figure*}

\renewcommand{\szx}{380}
\renewcommand{\szy}{320}
\renewcommand{\szsub}{0.115}
\renewcommand{\szax}{4*\szsub}
\renewcommand{\foldoi}{RealRK_B04_09_StackComparison_320_380}

\renewcommand{\vmin}{-7.73}
\renewcommand{\vmax}{20.90}
\pgfmathsetmacro{\step}{\vmin + (\vmax - \vmin)/4}
\renewcommand{\insep}{1} 

\begin{figure*}[t]
    \centering
    \begin{tikzpicture}[baseline=0pt]
    \begin{groupplot}[name = mygroup,
     group style = {group size = 4 by 5,
     vertical sep=\insep, horizontal sep=\insep},
     every plot/.style={width = {\szsub\textwidth},height = {\szy/\szx*\szsub\textwidth},axis equal image, scale only axis,
     }
    xmin = 0,xmax = \szx, ymin = 0, ymax = \szy,
    enlargelimits=false,
    axis equal image,
    scale only axis,
    width = {\szsub\textwidth},
    height = {\szy/\szx*\szsub\textwidth},
    hide axis,
    title style = {anchor=base}
    ]
    
    \nextgroupplot[ylabel = {\footnotesize \textbf{Frame 1}},
    y label style={at={(axis description cs:0.00001*\szx,.5)},anchor=south},
    yticklabels = {,,},
    xticklabels = {,,},
    hide axis = false,
    title = {\textbf{LS}},
        ]
    \addplot graphics[xmin = 0, xmax = \szx, ymin = 0, ymax = \szy] {Figures/\foldoi/LS1_1};
    
    \draw[-{Latex[length=1.25mm]}, thick,red] (0.1*\szx, 0.07*\szy)-- (0.2*\szx, 0.17*\szy);
    \draw[-{Latex[length=1.25mm]}, thick,red] ( 0.05*\szx, 0.89*\szy)-- (0.15*\szx, 0.79*\szy);
    
   \nextgroupplot[title = {\textbf{IRTV}},
        ]
    \addplot graphics[xmin = 0, xmax = \szx, ymin = 0, ymax = \szy] {Figures/\foldoi/IRTV1_1};
    \draw[-{Latex[length=1.25mm]}, thick,red] (0.1*\szx, 0.07*\szy)-- (0.2*\szx, 0.17*\szy);
    \draw[-{Latex[length=1.25mm]}, thick,red] ( 0.05*\szx, 0.89*\szy)-- (0.15*\szx, 0.79*\szy);
    \nextgroupplot[title = {\textbf{PUMA}},
        ]
    \addplot graphics[xmin = 0, xmax = \szx, ymin = 0, ymax = \szy] {Figures/\foldoi/PUMA1_1};
    \draw[-{Latex[length=1.25mm]}, thick,red] (0.1*\szx, 0.07*\szy)-- (0.2*\szx, 0.17*\szy);
    \draw[-{Latex[length=1.25mm]}, thick,red] ( 0.05*\szx, 0.89*\szy)-- (0.15*\szx, 0.79*\szy);
    \nextgroupplot[title = {\textbf{PUDIP}},
        ]
    \addplot graphics[xmin = 0, xmax = \szx, ymin = 0, ymax = \szy] {Figures/\foldoi/DIP1_1};
    \draw[-{Latex[length=1.25mm]}, thick,red] (0.1*\szx, 0.07*\szy)-- (0.2*\szx, 0.17*\szy);
    \draw[-{Latex[length=1.25mm]}, thick,red] ( 0.05*\szx, 0.89*\szy)-- (0.15*\szx, 0.79*\szy);
    \nextgroupplot[
    ylabel = {\footnotesize \textbf{Frame 3}},
    y label style={at={(axis description cs:0.00001*\szx,.5)},anchor=south},
    yticklabels = {,,},
    xticklabels = {,,},
    hide axis = false,
        ]
    \addplot graphics[xmin = 0, xmax = \szx, ymin = 0, ymax = \szy] {Figures/\foldoi/LS1_3};
    \draw[-{Latex[length=1.25mm]}, thick,red] (0.1*\szx, 0.07*\szy)-- (0.2*\szx, 0.17*\szy);
    \draw[-{Latex[length=1.25mm]}, thick,red] ( 0.05*\szx, 0.89*\szy)-- (0.15*\szx, 0.79*\szy);
    \nextgroupplot
    \addplot graphics[xmin = 0, xmax = \szx, ymin = 0, ymax = \szy] {Figures/\foldoi/IRTV1_3};
    \draw[-{Latex[length=1.25mm]}, thick,red] (0.1*\szx, 0.07*\szy)-- (0.2*\szx, 0.17*\szy);
    \draw[-{Latex[length=1.25mm]}, thick,red] ( 0.05*\szx, 0.89*\szy)-- (0.15*\szx, 0.79*\szy);
    \nextgroupplot
    \addplot graphics[xmin = 0, xmax = \szx, ymin = 0, ymax = \szy] {Figures/\foldoi/PUMA1_3};
    \draw[-{Latex[length=1.25mm]}, thick,red] (0.1*\szx, 0.07*\szy)-- (0.2*\szx, 0.17*\szy);
    \draw[-{Latex[length=1.25mm]}, thick,red] ( 0.05*\szx, 0.89*\szy)-- (0.15*\szx, 0.79*\szy);
    
    \nextgroupplot
    \addplot graphics[xmin = 0, xmax = \szx, ymin = 0, ymax = \szy] {Figures/\foldoi/DIP1_3};
    \draw[-{Latex[length=1.25mm]}, thick,red] (0.1*\szx, 0.07*\szy)-- (0.2*\szx, 0.17*\szy);
    \draw[-{Latex[length=1.25mm]}, thick,red] ( 0.05*\szx, 0.89*\szy)-- (0.15*\szx, 0.79*\szy);
    \nextgroupplot[
    ylabel = {\footnotesize \textbf{Frame 5}},
    y label style={at={(axis description cs:0.00001*\szx,.5)},anchor=south},
    yticklabels = {,,},
    xticklabels = {,,},
    hide axis = false,
        ]
    \addplot graphics[xmin = 0, xmax = \szx, ymin = 0, ymax = \szy] {Figures/\foldoi/LS1_5};
    \draw[-{Latex[length=1.25mm]}, thick,red] (0.1*\szx, 0.07*\szy)-- (0.2*\szx, 0.17*\szy);
    \draw[-{Latex[length=1.25mm]}, thick,red] ( 0.05*\szx, 0.89*\szy)-- (0.15*\szx, 0.79*\szy);
    \nextgroupplot
    \addplot graphics[xmin = 0, xmax = \szx, ymin = 0, ymax = \szy] {Figures/\foldoi/IRTV1_5};
    \draw[-{Latex[length=1.25mm]}, thick,red] (0.1*\szx, 0.07*\szy)-- (0.2*\szx, 0.17*\szy);
    \draw[-{Latex[length=1.25mm]}, thick,red] ( 0.05*\szx, 0.89*\szy)-- (0.15*\szx, 0.79*\szy);
    \nextgroupplot
    \addplot graphics[xmin = 0, xmax = \szx, ymin = 0, ymax = \szy] {Figures/\foldoi/PUMA1_5};
    \draw[-{Latex[length=1.25mm]}, thick,red] (0.1*\szx, 0.07*\szy)-- (0.2*\szx, 0.17*\szy);
    \draw[-{Latex[length=1.25mm]}, thick,red] ( 0.05*\szx, 0.89*\szy)-- (0.15*\szx, 0.79*\szy);
    \nextgroupplot
    \addplot graphics[xmin = 0, xmax = \szx, ymin = 0, ymax = \szy] {Figures/\foldoi/DIP1_5};
    \draw[-{Latex[length=1.25mm]}, thick,red] (0.1*\szx, 0.07*\szy)-- (0.2*\szx, 0.17*\szy);
    \draw[-{Latex[length=1.25mm]}, thick,red] ( 0.05*\szx, 0.89*\szy)-- (0.15*\szx, 0.79*\szy);

    \nextgroupplot[
    ylabel = {\footnotesize \textbf{Frame 7}},
    y label style={at={(axis description cs:0.00001*\szx,.5)},anchor=south},
    yticklabels = {,,},
    xticklabels = {,,},
    hide axis = false,
        ]
    \addplot graphics[xmin = 0, xmax = \szx, ymin = 0, ymax = \szy] {Figures/\foldoi/LS1_7};
    \draw[-{Latex[length=1.25mm]}, thick,red] (0.1*\szx, 0.07*\szy)-- (0.2*\szx, 0.17*\szy);
    \draw[-{Latex[length=1.25mm]}, thick,red] ( 0.05*\szx, 0.89*\szy)-- (0.15*\szx, 0.79*\szy);
    \nextgroupplot
    \addplot graphics[xmin = 0, xmax = \szx, ymin = 0, ymax = \szy] {Figures/\foldoi/IRTV1_7};
    \draw[-{Latex[length=1.25mm]}, thick,red] (0.1*\szx, 0.07*\szy)-- (0.2*\szx, 0.17*\szy);
    \draw[-{Latex[length=1.25mm]}, thick,red] ( 0.05*\szx, 0.89*\szy)-- (0.15*\szx, 0.79*\szy);
    \nextgroupplot\addplot graphics[xmin = 0, xmax = \szx, ymin = 0, ymax = \szy] {Figures/\foldoi/PUMA1_7};
    \draw[-{Latex[length=1.25mm]}, thick,red] (0.1*\szx, 0.07*\szy)-- (0.2*\szx, 0.17*\szy);
    \draw[-{Latex[length=1.25mm]}, thick,red] ( 0.05*\szx, 0.89*\szy)-- (0.15*\szx, 0.79*\szy);
    \nextgroupplot\addplot graphics[xmin = 0, xmax = \szx, ymin = 0, ymax = \szy] {Figures/\foldoi/DIP1_7_new};
    \draw[-{Latex[length=1.25mm]}, thick,red] (0.1*\szx, 0.07*\szy)-- (0.2*\szx, 0.17*\szy);
    \draw[-{Latex[length=1.25mm]}, thick,red] ( 0.05*\szx, 0.89*\szy)-- (0.15*\szx, 0.79*\szy);
    \nextgroupplot[
    ylabel = {\footnotesize \textbf{Frame 9}},
    y label style={at={(axis description cs:0.00001*\szx,.5)},anchor=south},
    yticklabels = {,,},
    xticklabels = {,,},
    hide axis = false,
        ]
    \addplot graphics[xmin = 0, xmax = \szx, ymin = 0, ymax = \szy] {Figures/\foldoi/LS1_9};
    \draw[-{Latex[length=1.25mm]}, thick,red] (0.1*\szx, 0.07*\szy)-- (0.2*\szx, 0.17*\szy);
    \draw[-{Latex[length=1.25mm]}, thick,red] ( 0.05*\szx, 0.89*\szy)-- (0.15*\szx, 0.79*\szy);
    \nextgroupplot
    \addplot graphics[xmin = 0, xmax = \szx, ymin = 0, ymax = \szy] {Figures/\foldoi/IRTV1_9};
    \draw[-{Latex[length=1.25mm]}, thick,red] (0.1*\szx, 0.07*\szy)-- (0.2*\szx, 0.17*\szy);
    \draw[-{Latex[length=1.25mm]}, thick,red] ( 0.05*\szx, 0.89*\szy)-- (0.15*\szx, 0.79*\szy);
    \nextgroupplot
    \addplot graphics[xmin = 0, xmax = \szx, ymin = 0, ymax = \szy] {Figures/\foldoi/PUMA1_9};
    \draw[-{Latex[length=1.25mm]}, thick,red] (0.1*\szx, 0.07*\szy)-- (0.2*\szx, 0.17*\szy);
    \draw[-{Latex[length=1.25mm]}, thick,red] ( 0.05*\szx, 0.89*\szy)-- (0.15*\szx, 0.79*\szy);
    \nextgroupplot
    \addplot graphics[xmin = 0, xmax = \szx, ymin = 0, ymax = \szy] {Figures/\foldoi/DIP1_9};
    \draw[-{Latex[length=1.25mm]}, thick,red] (0.1*\szx, 0.07*\szy)-- (0.2*\szx, 0.17*\szy);
    \draw[-{Latex[length=1.25mm]}, thick,red] ( 0.05*\szx, 0.89*\szy)-- (0.15*\szx, 0.79*\szy);
\end{groupplot}

\begin{axis}[at={($(group c1r5.south west) - (0,1em)$)}, anchor = north west,
    xmin = \vmin, xmax = \vmax, ymin = 0, ymax = 0.5,
    width=0,
    height = 0,
    scale only axis,
    enlargelimits=false,
    hide axis,
    axis equal image,
    colorbar horizontal,
    colormap/parula,
    colorbar style={
    at = {(0.025*\szax*\textwidth,0)},
        anchor = north west,
        xticklabel pos = lower,
        xmin = \vmin, xmax = \vmax,
        point meta min = \vmin,
        point meta max = \vmax,
        scale only axis,
        enlargelimits = false,
        scaled x ticks = true,
        width = {0.95*\szax*\textwidth},
    samples = 200,
    height = 0.15cm,
    xticklabel style={/pgf/number format/fixed, /pgf/number format/precision=0},
    xtick = {\vmin,\step,...,\vmax},
    xlabel={rad},
    xlabel style={yshift=0cm,xshift=0cm},
        }]
        \end{axis}
    \end{tikzpicture}
    \begin{tikzpicture}[baseline=0pt]
    \renewcommand{\foldoi}{RealRK_B04_09_StackComparison_320_380_seg_0.2}
    \begin{groupplot}[name = mygroup,
     group style = {group size = 4 by 5,
     vertical sep=\insep, horizontal sep=\insep},
     every plot/.style={width = {\szsub\textwidth},height = {\szy/\szx*\szsub\textwidth},axis equal image, scale only axis,
     }
    xmin = 0,xmax = \szx, ymin = 0, ymax = \szy,
    enlargelimits=false,
    axis equal image,
    scale only axis,
    width = {\szsub\textwidth},
    height = {\szy/\szx*\szsub\textwidth},
    hide axis,
    title style = {anchor=base}
    ]
    
    \nextgroupplot[
    y label style={at={(axis description cs:0.00001*\szx,.5)},anchor=south},
    yticklabels = {,,},
    xticklabels = {,,},
    hide axis = false,
    title = {\textbf{LS}},
        ]
    \addplot graphics[xmin = 0, xmax = \szx, ymin = 0, ymax = \szy] {Figures/\foldoi/LS1_1};
    \draw[-{Latex[length=1.25mm]}, thick,red] (0.1*\szx, 0.07*\szy)-- (0.2*\szx, 0.17*\szy);
    \draw[-{Latex[length=1.25mm]}, thick,red] ( 0.05*\szx, 0.89*\szy)-- (0.15*\szx, 0.79*\szy);
   \nextgroupplot[title = {\textbf{IRTV}},
        ]
    \addplot graphics[xmin = 0, xmax = \szx, ymin = 0, ymax = \szy] {Figures/\foldoi/IRTV1_1};
    \draw[-{Latex[length=1.25mm]}, thick,red] (0.1*\szx, 0.07*\szy)-- (0.2*\szx, 0.17*\szy);
    \draw[-{Latex[length=1.25mm]}, thick,red] ( 0.05*\szx, 0.89*\szy)-- (0.15*\szx, 0.79*\szy);
    \nextgroupplot[title = {\textbf{PUMA}},
        ]
    \addplot graphics[xmin = 0, xmax = \szx, ymin = 0, ymax = \szy] {Figures/\foldoi/PUMA1_1};
    \draw[-{Latex[length=1.25mm]}, thick,red] (0.1*\szx, 0.07*\szy)-- (0.2*\szx, 0.17*\szy);
    \draw[-{Latex[length=1.25mm]}, thick,red] ( 0.05*\szx, 0.89*\szy)-- (0.15*\szx, 0.79*\szy);
    \nextgroupplot[title = {\textbf{PUDIP}},
        ]
    \addplot graphics[xmin = 0, xmax = \szx, ymin = 0, ymax = \szy] {Figures/\foldoi/DIP1_1};
    \draw[-{Latex[length=1.25mm]}, thick,red] (0.1*\szx, 0.07*\szy)-- (0.2*\szx, 0.17*\szy);
    \draw[-{Latex[length=1.25mm]}, thick,red] ( 0.05*\szx, 0.89*\szy)-- (0.15*\szx, 0.79*\szy);
    \nextgroupplot[
    y label style={at={(axis description cs:0.00001*\szx,.5)},anchor=south},
    yticklabels = {,,},
    xticklabels = {,,},
    hide axis = false,
        ]
    \addplot graphics[xmin = 0, xmax = \szx, ymin = 0, ymax = \szy] {Figures/\foldoi/LS1_3};
    \draw[-{Latex[length=1.25mm]}, thick,red] (0.1*\szx, 0.07*\szy)-- (0.2*\szx, 0.17*\szy);
    \draw[-{Latex[length=1.25mm]}, thick,red] ( 0.05*\szx, 0.89*\szy)-- (0.15*\szx, 0.79*\szy);
    \nextgroupplot
    \addplot graphics[xmin = 0, xmax = \szx, ymin = 0, ymax = \szy] {Figures/\foldoi/IRTV1_3};
    \draw[-{Latex[length=1.25mm]}, thick,red] (0.1*\szx, 0.07*\szy)-- (0.2*\szx, 0.17*\szy);
    \draw[-{Latex[length=1.25mm]}, thick,red] ( 0.05*\szx, 0.89*\szy)-- (0.15*\szx, 0.79*\szy);
    \nextgroupplot
    \addplot graphics[xmin = 0, xmax = \szx, ymin = 0, ymax = \szy] {Figures/\foldoi/PUMA1_3};
    \draw[-{Latex[length=1.25mm]}, thick,red] (0.1*\szx, 0.07*\szy)-- (0.2*\szx, 0.17*\szy);
    \draw[-{Latex[length=1.25mm]}, thick,red] ( 0.05*\szx, 0.89*\szy)-- (0.15*\szx, 0.79*\szy);
    \nextgroupplot
    \addplot graphics[xmin = 0, xmax = \szx, ymin = 0, ymax = \szy] {Figures/\foldoi/DIP1_3};
    \draw[-{Latex[length=1.25mm]}, thick,red] (0.1*\szx, 0.07*\szy)-- (0.2*\szx, 0.17*\szy);
    \draw[-{Latex[length=1.25mm]}, thick,red] ( 0.05*\szx, 0.89*\szy)-- (0.15*\szx, 0.79*\szy);
    \nextgroupplot[
    y label style={at={(axis description cs:0.00001*\szx,.5)},anchor=south},
    yticklabels = {,,},
    xticklabels = {,,},
    hide axis = false,
        ]
    \addplot graphics[xmin = 0, xmax = \szx, ymin = 0, ymax = \szy] {Figures/\foldoi/LS1_5};
    \draw[-{Latex[length=1.25mm]}, thick,red] (0.1*\szx, 0.07*\szy)-- (0.2*\szx, 0.17*\szy);
    \draw[-{Latex[length=1.25mm]}, thick,red] ( 0.05*\szx, 0.89*\szy)-- (0.15*\szx, 0.79*\szy);
    \nextgroupplot
    \addplot graphics[xmin = 0, xmax = \szx, ymin = 0, ymax = \szy] {Figures/\foldoi/IRTV1_5};
    \draw[-{Latex[length=1.25mm]}, thick,red] (0.1*\szx, 0.07*\szy)-- (0.2*\szx, 0.17*\szy);
    \draw[-{Latex[length=1.25mm]}, thick,red] ( 0.05*\szx, 0.89*\szy)-- (0.15*\szx, 0.79*\szy);
    \nextgroupplot
    \addplot graphics[xmin = 0, xmax = \szx, ymin = 0, ymax = \szy] {Figures/\foldoi/PUMA1_5};
    \draw[-{Latex[length=1.25mm]}, thick,red] (0.1*\szx, 0.07*\szy)-- (0.2*\szx, 0.17*\szy);
    \draw[-{Latex[length=1.25mm]}, thick,red] ( 0.05*\szx, 0.89*\szy)-- (0.15*\szx, 0.79*\szy);
    \nextgroupplot
    \addplot graphics[xmin = 0, xmax = \szx, ymin = 0, ymax = \szy] {Figures/\foldoi/DIP1_5};
    \draw[-{Latex[length=1.25mm]}, thick,red] (0.1*\szx, 0.07*\szy)-- (0.2*\szx, 0.17*\szy);
    \draw[-{Latex[length=1.25mm]}, thick,red] ( 0.05*\szx, 0.89*\szy)-- (0.15*\szx, 0.79*\szy);
    \nextgroupplot[
    y label style={at={(axis description cs:0.00001*\szx,.5)},anchor=south},
    yticklabels = {,,},
    xticklabels = {,,},
    hide axis = false,
        ]
    \addplot graphics[xmin = 0, xmax = \szx, ymin = 0, ymax = \szy] {Figures/\foldoi/LS1_7};
    \draw[-{Latex[length=1.25mm]}, thick,red] (0.1*\szx, 0.07*\szy)-- (0.2*\szx, 0.17*\szy);
    \draw[-{Latex[length=1.25mm]}, thick,red] ( 0.05*\szx, 0.89*\szy)-- (0.15*\szx, 0.79*\szy);
    \nextgroupplot
    \addplot graphics[xmin = 0, xmax = \szx, ymin = 0, ymax = \szy] {Figures/\foldoi/IRTV1_7};
    \draw[-{Latex[length=1.25mm]}, thick,red] (0.1*\szx, 0.07*\szy)-- (0.2*\szx, 0.17*\szy);
    \draw[-{Latex[length=1.25mm]}, thick,red] ( 0.05*\szx, 0.89*\szy)-- (0.15*\szx, 0.79*\szy);
    \nextgroupplot\addplot graphics[xmin = 0, xmax = \szx, ymin = 0, ymax = \szy] {Figures/\foldoi/PUMA1_7};
    \draw[-{Latex[length=1.25mm]}, thick,red] (0.1*\szx, 0.07*\szy)-- (0.2*\szx, 0.17*\szy);
    \draw[-{Latex[length=1.25mm]}, thick,red] ( 0.05*\szx, 0.89*\szy)-- (0.15*\szx, 0.79*\szy);
    \nextgroupplot\addplot graphics[xmin = 0, xmax = \szx, ymin = 0, ymax = \szy] {Figures/\foldoi/DIP1_7_segnew};
    \draw[-{Latex[length=1.25mm]}, thick,red] (0.1*\szx, 0.07*\szy)-- (0.2*\szx, 0.17*\szy);
    \draw[-{Latex[length=1.25mm]}, thick,red] ( 0.05*\szx, 0.89*\szy)-- (0.15*\szx, 0.79*\szy);
    \nextgroupplot[
    y label style={at={(axis description cs:0.00001*\szx,.5)},anchor=south},
    yticklabels = {,,},
    xticklabels = {,,},
    hide axis = false,
        ]
    \addplot graphics[xmin = 0, xmax = \szx, ymin = 0, ymax = \szy] {Figures/\foldoi/LS1_9};
    \draw[-{Latex[length=1.25mm]}, thick,red] (0.1*\szx, 0.07*\szy)-- (0.2*\szx, 0.17*\szy);
    \draw[-{Latex[length=1.25mm]}, thick,red] ( 0.05*\szx, 0.89*\szy)-- (0.15*\szx, 0.79*\szy);
    \nextgroupplot
    \addplot graphics[xmin = 0, xmax = \szx, ymin = 0, ymax = \szy] {Figures/\foldoi/IRTV1_9};
    \draw[-{Latex[length=1.25mm]}, thick,red] (0.1*\szx, 0.07*\szy)-- (0.2*\szx, 0.17*\szy);
    \draw[-{Latex[length=1.25mm]}, thick,red] ( 0.05*\szx, 0.89*\szy)-- (0.15*\szx, 0.79*\szy);
    \nextgroupplot
    \addplot graphics[xmin = 0, xmax = \szx, ymin = 0, ymax = \szy] {Figures/\foldoi/PUMA1_9};
    \draw[-{Latex[length=1.25mm]}, thick,red] (0.1*\szx, 0.07*\szy)-- (0.2*\szx, 0.17*\szy);
    \draw[-{Latex[length=1.25mm]}, thick,red] ( 0.05*\szx, 0.89*\szy)-- (0.15*\szx, 0.79*\szy);
    \nextgroupplot
    \addplot graphics[xmin = 0, xmax = \szx, ymin = 0, ymax = \szy] {Figures/\foldoi/DIP1_9};
    \draw[-{Latex[length=1.25mm]}, thick,red] (0.1*\szx, 0.07*\szy)-- (0.2*\szx, 0.17*\szy);
    \draw[-{Latex[length=1.25mm]}, thick,red] ( 0.05*\szx, 0.89*\szy)-- (0.15*\szx, 0.79*\szy);
\end{groupplot}
\end{tikzpicture}
\caption{Time-lapse reconstructions for supplementary real data of size ($320 \times 380$) and their corresponding segmentation. Left: the images were saturated for visualization purpose. 
Right: segmentation of time-lapse reconstructions.
We thresholded at $20\%$ of the maximum value of the image.}
\label{fig:timelapsesupp2}
\end{figure*}

\renewcommand{\szx}{320}
\renewcommand{\szy}{300}
\renewcommand{\szsub}{0.115}
\renewcommand{\szax}{4*\szsub}
\renewcommand{\foldoi}{RealRK_C05_04_StackComparison_refocus_300_320}

\renewcommand{\vmin}{-11.77}
\renewcommand{\vmax}{16.43}
\pgfmathsetmacro{\step}{\vmin + (\vmax - \vmin)/4}
\renewcommand{\insep}{1} 

\begin{figure*}[t]
    \centering
    \begin{tikzpicture}[baseline=0pt]
    \begin{groupplot}[name = mygroup,
     group style = {group size = 4 by 5,
     vertical sep=\insep, horizontal sep=\insep},
     every plot/.style={width = {\szsub\textwidth},height = {\szy/\szx*\szsub\textwidth},axis equal image, scale only axis,
     }
    xmin = 0,xmax = \szx, ymin = 0, ymax = \szy,
    enlargelimits=false,
    axis equal image,
    scale only axis,
    width = {\szsub\textwidth},
    height = {\szy/\szx*\szsub\textwidth},
    hide axis,
    title style = {anchor=base}
    ]
    
    \nextgroupplot[ylabel = {\footnotesize \textbf{Frame 1}},
    y label style={at={(axis description cs:0.00001*\szx,.5)},anchor=south},
    yticklabels = {,,},
    xticklabels = {,,},
    hide axis = false,
    title = {\textbf{LS}},
        ]
    \addplot graphics[xmin = 0, xmax = \szx, ymin = 0, ymax = \szy] {Figures/\foldoi/LS1_1};

    \draw[-{Latex[length=1.25mm]}, thick,red] (0.65*\szx, 0.1*\szy)-- (0.55*\szx, 0.2*\szy);
    \draw[-{Latex[length=1.25mm]}, thick,red] ( 0.95*\szx, 0.91*\szy)-- (0.85*\szx, 0.81*\szy);
   \nextgroupplot[title = {\textbf{IRTV}},
        ]
    \addplot graphics[xmin = 0, xmax = \szx, ymin = 0, ymax = \szy] {Figures/\foldoi/IRTV1_1};
    \draw[-{Latex[length=1.25mm]}, thick,red] (0.65*\szx, 0.1*\szy)-- (0.55*\szx, 0.2*\szy);
    \draw[-{Latex[length=1.25mm]}, thick,red] ( 0.95*\szx, 0.91*\szy)-- (0.85*\szx, 0.81*\szy);

    \nextgroupplot[title = {\textbf{PUMA}},
        ]
    \addplot graphics[xmin = 0, xmax = \szx, ymin = 0, ymax = \szy] {Figures/\foldoi/PUMA1_1};
    \draw[-{Latex[length=1.25mm]}, thick,red] (0.65*\szx, 0.1*\szy)-- (0.55*\szx, 0.2*\szy);
    \draw[-{Latex[length=1.25mm]}, thick,red] ( 0.95*\szx, 0.91*\szy)-- (0.85*\szx, 0.81*\szy);
    \nextgroupplot[title = {\textbf{PUDIP}},
        ]
    \addplot graphics[xmin = 0, xmax = \szx, ymin = 0, ymax = \szy] {Figures/\foldoi/DIP1_1};
    \draw[-{Latex[length=1.25mm]}, thick,red] (0.65*\szx, 0.1*\szy)-- (0.55*\szx, 0.2*\szy);
    \draw[-{Latex[length=1.25mm]}, thick,red] ( 0.95*\szx, 0.91*\szy)-- (0.85*\szx, 0.81*\szy);
    \nextgroupplot[
    ylabel = {\footnotesize \textbf{Frame 3}},
    y label style={at={(axis description cs:0.00001*\szx,.5)},anchor=south},
    yticklabels = {,,},
    xticklabels = {,,},
    hide axis = false,
        ]
    \addplot graphics[xmin = 0, xmax = \szx, ymin = 0, ymax = \szy] {Figures/\foldoi/LS1_3};
    \draw[-{Latex[length=1.25mm]}, thick,red] (0.65*\szx, 0.1*\szy)-- (0.55*\szx, 0.2*\szy);
    \draw[-{Latex[length=1.25mm]}, thick,red] ( 0.95*\szx, 0.91*\szy)-- (0.85*\szx, 0.81*\szy);
    \nextgroupplot
    \addplot graphics[xmin = 0, xmax = \szx, ymin = 0, ymax = \szy] {Figures/\foldoi/IRTV1_3};
    \draw[-{Latex[length=1.25mm]}, thick,red] (0.65*\szx, 0.1*\szy)-- (0.55*\szx, 0.2*\szy);
    \draw[-{Latex[length=1.25mm]}, thick,red] ( 0.95*\szx, 0.91*\szy)-- (0.85*\szx, 0.81*\szy);
    \nextgroupplot
    \addplot graphics[xmin = 0, xmax = \szx, ymin = 0, ymax = \szy] {Figures/\foldoi/PUMA1_3};
    \draw[-{Latex[length=1.25mm]}, thick,red] (0.65*\szx, 0.1*\szy)-- (0.55*\szx, 0.2*\szy);
    \draw[-{Latex[length=1.25mm]}, thick,red] ( 0.95*\szx, 0.91*\szy)-- (0.85*\szx, 0.81*\szy);
    
    \nextgroupplot
    \addplot graphics[xmin = 0, xmax = \szx, ymin = 0, ymax = \szy] {Figures/\foldoi/DIP1_3};
    \draw[-{Latex[length=1.25mm]}, thick,red] (0.65*\szx, 0.1*\szy)-- (0.55*\szx, 0.2*\szy);
    \draw[-{Latex[length=1.25mm]}, thick,red] ( 0.95*\szx, 0.91*\szy)-- (0.85*\szx, 0.81*\szy);
    \nextgroupplot[
    ylabel = {\footnotesize \textbf{Frame 5}},
    y label style={at={(axis description cs:0.00001*\szx,.5)},anchor=south},
    yticklabels = {,,},
    xticklabels = {,,},
    hide axis = false,
        ]
    \addplot graphics[xmin = 0, xmax = \szx, ymin = 0, ymax = \szy] {Figures/\foldoi/LS1_5};
    \draw[-{Latex[length=1.25mm]}, thick,red] (0.65*\szx, 0.1*\szy)-- (0.55*\szx, 0.2*\szy);
    \draw[-{Latex[length=1.25mm]}, thick,red] ( 0.95*\szx, 0.91*\szy)-- (0.85*\szx, 0.81*\szy);
    \nextgroupplot
    \addplot graphics[xmin = 0, xmax = \szx, ymin = 0, ymax = \szy] {Figures/\foldoi/IRTV1_5};
    \draw[-{Latex[length=1.25mm]}, thick,red] (0.65*\szx, 0.1*\szy)-- (0.55*\szx, 0.2*\szy);
    \draw[-{Latex[length=1.25mm]}, thick,red] ( 0.95*\szx, 0.91*\szy)-- (0.85*\szx, 0.81*\szy);
    \nextgroupplot
    \addplot graphics[xmin = 0, xmax = \szx, ymin = 0, ymax = \szy] {Figures/\foldoi/PUMA1_5};
    \draw[-{Latex[length=1.25mm]}, thick,red] (0.65*\szx, 0.1*\szy)-- (0.55*\szx, 0.2*\szy);
    \draw[-{Latex[length=1.25mm]}, thick,red] ( 0.95*\szx, 0.91*\szy)-- (0.85*\szx, 0.81*\szy);
    \nextgroupplot
    \addplot graphics[xmin = 0, xmax = \szx, ymin = 0, ymax = \szy] {Figures/\foldoi/DIP1_5};
    \draw[-{Latex[length=1.25mm]}, thick,red] (0.65*\szx, 0.1*\szy)-- (0.55*\szx, 0.2*\szy);
    \draw[-{Latex[length=1.25mm]}, thick,red] ( 0.95*\szx, 0.91*\szy)-- (0.85*\szx, 0.81*\szy);

    \nextgroupplot[
    ylabel = {\footnotesize \textbf{Frame 7}},
    y label style={at={(axis description cs:0.00001*\szx,.5)},anchor=south},
    yticklabels = {,,},
    xticklabels = {,,},
    hide axis = false,
        ]
    \addplot graphics[xmin = 0, xmax = \szx, ymin = 0, ymax = \szy] {Figures/\foldoi/LS1_7};
    \draw[-{Latex[length=1.25mm]}, thick,red] (0.65*\szx, 0.1*\szy)-- (0.55*\szx, 0.2*\szy);
    \draw[-{Latex[length=1.25mm]}, thick,red] ( 0.95*\szx, 0.91*\szy)-- (0.85*\szx, 0.81*\szy);
    \nextgroupplot
    \addplot graphics[xmin = 0, xmax = \szx, ymin = 0, ymax = \szy] {Figures/\foldoi/IRTV1_7};
    \draw[-{Latex[length=1.25mm]}, thick,red] (0.65*\szx, 0.1*\szy)-- (0.55*\szx, 0.2*\szy);
    \draw[-{Latex[length=1.25mm]}, thick,red] ( 0.95*\szx, 0.91*\szy)-- (0.85*\szx, 0.81*\szy);
    \nextgroupplot\addplot graphics[xmin = 0, xmax = \szx, ymin = 0, ymax = \szy] {Figures/\foldoi/PUMA1_7};
    \draw[-{Latex[length=1.25mm]}, thick,red] (0.65*\szx, 0.1*\szy)-- (0.55*\szx, 0.2*\szy);
    \draw[-{Latex[length=1.25mm]}, thick,red] ( 0.95*\szx, 0.91*\szy)-- (0.85*\szx, 0.81*\szy);
    \nextgroupplot\addplot graphics[xmin = 0, xmax = \szx, ymin = 0, ymax = \szy] {Figures/\foldoi/DIP1_7};
    \draw[-{Latex[length=1.25mm]}, thick,red] (0.65*\szx, 0.1*\szy)-- (0.55*\szx, 0.2*\szy);
    \draw[-{Latex[length=1.25mm]}, thick,red] ( 0.95*\szx, 0.91*\szy)-- (0.85*\szx, 0.81*\szy);
    \nextgroupplot[
    ylabel = {\footnotesize \textbf{Frame 9}},
    y label style={at={(axis description cs:0.00001*\szx,.5)},anchor=south},
    yticklabels = {,,},
    xticklabels = {,,},
    hide axis = false,
        ]
    \addplot graphics[xmin = 0, xmax = \szx, ymin = 0, ymax = \szy] {Figures/\foldoi/LS1_9};
    \draw[-{Latex[length=1.25mm]}, thick,red] (0.65*\szx, 0.1*\szy)-- (0.55*\szx, 0.2*\szy);
    \draw[-{Latex[length=1.25mm]}, thick,red] ( 0.95*\szx, 0.91*\szy)-- (0.85*\szx, 0.81*\szy);
    \nextgroupplot
    \addplot graphics[xmin = 0, xmax = \szx, ymin = 0, ymax = \szy] {Figures/\foldoi/IRTV1_9};
    \draw[-{Latex[length=1.25mm]}, thick,red] (0.65*\szx, 0.1*\szy)-- (0.55*\szx, 0.2*\szy);
    \draw[-{Latex[length=1.25mm]}, thick,red] ( 0.95*\szx, 0.91*\szy)-- (0.85*\szx, 0.81*\szy);
    \nextgroupplot
    \addplot graphics[xmin = 0, xmax = \szx, ymin = 0, ymax = \szy] {Figures/\foldoi/PUMA1_9};
    \draw[-{Latex[length=1.25mm]}, thick,red] (0.65*\szx, 0.1*\szy)-- (0.55*\szx, 0.2*\szy);
    \draw[-{Latex[length=1.25mm]}, thick,red] ( 0.95*\szx, 0.91*\szy)-- (0.85*\szx, 0.81*\szy);
    \nextgroupplot
    \addplot graphics[xmin = 0, xmax = \szx, ymin = 0, ymax = \szy] {Figures/\foldoi/DIP1_9};
\draw[-{Latex[length=1.25mm]}, thick,red] (0.65*\szx, 0.1*\szy)-- (0.55*\szx, 0.2*\szy);
    \draw[-{Latex[length=1.25mm]}, thick,red] ( 0.95*\szx, 0.91*\szy)-- (0.85*\szx, 0.81*\szy);
\end{groupplot}

\begin{axis}[at={($(group c1r5.south west) - (0,1em)$)}, anchor = north west,
    xmin = \vmin, xmax = \vmax, ymin = 0, ymax = 0.5,
    width=0,
    height = 0,
    scale only axis,
    enlargelimits=false,
    hide axis,
    axis equal image,
    colorbar horizontal,
    colormap/parula,
    colorbar style={
    at = {(0.025*\szax*\textwidth,0)},
        anchor = north west,
        xticklabel pos = lower,
        xmin = \vmin, xmax = \vmax,
        point meta min = \vmin,
        point meta max = \vmax,
        scale only axis,
        enlargelimits = false,
        scaled x ticks = true,
        width = {0.95*\szax*\textwidth},
    samples = 200,
    height = 0.15cm,
    xticklabel style={/pgf/number format/fixed, /pgf/number format/precision=0},
    xtick = {\vmin,\step,...,\vmax},
    xlabel={rad},
    xlabel style={yshift=0cm,xshift=0cm},
        }]
        \end{axis}
    \end{tikzpicture}
    \begin{tikzpicture}[baseline=0pt]
    \renewcommand{\foldoi}{RealRK_C05_04_StackComparison_refocus_300_320_seg_0.20}
    
    \begin{groupplot}[name = mygroup,
     group style = {group size = 4 by 5,
     vertical sep=\insep, horizontal sep=\insep},
     every plot/.style={width = {\szsub\textwidth},height = {\szy/\szx*\szsub\textwidth},axis equal image, scale only axis,
     }
    xmin = 0,xmax = \szx, ymin = 0, ymax = \szy,
    enlargelimits=false,
    axis equal image,
    scale only axis,
    width = {\szsub\textwidth},
    height = {\szy/\szx*\szsub\textwidth},
    hide axis,
    title style = {anchor=base}
    ]
    
    \nextgroupplot[
    y label style={at={(axis description cs:0.00001*\szx,.5)},anchor=south},
    yticklabels = {,,},
    xticklabels = {,,},
    hide axis = false,
    title = {\textbf{LS}},
        ]
    \addplot graphics[xmin = 0, xmax = \szx, ymin = 0, ymax = \szy] {Figures/\foldoi/LS1_1};
    \draw[-{Latex[length=1.25mm]}, thick,red] (0.65*\szx, 0.1*\szy)-- (0.55*\szx, 0.2*\szy);
    \draw[-{Latex[length=1.25mm]}, thick,red] ( 0.95*\szx, 0.91*\szy)-- (0.85*\szx, 0.81*\szy);
   \nextgroupplot[title = {\textbf{IRTV}},
        ]
    \addplot graphics[xmin = 0, xmax = \szx, ymin = 0, ymax = \szy] {Figures/\foldoi/IRTV1_1};
    \draw[-{Latex[length=1.25mm]}, thick,red] (0.65*\szx, 0.1*\szy)-- (0.55*\szx, 0.2*\szy);
    \draw[-{Latex[length=1.25mm]}, thick,red] ( 0.95*\szx, 0.91*\szy)-- (0.85*\szx, 0.81*\szy);
    \nextgroupplot[title = {\textbf{PUMA}},
        ]
    \addplot graphics[xmin = 0, xmax = \szx, ymin = 0, ymax = \szy] {Figures/\foldoi/PUMA1_1};
    \draw[-{Latex[length=1.25mm]}, thick,red] (0.65*\szx, 0.1*\szy)-- (0.55*\szx, 0.2*\szy);
    \draw[-{Latex[length=1.25mm]}, thick,red] ( 0.95*\szx, 0.91*\szy)-- (0.85*\szx, 0.81*\szy);
    \nextgroupplot[title = {\textbf{PUDIP}},
        ]
    \addplot graphics[xmin = 0, xmax = \szx, ymin = 0, ymax = \szy] {Figures/\foldoi/DIP1_1};
    \draw[-{Latex[length=1.25mm]}, thick,red] (0.65*\szx, 0.1*\szy)-- (0.55*\szx, 0.2*\szy);
    \draw[-{Latex[length=1.25mm]}, thick,red] ( 0.95*\szx, 0.91*\szy)-- (0.85*\szx, 0.81*\szy);
    \nextgroupplot[
    y label style={at={(axis description cs:0.00001*\szx,.5)},anchor=south},
    yticklabels = {,,},
    xticklabels = {,,},
    hide axis = false,
        ]
    \addplot graphics[xmin = 0, xmax = \szx, ymin = 0, ymax = \szy] {Figures/\foldoi/LS1_3};
    \draw[-{Latex[length=1.25mm]}, thick,red] (0.65*\szx, 0.1*\szy)-- (0.55*\szx, 0.2*\szy);
    \draw[-{Latex[length=1.25mm]}, thick,red] ( 0.95*\szx, 0.91*\szy)-- (0.85*\szx, 0.81*\szy);
    \nextgroupplot
    \addplot graphics[xmin = 0, xmax = \szx, ymin = 0, ymax = \szy] {Figures/\foldoi/IRTV1_3};
    \draw[-{Latex[length=1.25mm]}, thick,red] (0.65*\szx, 0.1*\szy)-- (0.55*\szx, 0.2*\szy);
    \draw[-{Latex[length=1.25mm]}, thick,red] ( 0.95*\szx, 0.91*\szy)-- (0.85*\szx, 0.81*\szy);
    \nextgroupplot
    \addplot graphics[xmin = 0, xmax = \szx, ymin = 0, ymax = \szy] {Figures/\foldoi/PUMA1_3};
    \draw[-{Latex[length=1.25mm]}, thick,red] (0.65*\szx, 0.1*\szy)-- (0.55*\szx, 0.2*\szy);
    \draw[-{Latex[length=1.25mm]}, thick,red] ( 0.95*\szx, 0.91*\szy)-- (0.85*\szx, 0.81*\szy);
    \nextgroupplot
    \addplot graphics[xmin = 0, xmax = \szx, ymin = 0, ymax = \szy] {Figures/\foldoi/DIP1_3};
    \draw[-{Latex[length=1.25mm]}, thick,red] (0.65*\szx, 0.1*\szy)-- (0.55*\szx, 0.2*\szy);
    \draw[-{Latex[length=1.25mm]}, thick,red] ( 0.95*\szx, 0.91*\szy)-- (0.85*\szx, 0.81*\szy);
    \nextgroupplot[
    y label style={at={(axis description cs:0.00001*\szx,.5)},anchor=south},
    yticklabels = {,,},
    xticklabels = {,,},
    hide axis = false,
        ]
    \addplot graphics[xmin = 0, xmax = \szx, ymin = 0, ymax = \szy] {Figures/\foldoi/LS1_5};
    \draw[-{Latex[length=1.25mm]}, thick,red] (0.65*\szx, 0.1*\szy)-- (0.55*\szx, 0.2*\szy);
    \draw[-{Latex[length=1.25mm]}, thick,red] ( 0.95*\szx, 0.91*\szy)-- (0.85*\szx, 0.81*\szy);
    \nextgroupplot
    \addplot graphics[xmin = 0, xmax = \szx, ymin = 0, ymax = \szy] {Figures/\foldoi/IRTV1_5};
    \draw[-{Latex[length=1.25mm]}, thick,red] (0.65*\szx, 0.1*\szy)-- (0.55*\szx, 0.2*\szy);
    \draw[-{Latex[length=1.25mm]}, thick,red] ( 0.95*\szx, 0.91*\szy)-- (0.85*\szx, 0.81*\szy);
    \nextgroupplot
    \addplot graphics[xmin = 0, xmax = \szx, ymin = 0, ymax = \szy] {Figures/\foldoi/PUMA1_5};
    \draw[-{Latex[length=1.25mm]}, thick,red] (0.65*\szx, 0.1*\szy)-- (0.55*\szx, 0.2*\szy);
    \draw[-{Latex[length=1.25mm]}, thick,red] ( 0.95*\szx, 0.91*\szy)-- (0.85*\szx, 0.81*\szy);
    \nextgroupplot
    \addplot graphics[xmin = 0, xmax = \szx, ymin = 0, ymax = \szy] {Figures/\foldoi/DIP1_5};
    \draw[-{Latex[length=1.25mm]}, thick,red] (0.65*\szx, 0.1*\szy)-- (0.55*\szx, 0.2*\szy);
    \draw[-{Latex[length=1.25mm]}, thick,red] ( 0.95*\szx, 0.91*\szy)-- (0.85*\szx, 0.81*\szy);
    \nextgroupplot[
    y label style={at={(axis description cs:0.00001*\szx,.5)},anchor=south},
    yticklabels = {,,},
    xticklabels = {,,},
    hide axis = false,
        ]
    \addplot graphics[xmin = 0, xmax = \szx, ymin = 0, ymax = \szy] {Figures/\foldoi/LS1_7};
    \draw[-{Latex[length=1.25mm]}, thick,red] (0.65*\szx, 0.1*\szy)-- (0.55*\szx, 0.2*\szy);
    \draw[-{Latex[length=1.25mm]}, thick,red] ( 0.95*\szx, 0.91*\szy)-- (0.85*\szx, 0.81*\szy);
    \nextgroupplot
    \addplot graphics[xmin = 0, xmax = \szx, ymin = 0, ymax = \szy] {Figures/\foldoi/IRTV1_7};
    \draw[-{Latex[length=1.25mm]}, thick,red] (0.65*\szx, 0.1*\szy)-- (0.55*\szx, 0.2*\szy);
    \draw[-{Latex[length=1.25mm]}, thick,red] ( 0.95*\szx, 0.91*\szy)-- (0.85*\szx, 0.81*\szy);
    \nextgroupplot\addplot graphics[xmin = 0, xmax = \szx, ymin = 0, ymax = \szy] {Figures/\foldoi/PUMA1_7};
    \draw[-{Latex[length=1.25mm]}, thick,red] (0.65*\szx, 0.1*\szy)-- (0.55*\szx, 0.2*\szy);
    \draw[-{Latex[length=1.25mm]}, thick,red] ( 0.95*\szx, 0.91*\szy)-- (0.85*\szx, 0.81*\szy);
    \nextgroupplot\addplot graphics[xmin = 0, xmax = \szx, ymin = 0, ymax = \szy] {Figures/\foldoi/DIP1_7};
\draw[-{Latex[length=1.25mm]}, thick,red] (0.65*\szx, 0.1*\szy)-- (0.55*\szx, 0.2*\szy);
    \draw[-{Latex[length=1.25mm]}, thick,red] ( 0.95*\szx, 0.91*\szy)-- (0.85*\szx, 0.81*\szy);
    \nextgroupplot[
    y label style={at={(axis description cs:0.00001*\szx,.5)},anchor=south},
    yticklabels = {,,},
    xticklabels = {,,},
    hide axis = false,
        ]
    \addplot graphics[xmin = 0, xmax = \szx, ymin = 0, ymax = \szy] {Figures/\foldoi/LS1_9};
    \draw[-{Latex[length=1.25mm]}, thick,red] (0.65*\szx, 0.1*\szy)-- (0.55*\szx, 0.2*\szy);
    \draw[-{Latex[length=1.25mm]}, thick,red] ( 0.95*\szx, 0.91*\szy)-- (0.85*\szx, 0.81*\szy);
    \nextgroupplot
    \addplot graphics[xmin = 0, xmax = \szx, ymin = 0, ymax = \szy] {Figures/\foldoi/IRTV1_9};
    \draw[-{Latex[length=1.25mm]}, thick,red] (0.65*\szx, 0.1*\szy)-- (0.55*\szx, 0.2*\szy);
    \draw[-{Latex[length=1.25mm]}, thick,red] ( 0.95*\szx, 0.91*\szy)-- (0.85*\szx, 0.81*\szy);
    \nextgroupplot
    \addplot graphics[xmin = 0, xmax = \szx, ymin = 0, ymax = \szy] {Figures/\foldoi/PUMA1_9};
    \draw[-{Latex[length=1.25mm]}, thick,red] (0.65*\szx, 0.1*\szy)-- (0.55*\szx, 0.2*\szy);
    \draw[-{Latex[length=1.25mm]}, thick,red] ( 0.95*\szx, 0.91*\szy)-- (0.85*\szx, 0.81*\szy);
    \nextgroupplot
    \addplot graphics[xmin = 0, xmax = \szx, ymin = 0, ymax = \szy] {Figures/\foldoi/DIP1_9};
    \draw[-{Latex[length=1.25mm]}, thick,red] (0.65*\szx, 0.1*\szy)-- (0.55*\szx, 0.2*\szy);
    \draw[-{Latex[length=1.25mm]}, thick,red] ( 0.95*\szx, 0.91*\szy)-- (0.85*\szx, 0.81*\szy);
\end{groupplot}
\end{tikzpicture}
\caption{Time-lapse reconstructions for supplementary real data of size ($300 \times 320$) and their corresponding segmentation. Left: the images were saturated for visualization purpose. 
Right: segmentation of time-lapse reconstructions.
We thresholded at $20\%$ of the maximum value of the image.
}\label{fig:timelapsesupp3}
\end{figure*}


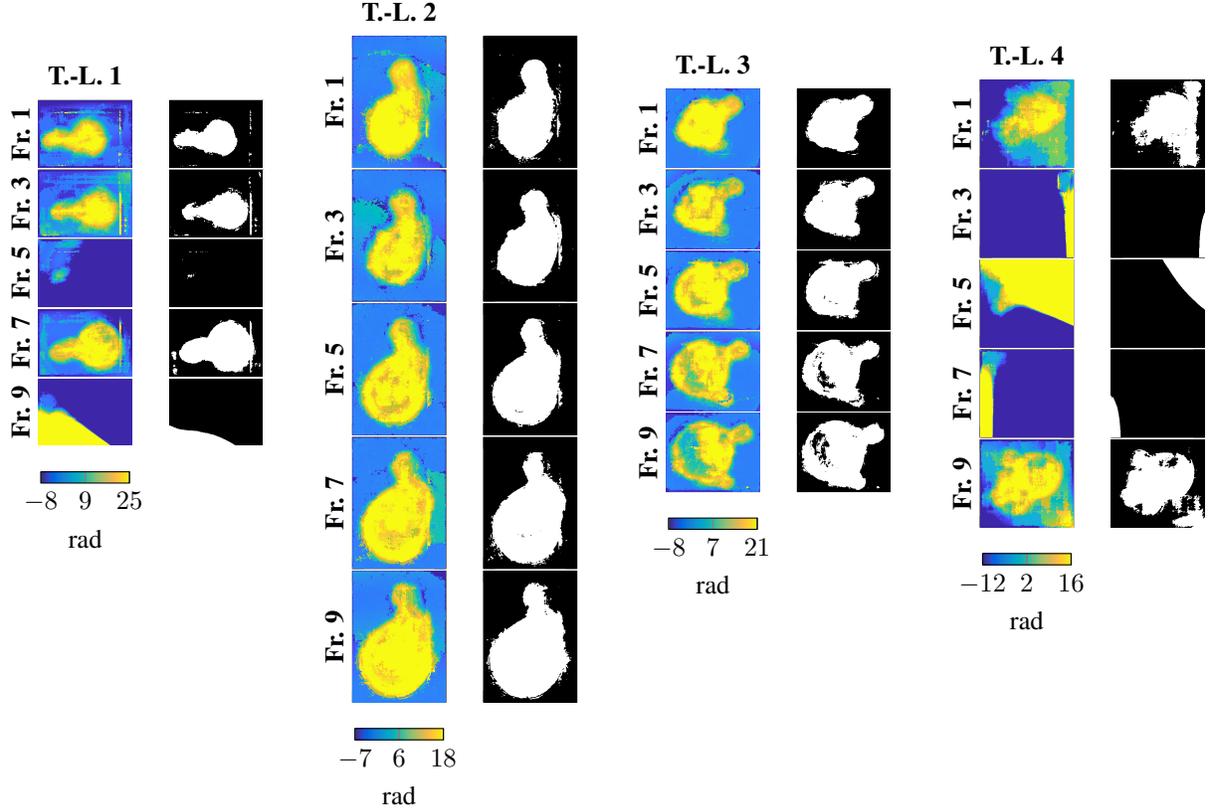
\begin{figure*}[t]
    \centering
    \renewcommand{\szx}{390}
    \renewcommand{\szy}{280}
    \renewcommand{\szsub}{0.075}
    \renewcommand{\szax}{\szsub}
    \renewcommand{\foldoi}{RealRK_B05_03_StackComparison_280_390}
    
    \renewcommand{\vmin}{-8}
    \renewcommand{\vmax}{25}
    \pgfmathsetmacro{\step}{\vmin + (\vmax - \vmin)/2}
    \renewcommand{\insep}{1} 
    \begin{tikzpicture}[baseline=0pt]
    \begin{groupplot}[name = mygroup1,
     group style = {group size = 1 by 5,
     vertical sep=\insep, horizontal sep=\insep},
     every plot/.style={width = {\szsub\textwidth},height = {\szy/\szx*\szsub\textwidth},axis equal image, scale only axis,
     }
    xmin = 0,xmax = \szx, ymin = 0, ymax = \szy,
    enlargelimits=false,
    axis equal image,
    scale only axis,
    width = {\szsub\textwidth},
    height = {\szy/\szx*\szsub\textwidth},
    hide axis,
    title style = {anchor=base}
    ]
    
    \nextgroupplot[
    y label style={at={(axis description cs:0.00001*\szx,.5)},anchor=south},
    yticklabels = {,,},
    xticklabels = {,,},
    hide axis = false,
    title = {\textbf{T.-L. 1}},
    ylabel = {\textbf{Fr. 1}}
        ]
    \addplot graphics[xmin = 0, xmax = \szx, ymin = 0, ymax = \szy] {Figures/\foldoi/PhaseNet1_1};

    \nextgroupplot[
    y label style={at={(axis description cs:0.00001*\szx,.5)},anchor=south},
    yticklabels = {,,},
    xticklabels = {,,},
    hide axis = false,
    ylabel = {\textbf{Fr. 3}}
        ]
    \addplot graphics[xmin = 0, xmax = \szx, ymin = 0, ymax = \szy] {Figures/\foldoi/PhaseNet1_3};
    \nextgroupplot[
    y label style={at={(axis description cs:0.00001*\szx,.5)},anchor=south},
    yticklabels = {,,},
    xticklabels = {,,},
    hide axis = false,
    ylabel = {\textbf{Fr. 5}}
        ]
    \addplot graphics[xmin = 0, xmax = \szx, ymin = 0, ymax = \szy] {Figures/\foldoi/PhaseNet1_5};

    \nextgroupplot[
    y label style={at={(axis description cs:0.00001*\szx,.5)},anchor=south},
    yticklabels = {,,},
    xticklabels = {,,},
    hide axis = false,
    ylabel = {\textbf{Fr. 7}}
        ]
    \addplot graphics[xmin = 0, xmax = \szx, ymin = 0, ymax = \szy] {Figures/\foldoi/PhaseNet1_7};
    
    \nextgroupplot[
    y label style={at={(axis description cs:0.00001*\szx,.5)},anchor=south},
    yticklabels = {,,},
    xticklabels = {,,},
    hide axis = false,
    ylabel = {\textbf{Fr. 9}}
        ]
    \addplot graphics[xmin = 0, xmax = \szx, ymin = 0, ymax = \szy] {Figures/\foldoi/PhaseNet1_9};
\end{groupplot}
\begin{axis}[at={($(group c1r5.south west) - (0,1em)$)}, anchor = north west,
    xmin = \vmin, xmax = \vmax, ymin = 0, ymax = 0.5,
    width=0,
    height = 0,
    scale only axis,
    enlargelimits=false,
    hide axis,
    axis equal image,
    colorbar horizontal,
    colormap/parula,
    colorbar style={
    at = {(0.025*\szax*\textwidth,0)},
        anchor = north west,
        xticklabel pos = lower,
        xmin = \vmin, xmax = \vmax,
        point meta min = \vmin,
        point meta max = \vmax,
        scale only axis,
        enlargelimits = false,
        scaled x ticks = true,
        width = {0.95*\szax*\textwidth},
    samples = 200,
    height = 0.15cm,
    xticklabel style={/pgf/number format/fixed, /pgf/number format/precision=0},
    xtick = {\vmin,\step,...,\vmax},
    xlabel={rad},
    xlabel style={yshift=0cm,xshift=0cm},
        }]
        \end{axis}
    \end{tikzpicture}
    \begin{tikzpicture}[baseline=0pt]
    \renewcommand{\foldoi}{RealRK_B05_03_StackComparison_280_390_seg_0.2}
    \begin{groupplot}[name = mygroup,
     group style = {group size = 1 by 5,
     vertical sep=\insep, horizontal sep=\insep},
     every plot/.style={width = {\szsub\textwidth},height = {\szy/\szx*\szsub\textwidth},axis equal image, scale only axis,
     }
    xmin = 0,xmax = \szx, ymin = 0, ymax = \szy,
    enlargelimits=false,
    axis equal image,
    scale only axis,
    width = {\szsub\textwidth},
    height = {\szy/\szx*\szsub\textwidth},
    hide axis,
    title style = {anchor=base}
    ]
    
    \nextgroupplot[
    y label style={at={(axis description cs:0.00001*\szx,.5)},anchor=south},
    yticklabels = {,,},
    xticklabels = {,,},
    hide axis = false,
        ]
    \addplot graphics[xmin = 0, xmax = \szx, ymin = 0, ymax = \szy] {Figures/\foldoi/PhaseNet1_1};
    \nextgroupplot[
    y label style={at={(axis description cs:0.00001*\szx,.5)},anchor=south},
    yticklabels = {,,},
    xticklabels = {,,},
    hide axis = false,
        ]
    \addplot graphics[xmin = 0, xmax = \szx, ymin = 0, ymax = \szy] {Figures/\foldoi/PhaseNet1_3};
    \nextgroupplot[
    y label style={at={(axis description cs:0.00001*\szx,.5)},anchor=south},
    yticklabels = {,,},
    xticklabels = {,,},
    hide axis = false,
        ]
    \addplot graphics[xmin = 0, xmax = \szx, ymin = 0, ymax = \szy] {Figures/\foldoi/PhaseNet1_5};
    \nextgroupplot[
    y label style={at={(axis description cs:0.00001*\szx,.5)},anchor=south},
    yticklabels = {,,},
    xticklabels = {,,},
    hide axis = false,
        ]
    \addplot graphics[xmin = 0, xmax = \szx, ymin = 0, ymax = \szy] {Figures/\foldoi/PhaseNet1_7};
    \nextgroupplot[
    y label style={at={(axis description cs:0.00001*\szx,.5)},anchor=south},
    yticklabels = {,,},
    xticklabels = {,,},
    hide axis = false,
        ]
    \addplot graphics[xmin = 0, xmax = \szx, ymin = 0, ymax = \szy] {Figures/\foldoi/PhaseNet1_9};
\end{groupplot}
\end{tikzpicture}
    \renewcommand{\szx}{270}
    \renewcommand{\szy}{380}
    \renewcommand{\szax}{\szsub}
    \renewcommand{\foldoi}{RealRK_B04_05_StackComparison_380_270}
    \renewcommand{\vmin}{-6.97}
    \renewcommand{\vmax}{18.32}
    \pgfmathsetmacro{\step}{\vmin + (\vmax - \vmin)/2}
    \renewcommand{\insep}{1} 
    \begin{tikzpicture}[baseline=0pt]
    \begin{groupplot}[name = mygroup1,
     group style = {group size = 1 by 5,
     vertical sep=\insep, horizontal sep=\insep},
     every plot/.style={width = {\szsub\textwidth},height = {\szy/\szx*\szsub\textwidth},axis equal image, scale only axis,
     }
    xmin = 0,xmax = \szx, ymin = 0, ymax = \szy,
    enlargelimits=false,
    axis equal image,
    scale only axis,
    width = {\szsub\textwidth},
    height = {\szy/\szx*\szsub\textwidth},
    hide axis,
    title style = {anchor=base}
    ]
    
    \nextgroupplot[
    y label style={at={(axis description cs:0.00001*\szx,.5)},anchor=south},
    yticklabels = {,,},
    xticklabels = {,,},
    hide axis = false,
    title = {\textbf{T.-L. 2}},
    ylabel = {\textbf{Fr. 1}}
        ]
    \addplot graphics[xmin = 0, xmax = \szx, ymin = 0, ymax = \szy] {Figures/\foldoi/PhaseNet1_1};

    \nextgroupplot[
    y label style={at={(axis description cs:0.00001*\szx,.5)},anchor=south},
    yticklabels = {,,},
    xticklabels = {,,},
    hide axis = false,
    ylabel = {\textbf{Fr. 3}}
        ]
    \addplot graphics[xmin = 0, xmax = \szx, ymin = 0, ymax = \szy] {Figures/\foldoi/PhaseNet1_3};
    \nextgroupplot[
    y label style={at={(axis description cs:0.00001*\szx,.5)},anchor=south},
    yticklabels = {,,},
    xticklabels = {,,},
    hide axis = false,
    ylabel = {\textbf{Fr. 5}}
        ]
    \addplot graphics[xmin = 0, xmax = \szx, ymin = 0, ymax = \szy] {Figures/\foldoi/PhaseNet1_5};

    \nextgroupplot[
    y label style={at={(axis description cs:0.00001*\szx,.5)},anchor=south},
    yticklabels = {,,},
    xticklabels = {,,},
    hide axis = false,
    ylabel = {\textbf{Fr. 7}}
        ]
    \addplot graphics[xmin = 0, xmax = \szx, ymin = 0, ymax = \szy] {Figures/\foldoi/PhaseNet1_7};
    
    \nextgroupplot[
    y label style={at={(axis description cs:0.00001*\szx,.5)},anchor=south},
    yticklabels = {,,},
    xticklabels = {,,},
    hide axis = false,
    ylabel = {\textbf{Fr. 9}}
        ]
    \addplot graphics[xmin = 0, xmax = \szx, ymin = 0, ymax = \szy] {Figures/\foldoi/PhaseNet1_9};
\end{groupplot}
\begin{axis}[at={($(group c1r5.south west) - (0,1em)$)}, anchor = north west,
    xmin = \vmin, xmax = \vmax, ymin = 0, ymax = 0.5,
    width=0,
    height = 0,
    scale only axis,
    enlargelimits=false,
    hide axis,
    axis equal image,
    colorbar horizontal,
    colormap/parula,
    colorbar style={
    at = {(0.025*\szax*\textwidth,0)},
        anchor = north west,
        xticklabel pos = lower,
        xmin = \vmin, xmax = \vmax,
        point meta min = \vmin,
        point meta max = \vmax,
        scale only axis,
        enlargelimits = false,
        scaled x ticks = true,
        width = {0.95*\szax*\textwidth},
    samples = 200,
    height = 0.15cm,
    xticklabel style={/pgf/number format/fixed, /pgf/number format/precision=0},
    xtick = {\vmin,\step,...,\vmax},
    xlabel={rad},
    xlabel style={yshift=0cm,xshift=0cm},
        }]
        \end{axis}
    \end{tikzpicture}
    \begin{tikzpicture}[baseline=0pt]
    \renewcommand{\foldoi}{RealRK_B04_05_StackComparison_380_270_seg_0.2}
    \begin{groupplot}[name = mygroup,
     group style = {group size = 1 by 5,
     vertical sep=\insep, horizontal sep=\insep},
     every plot/.style={width = {\szsub\textwidth},height = {\szy/\szx*\szsub\textwidth},axis equal image, scale only axis,
     }
    xmin = 0,xmax = \szx, ymin = 0, ymax = \szy,
    enlargelimits=false,
    axis equal image,
    scale only axis,
    width = {\szsub\textwidth},
    height = {\szy/\szx*\szsub\textwidth},
    hide axis,
    title style = {anchor=base}
    ]
    
    \nextgroupplot[
    y label style={at={(axis description cs:0.00001*\szx,.5)},anchor=south},
    yticklabels = {,,},
    xticklabels = {,,},
    hide axis = false,
        ]
    \addplot graphics[xmin = 0, xmax = \szx, ymin = 0, ymax = \szy] {Figures/\foldoi/PhaseNet1_1};
    \nextgroupplot[
    y label style={at={(axis description cs:0.00001*\szx,.5)},anchor=south},
    yticklabels = {,,},
    xticklabels = {,,},
    hide axis = false,
        ]
    \addplot graphics[xmin = 0, xmax = \szx, ymin = 0, ymax = \szy] {Figures/\foldoi/PhaseNet1_3};
    \nextgroupplot[
    y label style={at={(axis description cs:0.00001*\szx,.5)},anchor=south},
    yticklabels = {,,},
    xticklabels = {,,},
    hide axis = false,
        ]
    \addplot graphics[xmin = 0, xmax = \szx, ymin = 0, ymax = \szy] {Figures/\foldoi/PhaseNet1_5};
    \nextgroupplot[
    y label style={at={(axis description cs:0.00001*\szx,.5)},anchor=south},
    yticklabels = {,,},
    xticklabels = {,,},
    hide axis = false,
        ]
    \addplot graphics[xmin = 0, xmax = \szx, ymin = 0, ymax = \szy] {Figures/\foldoi/PhaseNet1_7};
    \nextgroupplot[
    y label style={at={(axis description cs:0.00001*\szx,.5)},anchor=south},
    yticklabels = {,,},
    xticklabels = {,,},
    hide axis = false,
        ]
    \addplot graphics[xmin = 0, xmax = \szx, ymin = 0, ymax = \szy] {Figures/\foldoi/PhaseNet1_9};
\end{groupplot}
\end{tikzpicture}
    \renewcommand{\szx}{380}
    \renewcommand{\szy}{320}
    \renewcommand{\szax}{\szsub}
    \renewcommand{\foldoi}{RealRK_B04_09_StackComparison_320_380}
    \renewcommand{\vmin}{-7.73}
    \renewcommand{\vmax}{20.90}
    \pgfmathsetmacro{\step}{\vmin + (\vmax - \vmin)/2}
    \renewcommand{\insep}{1} 
    \begin{tikzpicture}[baseline=0pt]
    \begin{groupplot}[name = mygroup1,
     group style = {group size = 1 by 5,
     vertical sep=\insep, horizontal sep=\insep},
     every plot/.style={width = {\szsub\textwidth},height = {\szy/\szx*\szsub\textwidth},axis equal image, scale only axis,
     }
    xmin = 0,xmax = \szx, ymin = 0, ymax = \szy,
    enlargelimits=false,
    axis equal image,
    scale only axis,
    width = {\szsub\textwidth},
    height = {\szy/\szx*\szsub\textwidth},
    hide axis,
    title style = {anchor=base}
    ]
    \nextgroupplot[
    y label style={at={(axis description cs:0.00001*\szx,.5)},anchor=south},
    yticklabels = {,,},
    xticklabels = {,,},
    hide axis = false,
    title = {\textbf{T.-L. 3}},
    ylabel = {\textbf{Fr. 1}}
        ]
    \addplot graphics[xmin = 0, xmax = \szx, ymin = 0, ymax = \szy] {Figures/\foldoi/PhaseNet1_1};

    \nextgroupplot[
    y label style={at={(axis description cs:0.00001*\szx,.5)},anchor=south},
    yticklabels = {,,},
    xticklabels = {,,},
    hide axis = false,
    ylabel = {\textbf{Fr. 3}}
        ]
    \addplot graphics[xmin = 0, xmax = \szx, ymin = 0, ymax = \szy] {Figures/\foldoi/PhaseNet1_3};
    \nextgroupplot[
    y label style={at={(axis description cs:0.00001*\szx,.5)},anchor=south},
    yticklabels = {,,},
    xticklabels = {,,},
    hide axis = false,
    ylabel = {\textbf{Fr. 5}}
        ]
    \addplot graphics[xmin = 0, xmax = \szx, ymin = 0, ymax = \szy] {Figures/\foldoi/PhaseNet1_5};

    \nextgroupplot[
    y label style={at={(axis description cs:0.00001*\szx,.5)},anchor=south},
    yticklabels = {,,},
    xticklabels = {,,},
    hide axis = false,
    ylabel = {\textbf{Fr. 7}}
        ]
    \addplot graphics[xmin = 0, xmax = \szx, ymin = 0, ymax = \szy] {Figures/\foldoi/PhaseNet1_7};
    
    \nextgroupplot[
    y label style={at={(axis description cs:0.00001*\szx,.5)},anchor=south},
    yticklabels = {,,},
    xticklabels = {,,},
    hide axis = false,
    ylabel = {\textbf{Fr. 9}}
        ]
    \addplot graphics[xmin = 0, xmax = \szx, ymin = 0, ymax = \szy] {Figures/\foldoi/PhaseNet1_9};
\end{groupplot}
\begin{axis}[at={($(group c1r5.south west) - (0,1em)$)}, anchor = north west,
    xmin = \vmin, xmax = \vmax, ymin = 0, ymax = 0.5,
    width=0,
    height = 0,
    scale only axis,
    enlargelimits=false,
    hide axis,
    axis equal image,
    colorbar horizontal,
    colormap/parula,
    colorbar style={
    at = {(0.025*\szax*\textwidth,0)},
        anchor = north west,
        xticklabel pos = lower,
        xmin = \vmin, xmax = \vmax,
        point meta min = \vmin,
        point meta max = \vmax,
        scale only axis,
        enlargelimits = false,
        scaled x ticks = true,
        width = {0.95*\szax*\textwidth},
    samples = 200,
    height = 0.15cm,
    xticklabel style={/pgf/number format/fixed, /pgf/number format/precision=0},
    xtick = {\vmin,\step,...,\vmax},
    xlabel={rad},
    xlabel style={yshift=0cm,xshift=0cm},
        }]
        \end{axis}
    \end{tikzpicture}
    \begin{tikzpicture}[baseline=0pt]
    \renewcommand{\foldoi}{RealRK_B04_09_StackComparison_320_380_seg_0.2}
    \begin{groupplot}[name = mygroup,
     group style = {group size = 1 by 5,
     vertical sep=\insep, horizontal sep=\insep},
     every plot/.style={width = {\szsub\textwidth},height = {\szy/\szx*\szsub\textwidth},axis equal image, scale only axis,
     }
    xmin = 0,xmax = \szx, ymin = 0, ymax = \szy,
    enlargelimits=false,
    axis equal image,
    scale only axis,
    width = {\szsub\textwidth},
    height = {\szy/\szx*\szsub\textwidth},
    hide axis,
    title style = {anchor=base}
    ]
    
    \nextgroupplot[
    y label style={at={(axis description cs:0.00001*\szx,.5)},anchor=south},
    yticklabels = {,,},
    xticklabels = {,,},
    hide axis = false,
        ]
    \addplot graphics[xmin = 0, xmax = \szx, ymin = 0, ymax = \szy] {Figures/\foldoi/PhaseNet1_1};
    \nextgroupplot[
    y label style={at={(axis description cs:0.00001*\szx,.5)},anchor=south},
    yticklabels = {,,},
    xticklabels = {,,},
    hide axis = false,
        ]
    \addplot graphics[xmin = 0, xmax = \szx, ymin = 0, ymax = \szy] {Figures/\foldoi/PhaseNet1_3};
    \nextgroupplot[
    y label style={at={(axis description cs:0.00001*\szx,.5)},anchor=south},
    yticklabels = {,,},
    xticklabels = {,,},
    hide axis = false,
        ]
    \addplot graphics[xmin = 0, xmax = \szx, ymin = 0, ymax = \szy] {Figures/\foldoi/PhaseNet1_5};
    \nextgroupplot[
    y label style={at={(axis description cs:0.00001*\szx,.5)},anchor=south},
    yticklabels = {,,},
    xticklabels = {,,},
    hide axis = false,
        ]
    \addplot graphics[xmin = 0, xmax = \szx, ymin = 0, ymax = \szy] {Figures/\foldoi/PhaseNet1_7};
    \nextgroupplot[
    y label style={at={(axis description cs:0.00001*\szx,.5)},anchor=south},
    yticklabels = {,,},
    xticklabels = {,,},
    hide axis = false,
        ]
    \addplot graphics[xmin = 0, xmax = \szx, ymin = 0, ymax = \szy] {Figures/\foldoi/PhaseNet1_9};
\end{groupplot}
\end{tikzpicture}
    \renewcommand{\szx}{320}
    \renewcommand{\szy}{300}
    \renewcommand{\szax}{\szsub}
    \renewcommand{\foldoi}{RealRK_C05_04_StackComparison_refocus_300_320}
    \renewcommand{\vmin}{-11.77}
    \renewcommand{\vmax}{16.43}
    \pgfmathsetmacro{\step}{\vmin + (\vmax - \vmin)/2}
    \renewcommand{\insep}{1} 
    \begin{tikzpicture}[baseline=0pt]
    \begin{groupplot}[name = mygroup1,
     group style = {group size = 1 by 5,
     vertical sep=\insep, horizontal sep=\insep},
     every plot/.style={width = {\szsub\textwidth},height = {\szy/\szx*\szsub\textwidth},axis equal image, scale only axis,
     }
    xmin = 0,xmax = \szx, ymin = 0, ymax = \szy,
    enlargelimits=false,
    axis equal image,
    scale only axis,
    width = {\szsub\textwidth},
    height = {\szy/\szx*\szsub\textwidth},
    hide axis,
    title style = {anchor=base}
    ]
    
    \nextgroupplot[
    y label style={at={(axis description cs:0.00001*\szx,.5)},anchor=south},
    yticklabels = {,,},
    xticklabels = {,,},
    hide axis = false,
    title = {\textbf{T.-L. 4}},
    ylabel = {\textbf{Fr. 1}}
        ]
    \addplot graphics[xmin = 0, xmax = \szx, ymin = 0, ymax = \szy] {Figures/\foldoi/PhaseNet1_1};

    \nextgroupplot[
    y label style={at={(axis description cs:0.00001*\szx,.5)},anchor=south},
    yticklabels = {,,},
    xticklabels = {,,},
    hide axis = false,
    ylabel = {\textbf{Fr. 3}}
        ]
    \addplot graphics[xmin = 0, xmax = \szx, ymin = 0, ymax = \szy] {Figures/\foldoi/PhaseNet1_3};
    \nextgroupplot[
    y label style={at={(axis description cs:0.00001*\szx,.5)},anchor=south},
    yticklabels = {,,},
    xticklabels = {,,},
    hide axis = false,
    ylabel = {\textbf{Fr. 5}}
        ]
    \addplot graphics[xmin = 0, xmax = \szx, ymin = 0, ymax = \szy] {Figures/\foldoi/PhaseNet1_5};

    \nextgroupplot[
    y label style={at={(axis description cs:0.00001*\szx,.5)},anchor=south},
    yticklabels = {,,},
    xticklabels = {,,},
    hide axis = false,
    ylabel = {\textbf{Fr. 7}}
        ]
    \addplot graphics[xmin = 0, xmax = \szx, ymin = 0, ymax = \szy] {Figures/\foldoi/PhaseNet1_7};
    
    \nextgroupplot[
    y label style={at={(axis description cs:0.00001*\szx,.5)},anchor=south},
    yticklabels = {,,},
    xticklabels = {,,},
    hide axis = false,
    ylabel = {\textbf{Fr. 9}}
        ]
    \addplot graphics[xmin = 0, xmax = \szx, ymin = 0, ymax = \szy] {Figures/\foldoi/PhaseNet1_9};
\end{groupplot}
\begin{axis}[at={($(group c1r5.south west) - (0,1em)$)}, anchor = north west,
    xmin = \vmin, xmax = \vmax, ymin = 0, ymax = 0.5,
    width=0,
    height = 0,
    scale only axis,
    enlargelimits=false,
    hide axis,
    axis equal image,
    colorbar horizontal,
    colormap/parula,
    colorbar style={
    at = {(0.025*\szax*\textwidth,0)},
        anchor = north west,
        xticklabel pos = lower,
        xmin = \vmin, xmax = \vmax,
        point meta min = \vmin,
        point meta max = \vmax,
        scale only axis,
        enlargelimits = false,
        scaled x ticks = true,
        width = {0.95*\szax*\textwidth},
    samples = 200,
    height = 0.15cm,
    xticklabel style={/pgf/number format/fixed, /pgf/number format/precision=0},
    xtick = {\vmin,\step,...,\vmax},
    xlabel={rad},
    xlabel style={yshift=0cm,xshift=0cm},
        }]
        \end{axis}
    \end{tikzpicture}
    \begin{tikzpicture}[baseline=0pt]
    \renewcommand{\foldoi}{RealRK_C05_04_StackComparison_refocus_300_320_seg_0.20}
    \begin{groupplot}[name = mygroup,
     group style = {group size = 1 by 5,
     vertical sep=\insep, horizontal sep=\insep},
     every plot/.style={width = {\szsub\textwidth},height = {\szy/\szx*\szsub\textwidth},axis equal image, scale only axis,
     }
    xmin = 0,xmax = \szx, ymin = 0, ymax = \szy,
    enlargelimits=false,
    axis equal image,
    scale only axis,
    width = {\szsub\textwidth},
    height = {\szy/\szx*\szsub\textwidth},
    hide axis,
    title style = {anchor=base}
    ]
    
    \nextgroupplot[
    y label style={at={(axis description cs:0.00001*\szx,.5)},anchor=south},
    yticklabels = {,,},
    xticklabels = {,,},
    hide axis = false,
        ]
    \addplot graphics[xmin = 0, xmax = \szx, ymin = 0, ymax = \szy] {Figures/\foldoi/PhaseNet1_1};
    \nextgroupplot[
    y label style={at={(axis description cs:0.00001*\szx,.5)},anchor=south},
    yticklabels = {,,},
    xticklabels = {,,},
    hide axis = false,
        ]
    \addplot graphics[xmin = 0, xmax = \szx, ymin = 0, ymax = \szy] {Figures/\foldoi/PhaseNet1_3};
    \nextgroupplot[
    y label style={at={(axis description cs:0.00001*\szx,.5)},anchor=south},
    yticklabels = {,,},
    xticklabels = {,,},
    hide axis = false,
        ]
    \addplot graphics[xmin = 0, xmax = \szx, ymin = 0, ymax = \szy] {Figures/\foldoi/PhaseNet1_5};
    \nextgroupplot[
    y label style={at={(axis description cs:0.00001*\szx,.5)},anchor=south},
    yticklabels = {,,},
    xticklabels = {,,},
    hide axis = false,
        ]
    \addplot graphics[xmin = 0, xmax = \szx, ymin = 0, ymax = \szy] {Figures/\foldoi/PhaseNet1_7};
    \nextgroupplot[
    y label style={at={(axis description cs:0.00001*\szx,.5)},anchor=south},
    yticklabels = {,,},
    xticklabels = {,,},
    hide axis = false,
        ]
    \addplot graphics[xmin = 0, xmax = \szx, ymin = 0, ymax = \szy] {Figures/\foldoi/PhaseNet1_9};
\end{groupplot}
\end{tikzpicture}
\caption{Time-lapse reconstructions by PhaseNet for all real data and corresponding segmentation.
For each panel of time-lapse measurements,
Left: the images were saturated for visualization purpose. 
Right: segmentation of time-lapse reconstructions.
We thresholded at $20\%$ of the maximum value of the image.
T.-L.: Time-Lapse; Fr.: Frame.
}\label{fig:timelapsePN}
\end{figure*}

We acquired other time-lapse measurements~(Figs.~\ref{fig:timelapsesupp1}-\ref{fig:timelapsesupp3}).
We observe that the unwrapped phases exhibit similar artifacts at the borders of the organoids.
The subsequent segmentation is also impacted, especially at the border, as pointed out by arrows.
We provide the reconstructions by PhaseNet in Fig.~\ref{fig:timelapsePN}.

\subsection{Relative Error Between Rewrapped and Wrapped Phase}

For real data, we use the relative error (Error) between the rewrapped phase $\mathcal{W}(\tilde{\mathbf{\Phi}})$ and the wrapped phase $\mathbf{\Psi}$ as an indicator of congruence with the measurements. Error is defined as
\begin{equation}
    \mathrm{Error}(\mathcal{W}(\tilde{\mathbf{\Phi}}),\mathbf{\Psi}) = \frac{\|\mathbf{\Psi}-\mathcal{W}(\tilde{\mathbf{\Phi}})\|_{2}}{\|\mathbf{\Psi}\|_{2}}.
\label{eq:ReError} 
\end{equation}

We provide in Table~\ref{tb:error} the relative error between the rewrapped and wrapped phase of the four experimental data for LS~\cite{ghiglia1994robust}, IRTV~\cite{kamilov2015isotropic}, PUMA~\cite{bioucas2007phase}, and our framework (PUDIP). This indicator shows that all methods are congruent with the measurements up to rounding errors.

\begin{table}[!tbp]
\setlength{\tabcolsep}{2.6pt}
\centering
\caption{Relative error between the rewrapped and the wrapped phase of four experimental data by different methods.}
\arrayrulecolor{black}
\begin{tabular} {r | rrrr} \toprule\toprule   
        &\multicolumn{1}{c}{LS} &\multicolumn{1}{c}{IRTV} &\multicolumn{1}{c}{PUMA}  &\multicolumn{1}{c}{PUDIP} \\
\midrule
Sample 1   &1.67$\times 10^{-31}$ &1.87$\times 10^{-31}$  &7.69$\times 10^{-31}$ &4.47$\times 10^{-14}$\\
Sample 2   &4.23$\times 10^{-31}$ &3.32$\times 10^{-31}$  &14.51$\times 10^{-31}$ &8.87$\times 10^{-14}$\\ 
Sample 3   &1.45$\times 10^{-31}$ &1.70$\times 10^{-31}$  &4.30$\times 10^{-31}$ &3.03$\times 10^{-14}$\\ 
Sample 4   &1.80$\times 10^{-31}$ &1.64$\times 10^{-31}$  &7.03$\times 10^{-31}$ &5.99$\times 10^{-14}$\\ 
\bottomrule\bottomrule
\end{tabular}
\label{tb:error}
\end{table}

\section{Training Dataset of PhaseNet}
\label{sec:phasenet}
\begin{figure*}[!tbp]
\centering
\includegraphics[width=1.0\textwidth]{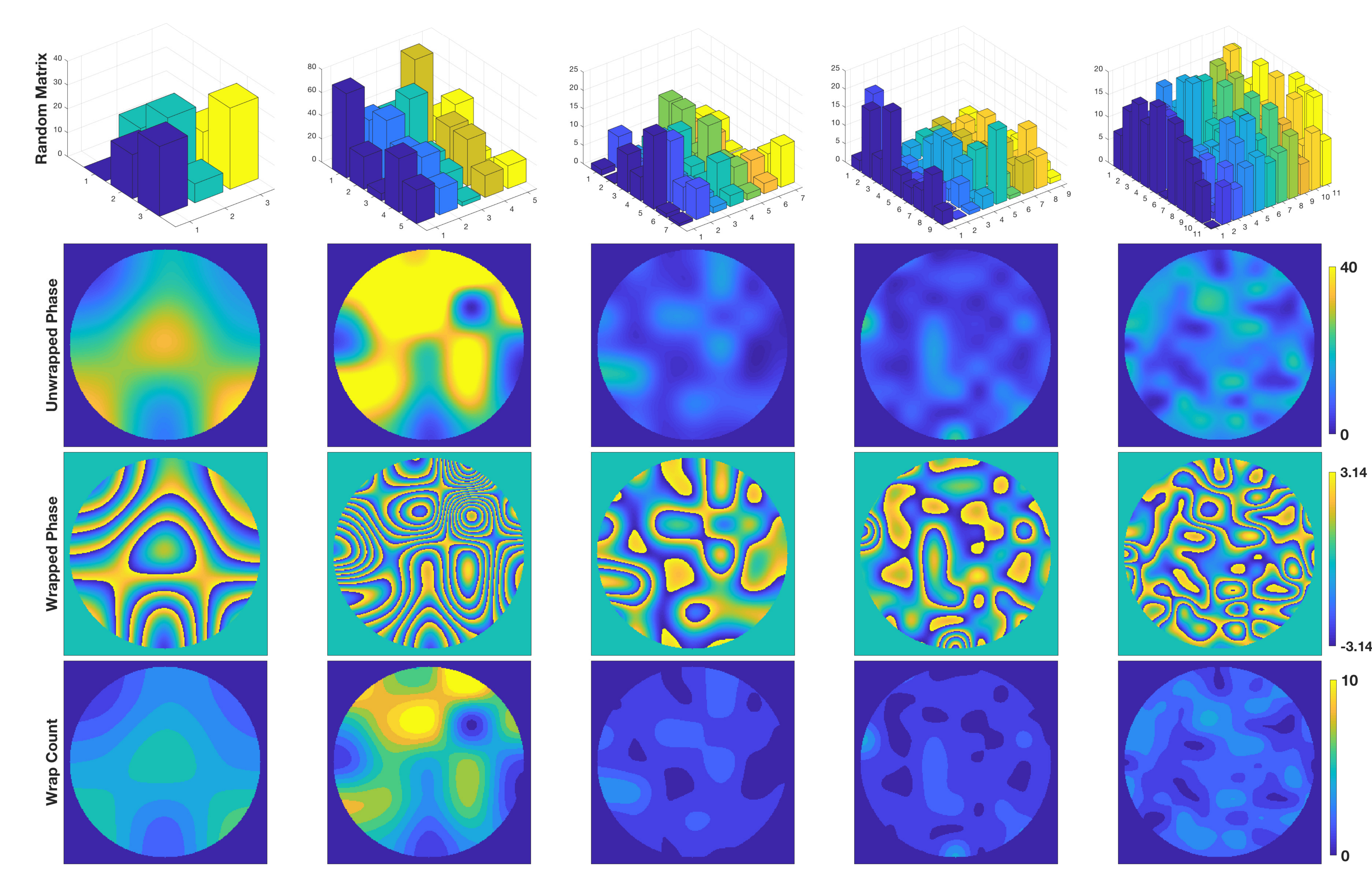}
\caption{Uniformly and Gaussian-distributed random square matrices (first row) and the corresponding unwrapped-phase images (second row), wrapped-phase images (third row), and the wrap-count images (last row). For training, the wrapped-phase images are the inputs and the wrap-count images are the ground-truths. From left to right, the size of the random matrix is ($3\times 3$), ($5\times 5$),  ($7\times 7$),  ($9\times 9$), and ($11\times 11$). The unwrapped-phase images vary in the range 0 to $40\pi$. The image size is ($256\times 256$).}
\label{fig:trainsample}
\end{figure*}

Samples of training data for PhaseNet~\cite{spoorthi2018phasenet} are shown in Fig.~\ref{fig:trainsample}. These were generated in two steps.
First, the elements of a square matrix whose size varies between $(3\times 3)$ and~\mbox{$(11 \times 11)$} were randomly generated
following a uniform distribution~$U(0,1)$ for half of the samples and a Gaussian distribution~$\mathcal{N}(0,1)$ followed by the subtraction of the minimum of the matrix for the other half.
Then, we multiplied the matrix by a scalar randomly generated following a uniform distribution~$U(3\pi,12\pi)$ and upsampled the matrix to a ($256 \times 256$) image using bicubic interpolation~\cite{li2001new}.
The obtained data had a maximum value ranging from $2\pi$ to $40\pi$.

We also computed the corresponding wrapped phase ($i.e.,$ the input to PhaseNet) and the wrap-count ($i.e.,$ the target for PhaseNet).
In Fig.~\ref{fig:trainsample}, the unwrapping task is much more complicated when the size of the random matrix or the maximum value of the object is larger.

\section{Additional metric: structural similarity}
For simulated data, we also compute the structural similarity (SSIM) between the reconstructed phase $\tilde{\mathbf{\Phi}}$ and ground-truth $\mathbf{\Phi}$ as
\begin{equation}
    \mathrm{SSIM}(\tilde{\mathbf{\Phi}},\mathbf{\Phi}) = \frac{(2\mu_{\mathbf{\Phi}}\mu_{\tilde{\mathbf{\Phi}}}+c_{1})(2\sigma_{{\mathbf{\Phi}}}\sigma_{\tilde{\mathbf{\Phi}}}+c_{2})}{(\mu_{{\mathbf{\Phi}}}^{2}+\mu_{\tilde{\mathbf{\Phi}}}^{2}+c_{1})(\sigma_{{\mathbf{\Phi}}}^{2}+\sigma_{\tilde{\mathbf{\Phi}}}^{2}+c_{2})},  
    \label{eq:SSIM}
\end{equation}
where $\mu_{{\mathbf{\Phi}}}$, $\mu_{\tilde{\mathbf{\Phi}}}$, $\sigma_{{\mathbf{\Phi}}}$, $\sigma_{\tilde{\mathbf{\Phi}}}$, and $\sigma_{{\mathbf{\Phi}}\tilde{\mathbf{\Phi}}}$ are the local means, standard deviations, and cross-covariance for images ${\mathbf{\Phi}}$, $\tilde{\mathbf{\Phi}}$, respectively. The regularization constants {$c_{1}=10^{-4}$ and $c_{2}=9\times 10^{-4}$ } avoid instabilities over image regions where the local mean or standard deviation is vanishing.

\section{Simulation Setup}
\label{sec:simsetup}
\subsection{Simulation of Phase Images of Organoid-like Sample}
\label{sec:fakeorganoid}
We simulated the acquisition of phase images of organoid-like samples.
We first created 3D volumes made of overlapping ellipsoids of uniform refractive index~$n_1$~(RI), to which
we added an external layer of RI~$n_2$.
Then, we simulated the propagation of a plane wave through the sample by using the beam propagation method~\cite{feit1988beam}.
We propagated the wave with a square voxel of length $0.2\micro\mathrm{m}$ in a square window of length~$102.4\micro\mathrm{m}$.
We refocused the complex total field at the center of the volume~($i.e.,$ free-space propagation).
We then downsampled the field to match the pixel size of the camera~($i.e.,~0.645\micro\mathrm{m}$).
Finally, we extracted the (wrapped) phase from the ratio between the total field and the incident field.
The straight-ray approximation~$\mathbf{\Phi}_{\mathrm{sr}}$ the expected phase~\cite{kak2002principles} as
\begin{equation}
    \mathbf{\Phi}_{\mathrm{sr}} = \frac{2\pi}{\lambda}\int_{-\infty}^{\infty} (n(x_1,x_2,x_3) - n_m)\mathrm{d}x_3,
\end{equation}
where $n(x_1,x_2,x_3):\mathbb{R}^3\rightarrow \mathbb{R}$ is the distribution of RI of the 3D volume, and $n_m > 0$ is the RI of the medium~($i.e.,$~water).

\subsection{Sharp Edges and Variable Concave Shapes}
\label{sec:simsetup1}

As shown in Fig.~\ref{fig:simulate1}, we first simulated one phase surface with the shape of an ellipse of radii~$80$ pixels and~$110$ pixels along the vertical and horizontal dimensions, respectively.
The ellipse was filled with a Gaussian function whose maximum is~$15$ and standard deviation is randomly generated~$\sigma \sim U(0.30,0.65)$.
The area outside the ellipse was set to~$0$.
We select the horizontal left-to-right direction as the x-axis and set the coordinate axes to be left-handed.
The ellipse was cropped with angles ranging from $0\degree$ to $270\degree$ with an increment of $45\degree$.
Similar to~\cite{bioucas2007phase,kamilov2015isotropic}, these croppings introduce a variety of discontinuities and shapes.

\subsection{Sharp Edges, Concave Shapes, and Variable Phase Maxima}
\label{sec:simsetup2}

For this numerical experiment~(Fig.~\ref{fig:simulate2}), we generated elliptical phase surfaces with radii $102$ pixels and $120$ pixels along the vertical and horizontal dimensions, respectively. The cropped part was kept constant, with an angle set at $135\degree$.
We scaled the phase so that its maximum was in the range of $6$ to $42$ with an increment of $6$. For this case, the high values induce several wrapping events. By controlling their number,we could tune the difficulty of the unwrapping task.

\subsection{Sharp Edges and Variable Heterogeneity}
\label{sec:simsetup3}
For this experiment, we generated one circular smooth object in the same way as for the dataset used to train PhaseNet (see Supplementary Note~S\ref{sec:phasenet}).
The samples have sharp edges and variable heterogeneity in a similar fashion to~\cite{wang2019one}~(Fig.~\ref{fig:simulate3_g}).
Its height varies from $15$ to $20$. The size of the square random matrices varies in the range of $(3\times 3)$ to $(11 \times 11)$ with an increment of $2$ on each side.
Larger matrices induce more complex objects.

\subsection{Sharp Edges and Variable Speckle Noise}
\label{sec:simsetup3}
In digital holography microscopy, the noise is mainly characterized by speckle noise~\cite{bianco2018strategies} that corrupts the image before the wrapping operation.
To assess the robustness of our method, we perturbed the (unwrapped) sample B (cropping angle $135\degree$) with speckle noise~\cite{goodman2007speckle}.
We added three levels of noise~$\{11.8,15.7,22.8\}$dB~(Fig.~\ref{fig:simulate_snr})
and computed the metrics with respect to the perturbed images.

\section{Parameter Setting of Simulated Experiments}
We performed $1,000$ iterations to optimize over the synthetic samples. 
The hyperparameters of the network were initialized to default values by PyTorch.
We used a learning rate of $0.01$ and we updated the weights $w_n$  
with $[\epsilon_{\mathrm{min}},\epsilon_{\mathrm{max}}]$.
We tried $[\epsilon_{\mathrm{min}},\epsilon_{\mathrm{max}}] = [0.1, 10]$, $[0.05, 20]$, $[0.02,50]$, and $[0.01, 100]$, choosing the best performance.
The weights were updated every $N_{w}=200$, $100$, $100$, and $100$ iterations for the samples~B, C, D, and  E, respectively.
As we randomly initialized the parameters of the network~$\boldsymbol{\theta}$, we repeated each experiment five times and report the average performance.

\section{Supplementary Reconstructions of Organoid-like Samples}
\label{subsec:simulateOrganoid}

We simulated supplementary organoid-like samples and unwrapped their corresponding wrapped phase. As shown in Fig.~\ref{fig:fakeorganoid2} and~\ref{fig:fakeorganoid3}, similar behaviors are observed. The slightly defocused parts are wrongly estimated by baseline methods, which impacts the whole unwrapping result.
The phase unwrapped by PUDIP matches the straight-ray approximation~$\mathbf{\Phi}_\mathrm{sr}$.

\renewcommand{\szx}{159}
\renewcommand{\szy}{159}
\renewcommand{\szsub}{0.15}
\renewcommand{\szax}{4*\szsub}
\renewcommand{\foldoi}{FakeOrganoid2_original}

\newcommand{\nvmin}{1.333}
\newcommand{\nvmax}{1.417}
\pgfmathsetmacro{\nstep}{\nvmin + (\nvmax - \nvmin)/1}
\renewcommand{\vmin}{0}
\renewcommand{\vmax}{22.96}
\pgfmathsetmacro{\step}{\vmin + (\vmax - \vmin)/3}
\renewcommand{\insep}{1} 

\begin{figure*}[t]
    \centering
    \begin{tikzpicture}
    \begin{groupplot}[
     group style = {group size = 6 by 2,
     vertical sep=\insep, horizontal sep=\insep},
     every plot/.style={width = {\szsub\textwidth},height = {\szy/\szx*\szsub\textwidth},axis equal image, scale only axis,
     }
    xmin = 0,xmax = \szx, ymin = 0, ymax = \szy,
    enlargelimits=false,
    axis equal image,
    scale only axis,
    width = {\szsub\textwidth},
    height = {\szy/\szx*\szsub\textwidth},
    hide axis,
    title style = {anchor=east,
    at={(axis cs:\szx,0)},
    color=white}
    ]
    \nextgroupplot[
    y label style={at={(axis description cs:0.00001*\szx,.5)},anchor=south},
    yticklabels = {,,},
    xticklabels = {,,},
    hide axis = false,
    title = {\textbf{XY}},
        ]
    \addplot graphics[xmin = 0, xmax = \szx, ymin = 0, ymax = \szy,] {Figures/\foldoi/nx};
    \draw[white,dotted,thick](0.5*\szx,0)--(0.5*\szx,\szy);
    \draw[white,dotted,thick](0,0.5*\szy)--(\szx,0.5*\szy);
    \nextgroupplot[
    y label style={at={(axis description cs:0.00001*\szx,.5)},anchor=south},
    yticklabels = {,,},
    xticklabels = {,,},
    hide axis = false,
    title = {\textbf{XZ}},
        ]
    \addplot graphics[xmin = 0, xmax = \szx, ymin = 0, ymax = \szy] {Figures/\foldoi/ny};
    \draw[white,dotted,thick](0,0.5*\szy)--(\szx,0.5*\szy);
    \nextgroupplot[
    y label style={at={(axis description cs:0.00001*\szx,.5)},anchor=south},
    yticklabels = {,,},
    xticklabels = {,,},
    hide axis = false,
    title = {\textbf{GA}},
        ]
    \addplot graphics[xmin = 0, xmax = \szx, ymin = 0, ymax = \szy] {Figures/\foldoi/GA};
    
    
    \nextgroupplot[title = {\textbf{LS}},
        ]
    \addplot graphics[xmin = 0, xmax = \szx, ymin = 0, ymax = \szy] {Figures/\foldoi/LS};
   \nextgroupplot[title = {\textbf{IRTV}},
        ]
    \addplot graphics[xmin = 0, xmax = \szx, ymin = 0, ymax = \szy] {Figures/\foldoi/IRTV};
    
    \nextgroupplot[
    y label style={at={(axis description cs:0.00001*\szx,.5)},anchor=south},
    yticklabels = {,,},
    xticklabels = {,,},
    hide axis = false,
    title = {\textbf{Wrapped Phase}},
        ]
    \addplot graphics[xmin = 0, xmax = \szx, ymin = 0, ymax = \szy] {Figures/\foldoi/wrapPhase};
    
    
    \nextgroupplot[
    y label style={at={(axis description cs:0.00001*\szx,.5)},anchor=south},
    yticklabels = {,,},
    xticklabels = {,,},
    hide axis = false,
    title = {\textbf{YZ}},
        ]
    \addplot graphics[xmin = 0, xmax = \szx, ymin = 0, ymax = \szy] {Figures/\foldoi/nz};
    \draw[white,dotted,thick](0.5*\szx,0)--(0.5*\szx,\szy);
    \nextgroupplot;
    
    \nextgroupplot[title = {\textbf{PUMA}},
        ]
    \addplot graphics[xmin = 0, xmax = \szx, ymin = 0, ymax = \szy] {Figures/\foldoi/PUMA};
    
    \nextgroupplot[title = {\textbf{PhaseNet}},
        ]
    \addplot graphics[xmin = 0, xmax = \szx, ymin = 0, ymax = \szy] {Figures/\foldoi/PhaseNet};

    \nextgroupplot[title = {\textbf{PUDIP}},
        ]
    \addplot graphics[xmin = 0, xmax = \szx, ymin = 0, ymax = \szy] {Figures/\foldoi/PUDIP};
    
    \nextgroupplot[title = {$\mathbf{\Phi}_{\mathrm{sr}}$},
        ]
    \addplot graphics[xmin = 0, xmax = \szx, ymin = 0, ymax = \szy] {Figures/\foldoi/linearap};
\end{groupplot}

\begin{axis}[at={($(group c6r1.north east) - (0,0.5em)$)}, anchor = north west,
    ymin = 0, ymax = 0.5, xmin = -3.14, xmax = 3.14,
    width = 0,
    height = 0,
    scale only axis,
    enlargelimits=false,
    hide axis,
    axis equal image,
    colorbar horizontal,
    colormap/parula,
    colorbar style={
    at = {(-0.5em,0.0)},
        anchor = north east,
        xticklabel pos = lower,
        xmin = -3.14, xmax = 3.14,
        point meta min = -3.14,
        point meta max = 3.14,
        scale only axis,
        enlargelimits = false,
        scaled x ticks = true,
        width = {0.075*\szax*\textwidth},
    samples = 200,
    height = 0.15cm,
    xticklabel style={/pgf/number format/fixed, /pgf/number format/precision=0},
    xtick = {-3.14,3.14},
    xticklabels = {{\color{white}$-\pi$}, {\color{white}$\pi$}},
    xlabel style={yshift=1em,xshift=0em}
        }]
\end{axis}

\begin{axis}[at={($(group c1r2.south east) - (0,1em)$)}, anchor = north west,
    xmin = \nvmin, xmax = \nvmax, ymin = 0, ymax = 0.5,
    width = 0,
    height = 0,
    scale only axis,
    enlargelimits=false,
    hide axis,
    axis equal image,
    colorbar horizontal,
    colormap/blackwhite,
    colorbar style={
    at = {(0.0*\szax*\textwidth,0.5)},
        anchor = north east,
        xticklabel pos = lower,
        xmin = \nvmin, xmax = \nvmax,
        point meta min = \nvmin,
        point meta max = \nvmax,
        scale only axis,
        enlargelimits = false,
        scaled x ticks = true,
        width = {0.25*\szax*\textwidth},
    samples = 200,
    height = 0.15cm,
    xticklabel style={/pgf/number format/fixed, /pgf/number format/precision=2},
    xtick = {\nvmin,\nstep,...,\nvmax},
    xlabel={rad},
    xlabel style={yshift=1em,xshift=0cm},
        }]
        \end{axis}

\begin{axis}[at={($(group c6r2.south east) - (0,1em)$)}, anchor = north west,
    xmin = \vmin, xmax = \vmax, ymin = 0, ymax = 0.5,
    width=0,
    height = 0,
    scale only axis,
    enlargelimits=false,
    hide axis,
    axis equal image,
    colorbar horizontal,
    colormap/parula,
    colorbar style={
    at = {(0.0*\szax*\textwidth,0.5)},
        anchor = north east,
        xticklabel pos = lower,
        xmin = \vmin, xmax = \vmax,
        point meta min = \vmin,
        point meta max = \vmax,
        scale only axis,
        enlargelimits = false,
        scaled x ticks = true,
        width = {0.95*\szax*\textwidth},
    samples = 200,
    height = 0.15cm,
    xticklabel style={/pgf/number format/fixed, /pgf/number format/precision=0},
    xtick = {\vmin,\step,...,\vmax},
    xlabel={rad},
    xlabel style={yshift=1em,xshift=0cm},
        }]
        \end{axis}
    \end{tikzpicture}
    
    \caption{Organoid-like reconstructions. The images were saturated for visualization purpose. The size of the unwrapped phase image is ($159 \times 159$).
    The first two columns are orthographic slices of the 3D distribution of refractive indices. All slices include the center of the volume. From the third to fifth column, the text gives the method used to unwrap. The wrapped phase and the straight-ray approximation~$\mathbf{\Phi}_{\mathrm{sr}}$ are displayed in the last column (from top to bottom).
    }
    \label{fig:fakeorganoid2}
\end{figure*}

\renewcommand{\szx}{159}
\renewcommand{\szy}{159}
\renewcommand{\szsub}{0.15}
\renewcommand{\szax}{4*\szsub}
\renewcommand{\foldoi}{FakeOrganoid3_original}

\renewcommand{\nvmin}{1.333}
\renewcommand{\nvmax}{1.417}
\pgfmathsetmacro{\nstep}{\nvmin + (\nvmax - \nvmin)/1}
\renewcommand{\vmin}{0}
\renewcommand{\vmax}{18.98}
\pgfmathsetmacro{\step}{\vmin + (\vmax - \vmin)/3}
\renewcommand{\insep}{1} 

\begin{figure*}[t]
    \centering
    \begin{tikzpicture}
    \begin{groupplot}[
     group style = {group size = 6 by 2,
     vertical sep=\insep, horizontal sep=\insep},
     every plot/.style={width = {\szsub\textwidth},height = {\szy/\szx*\szsub\textwidth},axis equal image, scale only axis,
     }
    xmin = 0,xmax = \szx, ymin = 0, ymax = \szy,
    enlargelimits=false,
    axis equal image,
    scale only axis,
    width = {\szsub\textwidth},
    height = {\szy/\szx*\szsub\textwidth},
    hide axis,
    title style = {anchor=east,
    at={(axis cs:\szx,0)},
    color=white}
    ]
    \nextgroupplot[
    y label style={at={(axis description cs:0.00001*\szx,.5)},anchor=south},
    yticklabels = {,,},
    xticklabels = {,,},
    hide axis = false,
    title = {\textbf{XY}},
        ]
    \addplot graphics[xmin = 0, xmax = \szx, ymin = 0, ymax = \szy,] {Figures/\foldoi/nx};
     \draw[white,dotted,thick](0.5*\szx,0)--(0.5*\szx,\szy);
    \draw[white,dotted,thick](0,0.5*\szy)--(\szx,0.5*\szy);
    \nextgroupplot[
    y label style={at={(axis description cs:0.00001*\szx,.5)},anchor=south},
    yticklabels = {,,},
    xticklabels = {,,},
    hide axis = false,
    title = {\textbf{XZ}},
        ]
    \addplot graphics[xmin = 0, xmax = \szx, ymin = 0, ymax = \szy] {Figures/\foldoi/ny};
    \draw[white,dotted,thick](0,0.5*\szy)--(\szx,0.5*\szy);

    \nextgroupplot[
    y label style={at={(axis description cs:0.00001*\szx,.5)},anchor=south},
    yticklabels = {,,},
    xticklabels = {,,},
    hide axis = false,
    title = {\textbf{GA}},
        ]
    \addplot graphics[xmin = 0, xmax = \szx, ymin = 0, ymax = \szy] {Figures/\foldoi/GA};
    
    
    \nextgroupplot[title = {\textbf{LS}},
        ]
    \addplot graphics[xmin = 0, xmax = \szx, ymin = 0, ymax = \szy] {Figures/\foldoi/LS};
   \nextgroupplot[title = {\textbf{IRTV}},
        ]
    \addplot graphics[xmin = 0, xmax = \szx, ymin = 0, ymax = \szy] {Figures/\foldoi/IRTV};
    
    \nextgroupplot[
    y label style={at={(axis description cs:0.00001*\szx,.5)},anchor=south},
    yticklabels = {,,},
    xticklabels = {,,},
    hide axis = false,
    title = {\textbf{Wrapped Phase}},
        ]
    \addplot graphics[xmin = 0, xmax = \szx, ymin = 0, ymax = \szy] {Figures/\foldoi/wrapPhase};
    
    
    \nextgroupplot[
    y label style={at={(axis description cs:0.00001*\szx,.5)},anchor=south},
    yticklabels = {,,},
    xticklabels = {,,},
    hide axis = false,
    title = {\textbf{YZ}},
        ]
    \addplot graphics[xmin = 0, xmax = \szx, ymin = 0, ymax = \szy] {Figures/\foldoi/nz};
    \draw[white,dotted,thick](0.5*\szx,0)--(0.5*\szx,\szy);
    \nextgroupplot;
    
    \nextgroupplot[title = {\textbf{PUMA}},
        ]
    \addplot graphics[xmin = 0, xmax = \szx, ymin = 0, ymax = \szy] {Figures/\foldoi/PUMA};
    
    \nextgroupplot[title = {\textbf{PhaseNet}},
        ]
    \addplot graphics[xmin = 0, xmax = \szx, ymin = 0, ymax = \szy] {Figures/\foldoi/PhaseNet};

    \nextgroupplot[title = {\textbf{PUDIP}},
        ]
    \addplot graphics[xmin = 0, xmax = \szx, ymin = 0, ymax = \szy] {Figures/\foldoi/PUDIP};
    
    \nextgroupplot[title = {$\mathbf{\Phi}_{\mathrm{sr}}$},
        ]
    \addplot graphics[xmin = 0, xmax = \szx, ymin = 0, ymax = \szy] {Figures/\foldoi/linearap};
\end{groupplot}

\begin{axis}[at={($(group c6r1.north east) - (0,0.5em)$)}, anchor = north east,
    ymin = 0, ymax = 0.5, xmin = -3.14, xmax = 3.14,
    width = 0,
    height = 0,
    scale only axis,
    enlargelimits=false,
    hide axis,
    axis equal image,
    colorbar horizontal,
    colormap/parula,
    colorbar style={
    at = {(-0.5em,0.0)},
        anchor = north east,
        xticklabel pos = lower,
        xmin = -3.14, xmax = 3.14,
        point meta min = -3.14,
        point meta max = 3.14,
        scale only axis,
        enlargelimits = false,
        scaled x ticks = true,
        width = {0.05*\szax*\textwidth},
    samples = 200,
    height = 0.15cm,
    xticklabel style={/pgf/number format/fixed, /pgf/number format/precision=0},
    xtick = {-3.14,3.14},
    xticklabels = {{\color{white}$-\pi$}, {\color{white}$\pi$}},
    xlabel style={yshift=1em,xshift=0em}
        }]
\end{axis}

\begin{axis}[at={($(group c1r2.south east) - (0,1em)$)}, anchor = north west,
    xmin = \nvmin, xmax = \nvmax, ymin = 0, ymax = 0.5,
    width = 0,
    height = 0,
    scale only axis,
    enlargelimits=false,
    hide axis,
    axis equal image,
    colorbar horizontal,
    colormap/blackwhite,
    colorbar style={
    at = {(0.0*\szax*\textwidth,0.5)},
        anchor = north east,
        xticklabel pos = lower,
        xmin = \nvmin, xmax = \nvmax,
        point meta min = \nvmin,
        point meta max = \nvmax,
        scale only axis,
        enlargelimits = false,
        scaled x ticks = true,
        width = {0.25*\szax*\textwidth},
    samples = 200,
    height = 0.15cm,
    xticklabel style={/pgf/number format/fixed, /pgf/number format/precision=2},
    xtick = {\nvmin,\nstep,...,\nvmax},
    xlabel={rad},
    xlabel style={yshift=1em,xshift=0cm},
        }]
        \end{axis}

\begin{axis}[at={($(group c6r2.south east) - (0,1em)$)}, anchor = north west,
    xmin = \vmin, xmax = \vmax, ymin = 0, ymax = 0.5,
    width=0,
    height = 0,
    scale only axis,
    enlargelimits=false,
    hide axis,
    axis equal image,
    colorbar horizontal,
    colormap/parula,
    colorbar style={
    at = {(0.0*\szax*\textwidth,0.5)},
        anchor = north east,
        xticklabel pos = lower,
        xmin = \vmin, xmax = \vmax,
        point meta min = \vmin,
        point meta max = \vmax,
        scale only axis,
        enlargelimits = false,
        scaled x ticks = true,
        width = {0.95*\szax*\textwidth},
    samples = 200,
    height = 0.15cm,
    xticklabel style={/pgf/number format/fixed, /pgf/number format/precision=0},
    xtick = {\vmin,\step,...,\vmax},
    xlabel={rad},
    xlabel style={yshift=1em,xshift=0cm},
        }]
        \end{axis}
    \end{tikzpicture}
    
    \caption{Organoid-like reconstructions. The images were saturated for visualization purpose. The size of the unwrapped phase image is ($159 \times 159$).
    The first two columns are orthographic slices of the 3D distribution of refractive indices. All slices include the center of the volume. From the third to fifth column, the text gives the method used to unwrap. The wrapped phase and the straight-ray approximation~$\mathbf{\Phi}_{\mathrm{sr}}$ are displayed in the last column (from top to bottom).
    }
    \label{fig:fakeorganoid3}
\end{figure*}

\section{Reconstructions of Sample B}
\label{subsec:simulateRec1}

\begin{figure*}[!tbp]
\centering
\includegraphics[width=0.95\textwidth]{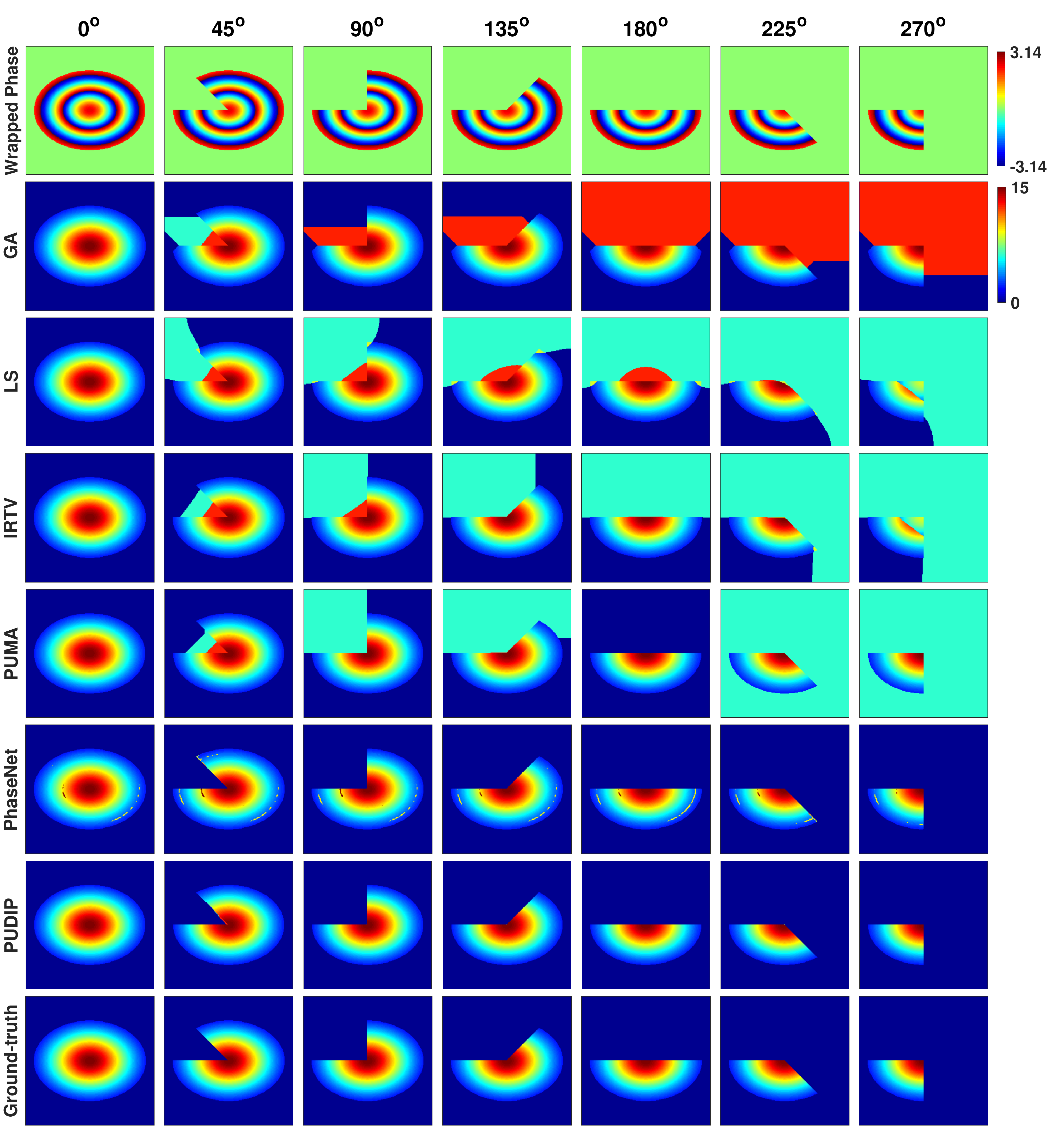}
\caption{Reconstructed unwrapped-phase images of simulated samples with diverse cropping angles (Sample B). From top to bottom: wrapped phase, results obtained by GA, LS, IRTV, PUMA, PhaseNet, and our approach (PUDIP).  The ground-truth images are presented in the last row.}
\label{fig:simulate1}
\end{figure*}

The reconstructions obtained by different methods for the sample B are shown in Fig.~\ref{fig:simulate1}. In this experiment, the cropping angle was gradually increased.
All methods except PhaseNet are able to recover the correct unwrapped phase in the absence of cropping.
When the phase image is cropped, all conventional methods lead to blocky errors, especially at large angles.
For the $180\degree$ case only, PUMA~\cite{bioucas2007phase} and the proposed method accurately unwrap the phase.

In general, CNN-based approaches perform better than the model-based methods.
Both PhaseNet and PUDIP are able to reconstruct more accurate shapes and values, especially over the cropped region.
For PhaseNet, we observe that the clustering-based postprocessing strongly improves the final results but still introduces undesirable values along the contours of clusters.
By contrast, our method recovers well the samples in all cases, including the few over which PhaseNet fails.
Moreover, the RSNR and SSIM shown in Tables~\ref{tb:RSNR1} and~\ref{tb:SSIM1} illustrate that PUDIP exhibits a significant improvement in nearly all cases.

\begin{table}[t]
\setlength{\tabcolsep}{3pt}
\centering
\caption{RSNR [dB] of the reconstructed-phase images versus the angle of cropping. The RSNR of our method (PUDIP) is the average of five experiments.\label{tb:RSNR1}}
\arrayrulecolor{black}
\begin{tabular}{r | cccccc}

\toprule\toprule   
Angle  & GA  & LS & IRTV & PUMA & Phas\text{e}N\text{e}t & PU\text{D}IP \\
\midrule
$0\degree$    &$\infty$ &$\infty$ &$\infty$ &$\infty$  &24.79 &$\infty$\\ 
$45\degree$   &6.80 &5.15 &8.61 &10.20 &14.10 &\textbf{15.99}\\ 
$90\degree$   &2.70 &2.86 &3.15 &3.87  &22.14 &\textbf{37.75}\\ 
$135\degree$  &-0.56 &1.32 &2.46 &2.06  &22.01 &\textbf{43.52}\\ 
$180\degree$  &-5.15 &-0.13 &0.84 &$\infty$  &19.33 &$\infty$\\ 
$225\degree$  &-6.70 &-0.43 &-0.24 &2.21  &19.96 &\textbf{41.44}\\ 
$270\degree$  &-8.00 &-1.85 &-1.66 &2.01  &21.23 &$\infty$\\
\bottomrule\bottomrule
\end{tabular}
\end{table}

\begin{table}[!tbp]
\setlength{\tabcolsep}{3pt}
\centering
\caption{SSIM of the reconstructed-phase images versus the angle of cropping. The SSIM of our method (PUDIP) is the average of five experiments. \label{tb:SSIM1}}
\arrayrulecolor{black}
\begin{tabular} {r | cccccc}
\toprule\toprule 
Angle  &GA  &LS &IRTV &PUMA &PhaseNet &PUDIP \\
\midrule
$0\degree$    &\textbf{1.0000} &\textbf{1.0000} &\textbf{1.0000} &\textbf{1.0000}  &0.9799 &\textbf{1.0000}\\ 
$45\degree$   &0.8975 &0.8346 &0.9429 &0.9595  &0.9680 &\textbf{0.9866}\\ 
$90\degree$   &0.9074 &0.7180 &0.7337 &0.7418  &0.9772 &\textbf{0.9995}\\ 
$135\degree$  &0.8360 &0.5716 &0.6510 &0.5576  &0.9769 &\textbf{1.0000} \\ 
$180\degree$  &0.4863 &0.4772 &0.4893 &\textbf{1.0000}  &0.9771 &\textbf{1.0000} \\ 
$225\degree$  &0.4269 &0.3411 &0.3225 &0.1183  &0.9858 &\textbf{1.0000} \\ 
$270\degree$  &0.3655 &0.2395 &0.2246 &0.0838  &0.9907 &\textbf{1.0000} \\ 
\bottomrule\bottomrule
\end{tabular}
\end{table}

\section{Reconstructions of Sample C}
\label{subsec:simulateRec2}
\begin{figure*}[!tbp]
\centering
\includegraphics[width=0.95\textwidth]{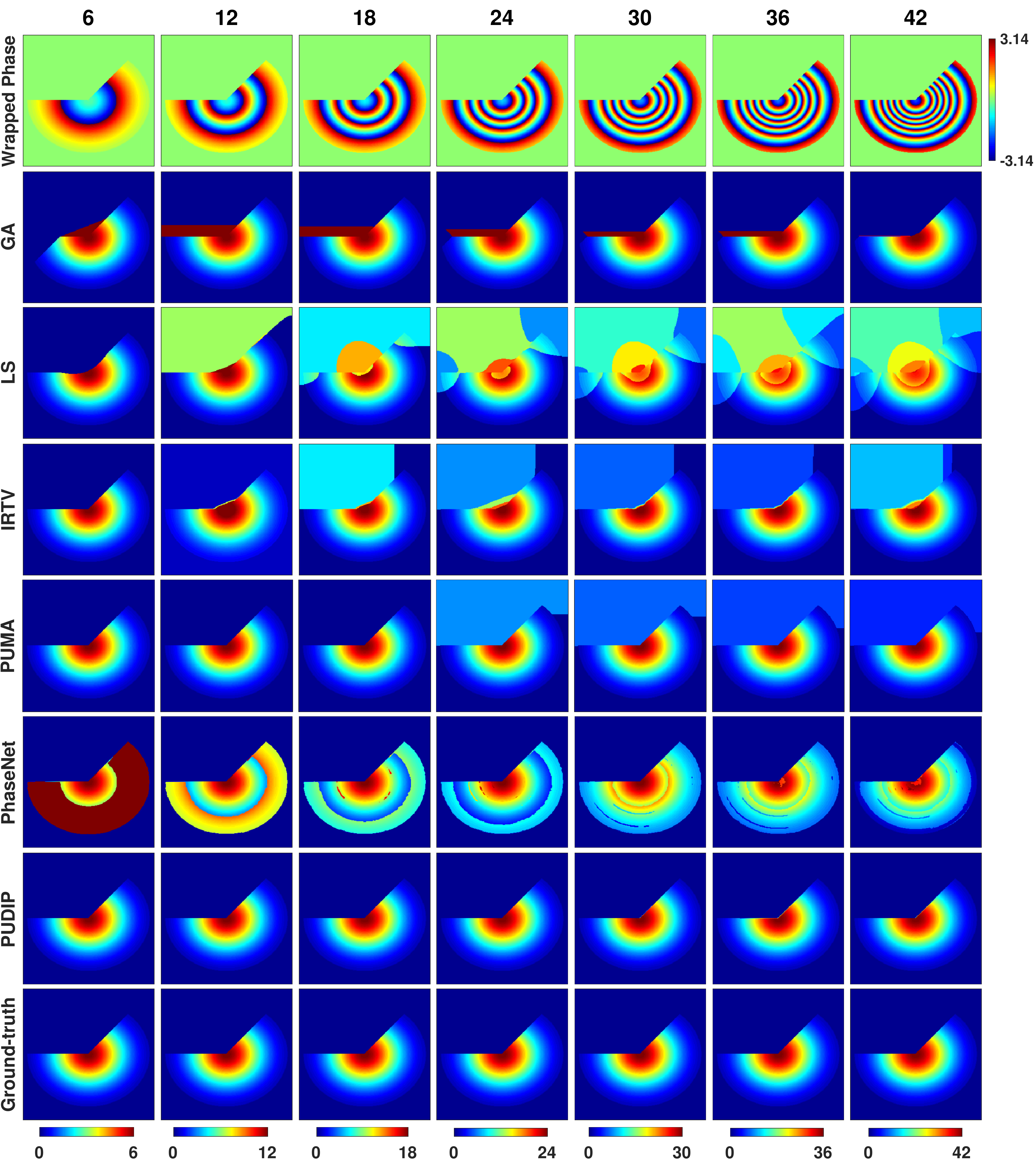}
\caption{Reconstructed unwrapped-phase images of simulated samples with diverse maximal values
(Sample C). From top to bottom: wrapped phase, results obtained by GA, LS, IRTV, PUMA, PhaseNet, and our approach (PUDIP). The ground-truth images are presented in the last row.}
\label{fig:simulate2}
\end{figure*}

We obtained similar results with the second experiment in which we increased the maximum value instead~(Fig.~\ref{fig:simulate2}).
When the height is low (first and second columns), IRTV~\cite{kamilov2015isotropic} and PUMA perform well.
When the height is higher, all the other baseline methods wrongly estimate large portions of the images.

PhaseNet always fails to recover the phase, which points out the sensitivity of this supervised-learning method to the mismatch between the training and testing set.
On the contrary, our learning framework always unwraps the phase with few errors and without prior training.
The RSNR and SSIM again corroborate these observations~(Tables~\ref{tb:RSNR2} and~\ref{tb:SSIM2}).

\begin{table}[!tbp]
\setlength{\tabcolsep}{3pt}
\centering
\caption{RSNR [dB] of the reconstructed-phase images versus the maximal value. The RSNR of our method (PUDIP) is the average of five experiments.}
\arrayrulecolor{black}
\begin{tabular} {r | cccccc}

\toprule\toprule   
Max value  &GA  &LS &IRTV &PUMA &Phas\text{e}N\text{e}t & PU\text{D}IP \\
\midrule
6   &5.69    &13.12  &$\infty$  &$\infty$  &-5.71  &$\infty$  \\
12  &1.02    &-0.39  &11.31  &$\infty$  &1.25  &$\infty$  \\
18  &1.45    &1.20   &3.22   &$\infty$  &5.62  &$\infty$  \\
24  &3.85    &0.21   &4.99   &5.69  &8.95  &\textbf{78.54}  \\
30  &5.20    &1.04   &7.38   &7.62  &8.35  &\textbf{28.53}  \\
36  &4.62    &0.48   &8.71   &9.18  &10.13  &\textbf{25.70}  \\
42  &14.53   &0.94   &4.37   &10.52  &12.46  &\textbf{27.74}  \\
\bottomrule\bottomrule
\end{tabular}
\label{tb:RSNR2}
\end{table}

\begin{table}[!tbp]
\setlength{\tabcolsep}{2.5pt}
\centering
\caption{SSIM of the reconstructed-phase images versus the maximal value. The SSIM of our method (PUDIP) is the average of five experiments.}
\arrayrulecolor{black}
\begin{tabular} {r | cccccc} \toprule\toprule   
Max height  &GA  &LS &IRTV &PUMA &PhaseNet &PUDIP\\
\midrule
6   &0.9299    &0.9834   &\textbf{1.0000}  &\textbf{1.0000}  &0.7105  &\textbf{1.0000}  \\
12  &0.9258    &0.5989   &0.3616   &\textbf{1.0000}  &0.7788  &\textbf{1.0000}   \\
18  &0.9311    &0.5539   &0.6481   &\textbf{1.0000}  &0.8067  &\textbf{1.0000}   \\
24  &0.9453    &0.5312   &0.6411   &0.5873   &0.8298  &\textbf{0.9990}  \\
30  &0.9551    &0.5160   &0.6435   &0.5866   &0.8168  &\textbf{0.9977}  \\
36  &0.9532    &0.5044   &0.6416   &0.5796   &0.8224  &\textbf{0.9957}  \\
42  &0.9782    &0.4951   &0.6364   &0.5784   &0.8597  &\textbf{0.9959}  \\
\bottomrule\bottomrule
\end{tabular}
\label{tb:SSIM2}
\end{table}

\section{Reconstructions of Sample D}
\label{subsec:simulateRec3}
\begin{figure*}[!tbp]
\centering
\includegraphics[width=0.85\textwidth]{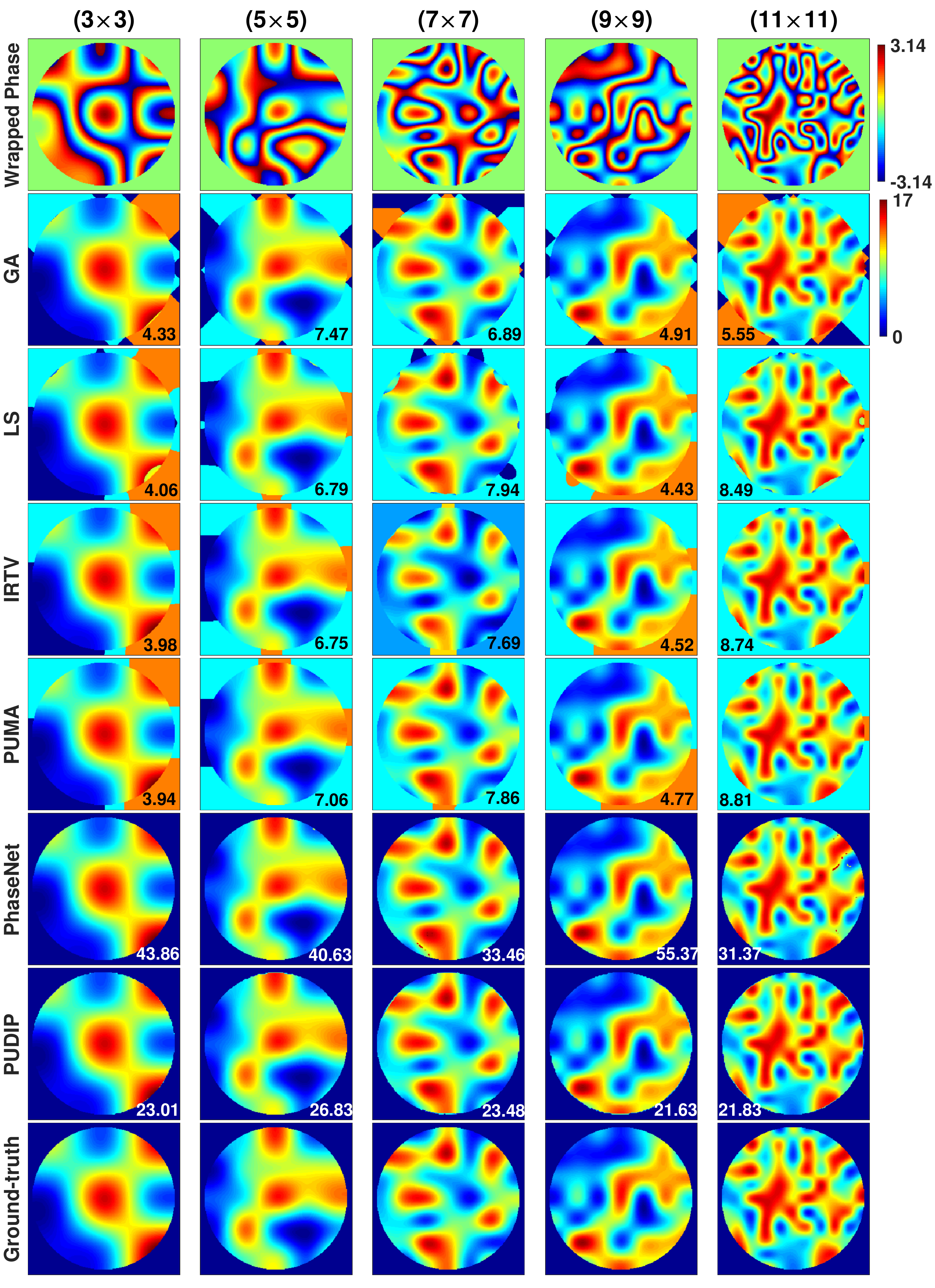}
\caption{Unwrapped-phase images of simulated samples with diverse random distributions (Sample D-1). From top to bottom: wrapped phase, results obtained by  GA, LS, IRTV, PUMA,  PhaseNet, and our approach (PUDIP).  The ground-truth images are presented in the last row. The numbers give the corresponding RSNR [dB].}
\label{fig:simulate3_g}
\end{figure*}

\begin{figure*}[!tbp]
\centering
\includegraphics[width=0.85\textwidth]{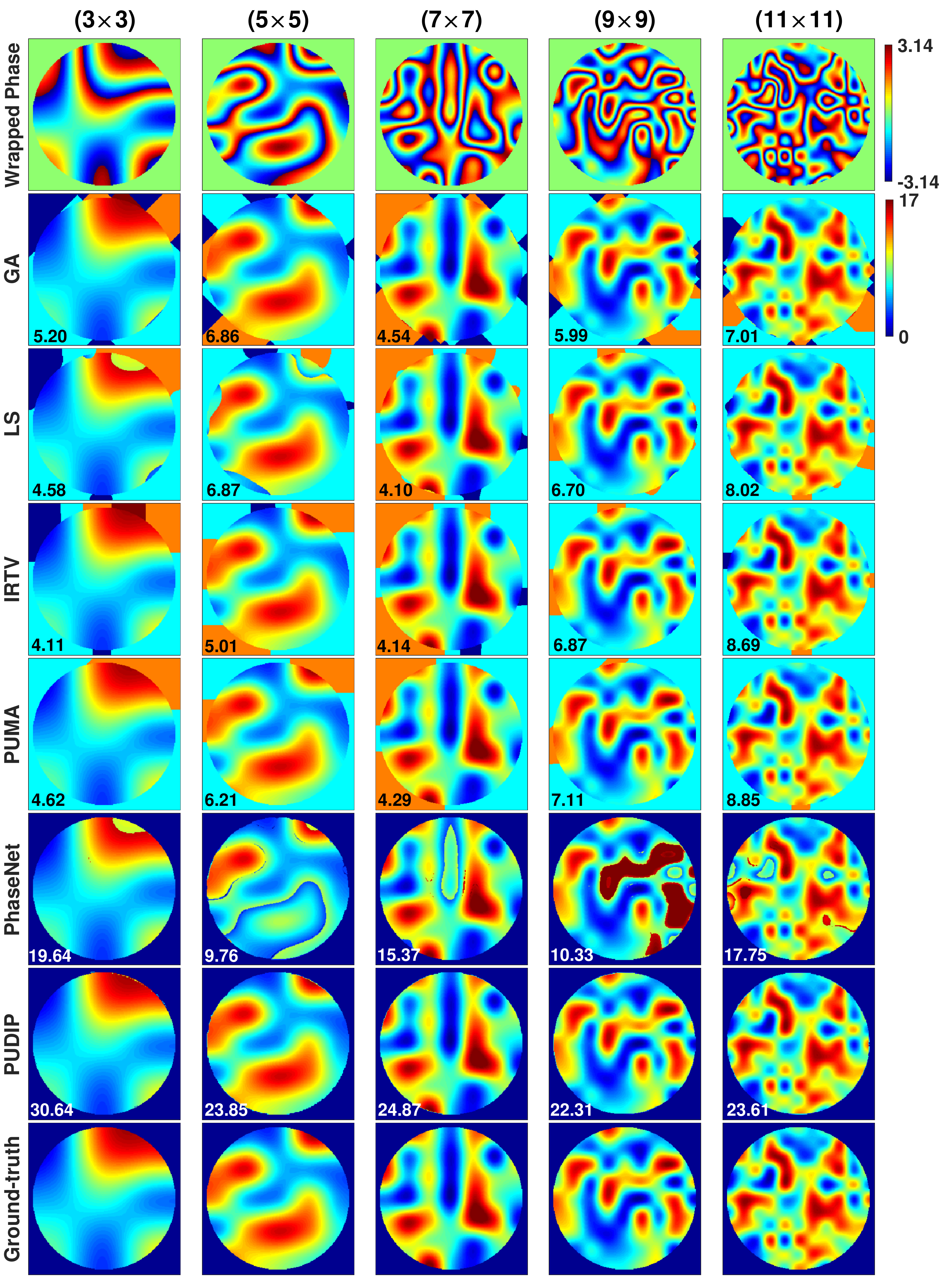}
\caption{Unwrapped-phase images of simulated samples with diverse random distributions (Sample D-2). From top to bottom: wrapped phase, results obtained by  GA, LS, IRTV, PUMA,  PhaseNet, and our approach (PUDIP).  The ground-truth images are presented in the last row. The numbers give the corresponding RSNR [dB].}
\label{fig:simulate3_b}
\end{figure*}

\begin{figure*}[!tbp]
\centering
\includegraphics[width=0.85\textwidth]{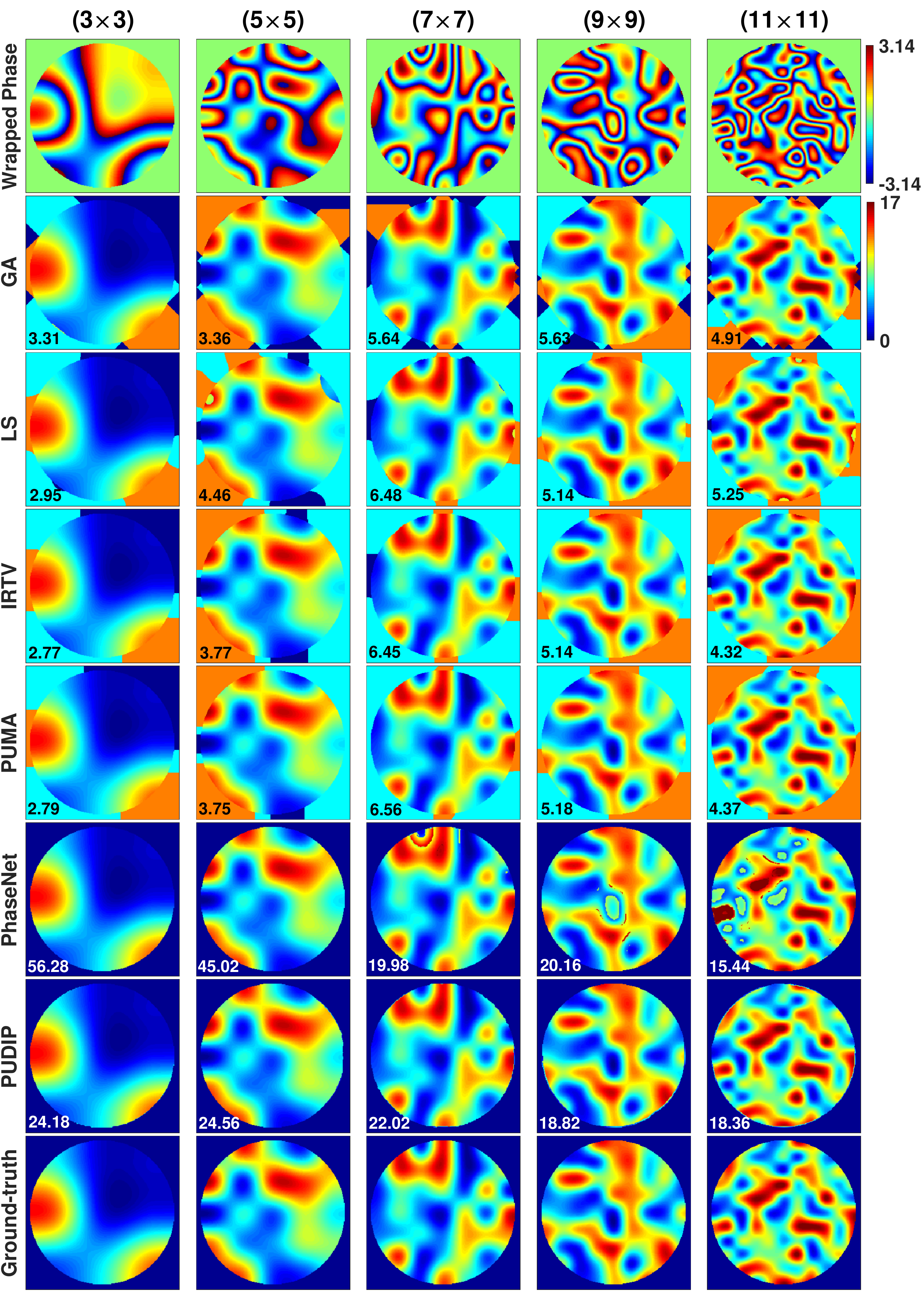}
\caption{Unwrapped-phase images of simulated samples with diverse random distributions (Sample D-3). From top to bottom: wrapped phase, results obtained by  GA, LS, IRTV, PUMA,  PhaseNet, and our approach (PUDIP).  The ground-truth images are presented in the last row. The numbers give the corresponding RSNR [dB].}
\label{fig:simulate3_n1}
\end{figure*}

\begin{figure*}[!tbp]
\centering
\includegraphics[width=0.85\textwidth]{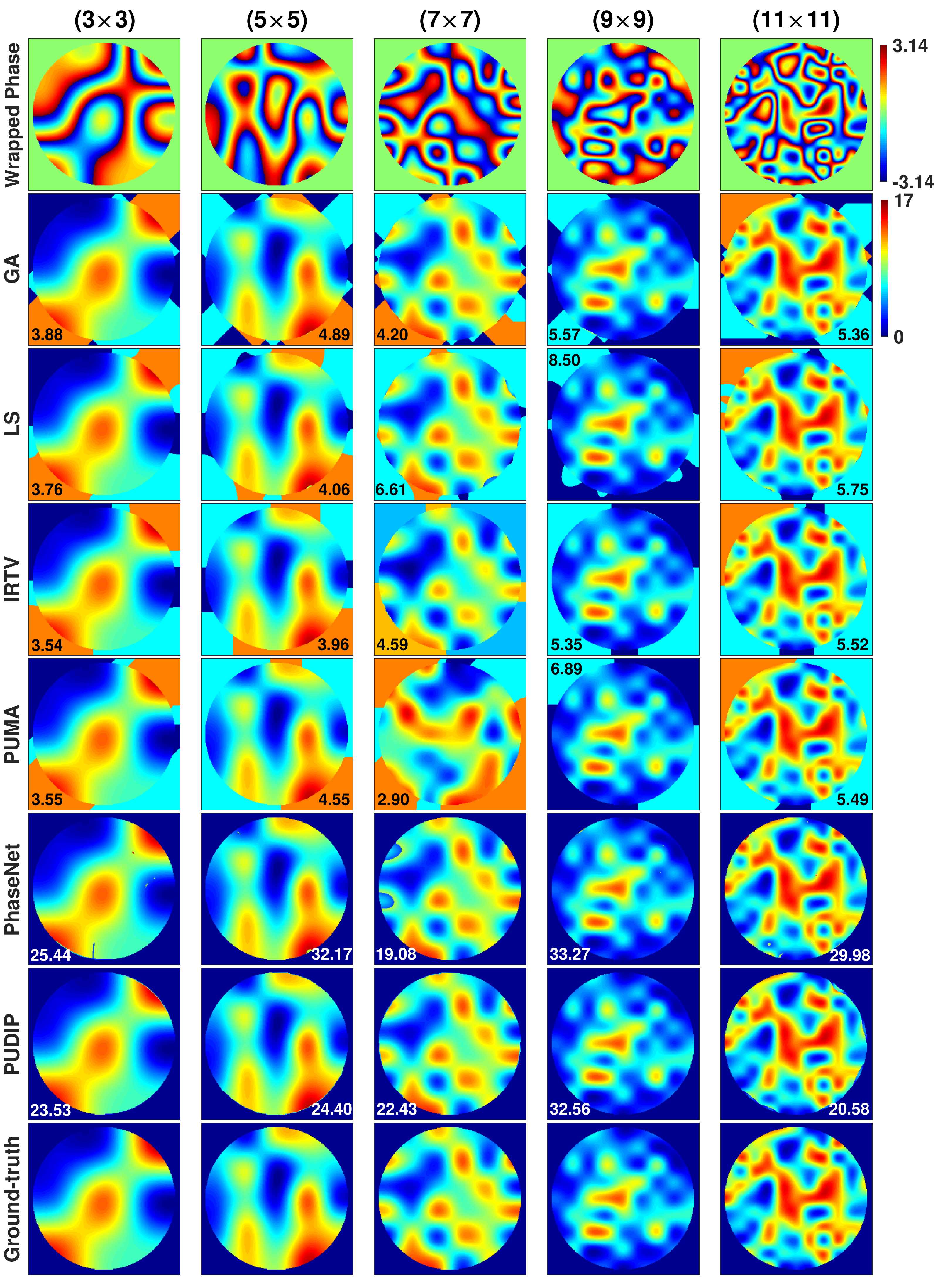}
\caption{Unwrapped-phase images of simulated samples with diverse random distributions (Sample D-4). From top to bottom: wrapped phase, results obtained by  GA, LS, IRTV, PUMA,  PhaseNet, and our approach (PUDIP).  The ground-truth images are presented in the last row. The numbers give the corresponding RSNR [dB].}
\label{fig:simulate3_n2}
\end{figure*}

As the samples are randomly generated, we simulated the sample D four times.
In general, model-based methods fail to restore the correct background in most cases, as well as the inner structures for several samples~(see second to fifth rows of Figs.~\ref{fig:simulate3_g}-\ref{fig:simulate3_n2}).
By contrast, both PhaseNet and PUDIP yield better phase reconstructions for different configurations (see sixth to seventh rows of Figs.~\ref{fig:simulate3_g}-\ref{fig:simulate3_n2}).
PhaseNet is quantitatively more accurate than PUDIP since there is no mismatch between the training and testing sets~(Table~\ref{tb:RSNR3}).
As reported in~\cite{ulyanov2018deep}, supervised schemes tend to outperform unsupervised approaches when the training and testing sets are consistent.
It is noteworthy that PUDIP commits errors only at the border of the disk and that the large discrepancy in the RSNR between PhaseNet and PUDIP mainly comes from the fact that any error is likely to be a multiple of $2\pi$. The SSIM~(Table~\ref{tb:SSIM3}) metric is less sensitive to isolated erroneous cases and the discrepancy is much smaller.
However, for some samples, PhaseNet wrongly estimates the phase over a large area inside the object~(sixth row of Fig.~\ref{fig:simulate3_b}).
Our method is more stable in its ability to unwrap the phase due to its feedback mechanism.

\begin{table}[!tbp]
\setlength{\tabcolsep}{3pt}
\centering
\caption{RSNR [dB] of the reconstructed-phase images versus the size of the random matrix. The metric is averaged over four samples for each size.
For each sample, we repeated five times the reconstructions of our method.
The reported RSNR of PUDIP is then the average of twenty experiments for each size.}
\arrayrulecolor{black}
\begin{tabular} {r | cccccc} \toprule\toprule   
Matrix size  &GA  &LS &IRTV &PUMA &PhaseNet &PUDIP\\
\midrule
($3 \times 3$)   &4.18    &3.84  &3.60  &3.72  &\textbf{36.30}  &21.94  \\
($5 \times 5$)  &5.57    &5.55  &4.87  &5.39  &\textbf{31.89}  &21.51  \\
($7 \times 7$)  &5.32    &6.28   &5.72   &5.41  &\textbf{21.97}  &19.98  \\
($9 \times 9$)  &5.53    &6.19   &5.47   &5.99  &\textbf{39.71}  &20.80  \\
($11 \times 11$)  &5.71    &6.88   &6.82   &6.88  &\textbf{23.63}  &18.65  \\
\bottomrule\bottomrule
\end{tabular}
\label{tb:RSNR3}
\end{table}

\begin{table}[!tbp]
\setlength{\tabcolsep}{3pt}
\centering
\caption{SSIM of the reconstructed-phase images versus the size of the random matrix. The metric is averaged over four samples for each size.
For each sample, we repeated five times the reconstructions of our method.
The reported SSIM of PUDIP is then the average of twenty experiments for each size.}

\arrayrulecolor{black}
\begin{tabular} {r | cccccc} \toprule\toprule   
Matrix size   &GA  &LS &IRTV &PUMA &PhaseNet &PUDIP\\
\midrule
($3 \times 3$)    &0.7361  &0.7222  &0.7253  &0.7065  &\textbf{0.9920}  &0.9699\\ 
($5 \times 5$)    &0.6828  &0.6506  &0.6592  &0.6478  &0.9567  &\textbf{0.9588}  \\
($7 \times 7$)    &0.6636  &0.6495  &0.6348  &0.6403  &\textbf{0.9576}  &0.9530 \\
($9 \times 9$)    &0.6511  &0.7020  &0.6579  &0.6872  &\textbf{0.9637}  &0.9294  \\
($11 \times 11$)  &0.6532  &0.6481  &0.6574  &0.6557  &0.9234  &\textbf{0.9344}  \\
\bottomrule\bottomrule
\end{tabular}
\label{tb:SSIM3}
\end{table}

\section{Reconstructions of Sample E}
\label{subsec:simulateRec4}
\begin{figure*}[!tbp]
\centering
\includegraphics[width=1.0\textwidth]{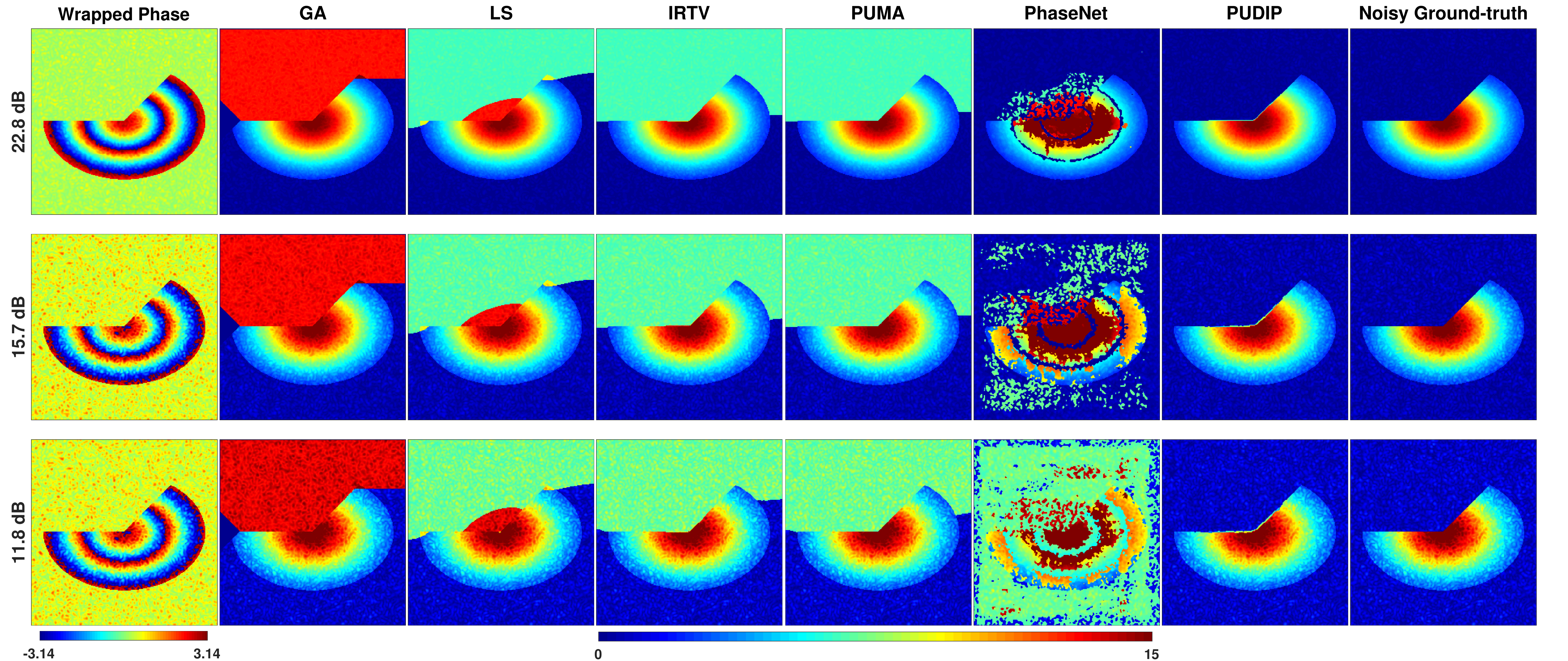}
\caption{Reconstructed unwrapped-phase images of simulated samples with diverse speckle noise (Sample E). From left to right: wrapped phase, results obtained by GA, LS, IRTV, PUMA,  PhaseNet, and our approach (PUDIP). The noisy ground-truth images are presented in the last column.}
\label{fig:simulate_snr}
\end{figure*}

We provide the reconstructions of the sample E in Fig.~\ref{fig:simulate_snr}.
The performances of the baseline methods are affected by the structured noise and fail to correctly unwrap the images~(Tables~\ref{tb:RSNR4} and~\ref{tb:SSIM4}).
PUDIP is stable, in that it correctly unwraps the phase, at the possible exception of few pixels at the border.
It is noteworthy that the robustness to noise is different from denoising, since we do not target at reducing the noise during the unwrapping process. This happens to other methods as well. When unwrapping is successful, one can then denoise the recovered phase image with any state-of-the-art denoising algorithms.

\begin{table}[!tbp]
\setlength{\tabcolsep}{3pt}
\centering
\caption{RSNR [dB] of the reconstructed-phase images versus the noise level. The RSNR of our method (PUDIP) is the average of five experiments.}
\arrayrulecolor{black}
\begin{tabular} {r | cccccc} \toprule\toprule   
Noise level (dB)  &GA  &LS &IRTV &PUMA &PhaseNet &PUDIP\\
\midrule
22.80   &-3.58    &1.67  &2.32  &2.34  &3.24  &\textbf{20.51}  \\
15.70  &-3.21    &2.01  &2.84  &2.72  &0.95  &\textbf{20.94}  \\
11.82  &2.81    &2.36   &3.13   &3.13  &2.45  &\textbf{20.80}  \\
\bottomrule\bottomrule
\end{tabular}
\label{tb:RSNR4}
\end{table}

\begin{table}[!tbp]
\setlength{\tabcolsep}{3pt}
\centering
\caption{SSIM of the reconstructed-phase images versus the noise level. The SSIM of our method (PUDIP) is the average of five experiments.}

\arrayrulecolor{black}
\begin{tabular} {r | cccccc} \toprule\toprule   
SNR (dB)   &GA  &LS &IRTV &PUMA &PhaseNet &PUDIP\\
\midrule
22.80    &0.0072  &0.1488  &0.1619  &0.1638  &-0.2946  &\textbf{0.9895}\\ 
15.70    &0.0081  &0.1260  &0.1493  &0.1502  &-0.1402  &\textbf{0.9913}  \\
11.82    &0.0023  &0.1022  &0.1195  &0.1318  &0.2046  &\textbf{0.9905} \\
\bottomrule\bottomrule
\end{tabular}
\label{tb:SSIM4}
\end{table}

\bibliographystyle{unsrt}  
\bibliography{pudip}
